\documentclass[11pt]{article}
\usepackage[utf8]{inputenc}
\usepackage{amsfonts,amsmath,amsthm,amssymb,mathrsfs,graphicx,bm,verbatim,xcolor,textcomp}
\usepackage{tabularx,lipsum,ltablex,array,makecell,booktabs,caption,multirow,float,wasysym}
\usepackage[hang,flushmargin]{footmisc} 
\usepackage[letterpaper, margin=1.0 in]{geometry}
\usepackage[auth-lg]{authblk} 

\usepackage{lscape}
\usepackage{adjustbox}
\usepackage[american]{babel}
\usepackage[longnamesfirst,authoryear]{natbib}
\cleardoublepage
\normalbaselines 
\usepackage{hyperref,url}
\hypersetup{colorlinks,allcolors=blue}
\usepackage[capitalise]{cleveref}

\graphicspath{{./figures/}}

\newtheorem{theorem}{Theorem}[section]

\newtheorem{lemma}{Lemma}[section]
\theoremstyle{definition}

\theoremstyle{definition}
\newtheorem{remark}{Remark}[section]

\newtheorem{proposition}{Proposition}[section]
\newtheorem*{notation}{Notation}

\numberwithin{equation}{section}
\theoremstyle{definition}
\newtheorem{assumption}{Assumption}
\newtheorem*{assumptionHK}{Assumption A}
\newtheorem*{assumptionSI}{Assumption SI}
\newtheorem*{assumptionSD}{Assumption SD}

\newcommand{\Keywords}[1]{\par\noindent {\small{\em Keywords\/}: #1}} 
\newcommand{\JELclass}[1]{\par\noindent {\small{\em JEL classification Codes\/}: #1}} 

\begin{document}
\title{SEMIPARAMETRIC ESTIMATION OF DYNAMIC BINARY CHOICE PANEL DATA MODELS\thanks{We thank the Editor (Peter C.B. Phillips), the Co-Editor (Iv\'{a}n Fern\'{a}ndez-Val), and the two anonymous referees for their valuable comments and suggestions, which have significantly improved the quality of this paper.  We are grateful to Yonghong An, Shakeeb Khan, Arthur Lewbel, Takuya Ura, Hanghui Zhang, and Yichong Zhang for their insightful feedback and discussions. We also thank participants at the 2019 Asian Meeting of the Econometric Society, the 2019 Shanghai Workshop of Econometrics, and the 2022 Australasia Meeting of the Econometric Society for their helpful comments. Fu Ouyang acknowledges the financial support provided by the Faculty of Business, Economics, and Law (BEL) at the University of Queensland through the 2019 BEL New Staff Research Start-Up Grant. All errors are our responsibility. \smallskip}}

\author[1]{{\large Fu Ouyang}\thanks{Email: \href{mailto:f.ouyang@uq.edu.au}{f.ouyang@uq.edu.au}.}}
\author[2]{{\large Thomas Tao Yang}\thanks{Email: \href{mailto:tao.yang@anu.edu.au}{tao.yang@anu.edu.au}.}}
\affil[1]{School of Economics, University of Queensland}
\affil[2]{Research School of Economics, Australian National University}
\date{\today}
\maketitle

\begin{abstract}
\noindent
We propose a new approach to the semiparametric analysis of panel data binary choice models with fixed effects and dynamics (lagged dependent variables). The model under consideration has the same random utility framework as in \cite*{HonoreKyriazidou2000}. We demonstrate that, with additional serial dependence conditions on the process of deterministic utility and tail restrictions on the error distribution, the (point) identification of the model can proceed in two steps, and requires matching only the value of an index function of explanatory variables over time, rather than the value of each explanatory variable. Our identification method motivates an easily implementable, two-step maximum score (2SMS) procedure -- producing estimators whose rates of convergence, in contrast to \citeauthor{HonoreKyriazidou2000}'s (\citeyear{HonoreKyriazidou2000}) methods, are independent of the model dimension. We then analyze the asymptotic properties of the 2SMS procedure and propose bootstrap-based distributional approximations for inference. Evidence from Monte Carlo simulations indicates that our procedure performs satisfactorily in finite samples.
\par \bigskip
\JELclass{C14, C23, C35.}
\medskip
\Keywords{Semiparametric estimation; Binary choice model; Panel data; Fixed effects; Dynamics; Maximum score; Bootstrap}
\end{abstract}
\par\bigskip

\section{Introduction}

In this paper, we propose a two-step estimation method for panel data binary
choice models with fixed effects and dynamics. Specifically, we consider
binary choice models of the form
\begin{equation}
y_{it}=1\left[  x_{it}^{\prime}\beta+\gamma y_{it-1}+\alpha_{i}-\epsilon
_{it}>0\right]  ,i=1,...,n,t=1,...,T,\footnote{The identification approach and
estimation method presented in this paper can be applied to models with
unbalanced panels as long as the
unbalancedness is not the result of endogenous
attrition.} \label{eq:1.1}%
\end{equation}
where $T$ is small and $n$ is large, $x_{it}$ is a $K\times1$ vector of
(time-varying) explanatory variables,\footnote{Any time-invariant covariates
can be thought of as being part of the fixed effect $\alpha_{i}$.} $y_{it-1}$
is the lagged dependent variable, $\alpha_{i}$ represents a time-invariant,
individual-specific (fixed) effect, and $\epsilon_{it}$ is an idiosyncratic
error term. Both $\alpha_{i}$ and $\epsilon_{it}$ are unobservable to the
econometrician. Following \cite*{HonoreKyriazidou2000} (referred to as HK henceforth), we assume the strong exogeneity that $(x_{i1},...,x_{iT})\perp(\epsilon_{i1},...\epsilon_{iT})|\alpha_i$ and that $\epsilon_{it}$ are independent and identically distributed (i.i.d.) across $t$ conditional on $\alpha_i$. Interest centers on estimating the preference parameter
$\theta\equiv(\beta^{\prime},\gamma)^{\prime}$. $y_{i0}$ is assumed to be
observed, although the model is not specified in the initial period 0. In the
literature, the lagged terms $y_{it-1}$ and the fixed effect $\alpha_{i}$ are referred
to as the ``state dependence'' (see
\cite*{Heckman1981b, Heckman1981a}) and the \textquotedblleft unobservable
heterogeneity\textquotedblright, respectively. The co-existence of these two
terms complicates the identification and estimation of $\theta$, owing to the
multiple sources of persistence in $y_{it}$.

This paper resembles other panel data discrete response literature using fixed
effects methods, in that we impose no restrictions on the
distribution of $\alpha_{i}$, conditional on the observed explanatory
variables. \cite{ArellanoHonore2001} review early works on estimating $\beta$ in model (\ref{eq:1.1}) with no state dependence ($y_{it-1}$). \cite*{Chamberlain2010} shows that, outside of the logistic case, these static binary choice models have a zero information bound, and the identification requires that at least one of the observed covariates have unbounded support. In the presence of lagged dependent variables, various conditional maximum likelihood methods have been developed for variants of model (\ref{eq:1.1}) with logistic errors and at least four observations ($T\geq3$) per individual.\footnote{Throughout this paper, this means the data contain $y_{i0}$ and $(y_{i1},y_{i2},y_{i3},x_{i1},x_{i2},x_{i3})$ for each individual $i$.} \cite{honore2021identification} provide a comprehensive review of this literature.\footnote{Works such as \cite{BartolucciNigro2010, BartolucciNigro2012} and \cite{al2017exponential} study the estimation of model (\ref{eq:1.1}) under alternative specifications.} Several new methods have been proposed for dynamic Logit models based on moment conditions. Leading examples include \cite{HonoreWeidner2020}, \cite{dobronyi-gu-kim}, \cite{kitazawa2022transformations}, and \cite{dano2023transition}, among others.

HK is the first to consider the semiparametric identification and estimation
of model (\ref{eq:1.1}). They demonstrate that $\theta$ can be identified
if, in addition to assumptions analogous to those in \cite*{Manski1987}, all
explanatory variables are strictly exogenous, $\epsilon_{it}$'s are serially
independent, and $T\geq3$. However, the rate of convergence of their estimator decreases as the number of continuous
regressors increases, and is slower than the standard MS rate
derived by \cite*{KimPollard1990}.

There are several alternative fixed effects approaches to the semi- and nonparametric analysis of dynamic binary choice models. \cite{HonoreLewbel2002} propose an identification strategy that requires an exclusion restriction (excluded regressor). \cite{ChenEtal2019} show that the exclusion restriction in \cite{HonoreLewbel2002} implicitly assume (conditional) serial independence of the excluded regressor. \cite{Williams2019} studies the nonparametric identification of dynamic binary choice models that satisfy certain exclusion restrictions. In the absence of excluded regressors, some recent works, such as \cite{KhanEtal2020} and \cite{Aristodemou2021}, characterize the (sharp) identified set for $\theta$ under mild conditions. We refer interested readers to the survey article by \cite{honore2021identification} and Chapter 7 in \cite{Hsiaobook} for a detailed review of this literature.

This paper takes one step in the direction of HK's semiparametric estimator, in the sense that we provide
sufficient conditions under which model (\ref{eq:1.1}) can be identified and
estimated, without needing to match each of the explanatory variables
over time, provided we have at least five observations per individual are
observed (i.e., $T\geq4$).\footnote{That is, at least $y_{i0}$ and $(y_{i1},y_{i2},y_{i3},y_{i4},x_{i1},x_{i2},x_{i3},x_{i4})$ are observed for each individual $i$. This is a restriction on the minimum panel length, which is satisfied for many longitudinal panel data sets.} The key insight here is that the identification
of $\theta$ can proceed in two steps. First, $\beta$ can be identified based
on sequences of $\left\{y_{it}\right\}  $, for which $y_{is-1}=y_{it-1}$ and
$y_{is+1}=y_{it+1}$ for some $1\leq s< t\leq T-1$ with $t\geq s+2$ (e.g., in the simplest case where $T=4$, $\beta$ is identified based on observations with $y_0=y_2=y_4$), if the
distribution of explanatory variables $x_{it}$ satisfies certain serial
dependence and stochastic dominance restrictions. Then, using the identified
$\beta$, $\gamma$ can be identified by simply matching
$x_{it}^{\prime}\beta$ over time. We propose an estimation procedure for $\beta$ and $\gamma$, establish the asymptotics for our estimators, and provide an inference method that uses the bootstrap. We investigate their finite-sample
properties using Monte Carlo experiments.

As demonstrated by \cite*{HonoreTamer2006}, matching exogenous utilities over
time seems to be essential for the point identification in ``distribution-free'' dynamic discrete choice
models.\footnote{More precisely,
\cite*{HonoreTamer2006} provided examples of point identification failure when it
is impossible to match $x_{it}$ over time.} However, the approach proposed here involves matching an
identified linear combination of $x_{it}$, rather than HK's matching each
component of $x_{it}$. Consequently, in contrast to the results presented by
HK, the rates of convergence of our proposed estimators are independent of the
dimension of the regressor space, making our approach
particularly useful for models with a higher dimensional design.

It is known that panel data binary choice models with unobserved
heterogeneity and dynamics can be estimated using the random effects or
correlated random coefficients approach. Examples include
\cite*{ArellanoCarrasco2003}, \cite*{Wooldridge2005}, and
\cite*{HonoreTamer2006}.
In addition to preference parameters, these approaches often allow the
econometrician to calculate other quantities of interest, such as choice
probabilities and marginal effects. However, these approaches require the
specification of the statistical relation between the explanatory variables
and $\alpha_{i}$. Further, they require one to specify the distribution
of $y_{i0}$, conditional on the observed explanatory variables and $\alpha
_{i}$, which raises the so-called initial condition problem. Conversely, the
fixed effects approaches attempt to estimate preference parameters without
making these subtle specifications. Finally, there is also literature exploring the identification and
estimation of various partial effects in panel data models; see, for example,
\cite*{AltonjiMatzkin2005}, \cite*{ChernozhukovEtal2013}, and more recent advancements by
\cite*{Torgovitsky2019}, \cite*{aguirregabiria2021identification}, \cite{dobronyi-gu-kim}, \cite*{davezies2021identification}, and \cite*{liu2021identification}, among others.

Dynamic binary choice models have a wide range of applications, including the study of labor force
participation (\cite*{DamrongplasitEtal2018}), poverty dynamics (\cite*{Biewen2009}),
health status (\cite*{Halliday2008}),
educational attainment (\cite*{CameronHeckman1998, CameronHeckman2001}), stock
market participation (\cite*{AlessieEtal2004}), brand
loyalty (\cite*{ChintaguntaEtal2001}), welfare participation (\cite*{ChayEtal1999}), and
firm behavior (\cite*{KerrEtal2014}), among others. Most
applications typically employ parametric forms of the model
(\ref{eq:1.1}), such as Logit and Probit, or random effects assumptions. The
robustness from the distribution-free and fixed effects specification makes
the approach proposed here a competitive alternative to existing
parametric and random effects methods. Note that the theoretical validity of our approach relies on certain restrictions on the serial dependence of explanatory variables. Thus, before applying our method, we suggest that applied researchers detrend and seasonally adjust $x_{it}$ in a way that makes them resemble ``white noise'' conditional on $\alpha_i$.\footnote{Monte Carlo results show that relaxing these restrictions does not significantly affect our estimators' finite-sample performances; see Section \ref{sec_simulation} for a more detailed discussion.}

The remainder of this paper is organized as follows. Section
\ref{sec_identification} establishes the identification of $\theta$ under
different sets of sufficient conditions, based on which, a two-step maximum score (2SMS) procedure is proposed in Section \ref{sec_estimation}. Sections \ref{sec_asymptotics} and
\ref{SEC:inferences} derive the asymptotic properties of the 2SMS estimator and
propose bootstrap-based inference methods, respectively. We present the results of Monte
Carlo experiments in Section \ref{sec_simulation} that examine the
finite-sample performance of our proposed method. Section
\ref{sec_conclusion} concludes the paper. We prove the main theorems and present the main simulation results in the \hyperref[appendix]{Appendices}. The Online Supplement to this paper includes proofs of all technical lemmas, technical details for the bootstrap inference, and results for supplementary simulation studies.

For ease of reference, we next describe the notation maintained throughout this paper.

\begin{notation}
All vectors are column vectors. $\mathbb{R}^{p}$ is a $p$-dimensional Euclidean space equipped with the
Euclidean norm $\Vert\cdot\Vert_{2}$. We reserve the letter $i\in\mathcal{N}\equiv\{1,...,n\}$ for indexing individuals,
and the letters $s,t\in\mathcal{T}\equiv\{1,...,T\}$ for indexing time periods. An
observation is indexed by $(i,t)$. Vector $x_{its}$ denotes $x_{it}-x_{is}$.
The first element of $x_{its}$ is denoted by $x_{its,1}$ and the sub-vector
comprising its remaining elements is denoted by $\tilde{x}_{its}$. As is common in the panel data literature, we use the notation $\xi^{t}$ to denote $\left(
\xi_{1}^{\prime},...,\xi_{t}^{\prime}\right)  ^{\prime}$. For example,
suppressing the subscript $i$, $y^{t}\equiv(y_{1},...,y_{t})^{\prime},$ a $t\times1$
vector. $F_{\zeta|\cdot}$ and $f_{\zeta|\cdot}$ denote, respectively, the
conditional cumulative distribution function (CDF) and probability density
function (PDF) of a random vector $\zeta$ conditional on $\cdot$. For two
random vectors, $u$ and $v$, the notation $u\overset{d}{=}v|\cdot$ means that
$u$ and $v$ have identical distributions, conditional on $\cdot$, and $u\perp
v|\cdot$ means that $u$ and $v$ are independent, conditional on $\cdot$. We use
$P(\cdot)$ and $\mathbb{E}[\cdot]$ to denote probability and expectation,
respectively. Function $1[\cdot]$ is an indicator function, equal to one
when the event in the brackets is true, and zero otherwise. Function
$sgn(\cdot)$ denotes the sign function, equal to 1 when $\cdot$ is positive,
0 when $\cdot$ is 0, and $-1$ when $\cdot$ is negative$.$ Symbols $\setminus$,
$^{\prime}$, $\propto$, $\Leftrightarrow$, $\overset{d}{\rightarrow}$, and
$\overset{P}{\rightarrow}$ represent set difference, matrix transposition,
proportionality, ``if and only if'',
convergence in distribution, and convergence in probability, respectively. For
any (random) positive sequences, $\{a_{n}\}$ and $\{b_{n}\}$, $a_{n}=O(b_{n})$
($O_{P}(b_{n})$) means that $a_{n}/b_{n}$ is bounded (bounded in probability),
and $a_{n}=o(b_{n})$ ($o_{P}(b_{n})$) means that $a_{n}/b_{n}\rightarrow0$
($a_{n}/b_{n}\overset{P}{\rightarrow}0$).
\end{notation}

\section{Identification}\label{sec_identification} 
This section provides sufficient conditions for
identifying the parameter $\theta$ with no need to match observed
covariates $x_{it}$ over time. Under these assumptions, we derive a set of
identification inequalities that can be taken to data for (point) estimation
and inference on the parameter $\theta$.

To simplify the notation, we suppress
the subscript $i$ in the rest of this paper whenever it is clear from the
context that all variables relate to each individual. Suppose that a random sample from a population of independent individuals\footnote{Here, the term ``independent individuals'' refers to the assumption that $(\alpha_i, y_{i0}, x_{i1},...,x_{iT}, \epsilon_{i1},...,\epsilon_{iT})$ is independently distributed across $i$.} is
observed for $T+1$ ($=|\mathcal{T}\cup\{0\}|$) periods. Recall that, for all
$t\in\mathcal{T}$,
\begin{equation}
y_{t}=1\left[  x_{t}^{\prime}\beta+\gamma y_{t-1}+\alpha-\epsilon
_{t}>0\right]  .\label{model}%
\end{equation}
Note that the model is incomplete, in the sense that it does not specify the
relationship between $y_{0}$ and $(x^{T},\alpha,\epsilon^{T})$. This is known
as the initial condition problem in panel data literature. This paper uses
a fixed effects approach, in which we attempt to estimate $\theta = (\beta',\gamma)'$ without
making any assumptions on the distribution of $\alpha$, conditional on
explanatory variables. This helps us to avoid explicitly specifying the
functional form of $p_{0}(x^{T},\alpha)\equiv P(y_{0}=1|x^{T},\alpha)$, and thus 
circumvents the initial condition problem.

As mentioned, we impose no restriction on $F_{\alpha|x^{T}}$, but place
the following restrictions on observed covariates $x^{T}$ and unobserved
idiosyncratic errors $\epsilon^{T}$:

\begin{assumptionHK}\label{Assumption:HK}
For all $\alpha$ and $s,t\in\mathcal{T}$,

\begin{enumerate}
\item[(a)] (i) $\epsilon^{T}\perp(x^{T},y_{0})|\alpha$, (ii) $\epsilon
_{s}\perp\epsilon_{t}|\alpha$, and (iii) $\epsilon_{s}\overset{d}{=}%
\epsilon_{t}|\alpha$.

\item[(b)] $F_{\epsilon_{t}|\alpha}$ is absolutely continuous with PDF
$f_{\epsilon_{t}|\alpha}$ and support $\mathbb{R}$.

\item[(c)] (i) One of the regressors, without loss of generality (w.l.o.g.)
$x_{ts,1},$ has almost everywhere (a.e.) positive probability density on
$\mathbb{R}$, conditional on $\tilde{x}_{ts}$ and $\alpha$, and (ii) the
coefficient $\beta_{1}$ on $x_{ts,1}$ is nonzero.

\item[(d)] The support $\mathcal{X}_{ts}$ of $F_{x_{ts}|\alpha}$ is not
contained in any proper linear subspace of $\mathbb{R}^{K}$.

\item[(e)] $\theta=(\beta^{\prime},\gamma)^{\prime}\in\mathcal{B}%
\times\text{int}(\mathcal{R})$, where $\mathcal{B}\equiv\{b=(b_{1}%
,...,b_{K})^{\prime}\in\mathbb{R}^{K}|\Vert b\Vert_{2}=1\}$ and $\mathcal{R}$
is a compact subset of $\mathbb{R}$ with a non-empty interior.
\end{enumerate}
\end{assumptionHK}

Assumption \hyperref[Assumption:HK]{A} places the same set of restrictions on
the joint distribution of $(x^{T},\alpha,\epsilon^{T})$ as in HK. While not stated
explicitly in their Theorem 4, HK use Assumption
\hyperref[Assumption:HK]{A}(a), the exogeneity of $(x^{T},y_{0})$ and serial
independence of $\{\epsilon_{t}\}$, conditional on $\alpha$, to derive the
moment inequalities for the identification. Note that Assumption
\hyperref[Assumption:HK]{A}(a) implies that the fixed effects $\alpha$ pick
up two types of dependence in the model: the dependence over time in the
unobservables, and the dependence between explanatory variables and
unobservables. As a result, in model (\ref{model}), $\epsilon_{t}$ is
independent of $\left(  x^{T},y^{t-1}\right)  $, conditional on $\alpha$.
Furthermore, Assumption \hyperref[Assumption:HK]{A}(a) is a special case of the
group homogeneity restriction, $\epsilon_{s}\overset{d}{=}\epsilon_{t}|(x_{s},x_{t},\alpha)
$, imposed in \cite*{Manski1987},
\cite*{PakesPorter2016}, and \cite*{ShiEtal2018} for identifying static discrete choice models (without controlling
the lagged term $y_{t-1}$ in the model). Thus, we can suppress
the time subscript $t$ in $F_{\epsilon_{t}|\alpha}$ and $f_{\epsilon
_{t}|\alpha}$ in the rest of this paper without ambiguity. Assumption
\hyperref[Assumption:HK]{A}(b) is a regularity condition that ensures that both
$y_{s}\neq y_{t}$ and $y_{s}=y_{t}$ occur with positive probabilities for all
$\alpha$ and $s,t\in\mathcal{T}$.

It is known and documented in the relevant literature (see, e.g., Lemma 1 of \cite*{Manski1985}) that to
establish the point identification of the parameter $\theta$ in a distribution-free setting, $x_{t}$ also needs to satisfy certain
regularity conditions. Assumption \hyperref[Assumption:HK]{A}(c) requires the
existence of a relevant, continuous regressor, with large support, which is a
standard restriction imposed in MS-type estimators. Assumption
\hyperref[Assumption:HK]{A}(d) is the familiar full-rank condition.
Assumptions \hyperref[Assumption:HK]{A}(c) and \hyperref[Assumption:HK]{A}(d) are identical to Assumption 2 of \cite*{Manski1987}.

Assumption \hyperref[Assumption:HK]{A}(e) is for scale normalization and
parameter space. This is a typical practice for discrete choice models,
because the identification of $\theta$ is only up to scale. In the
semiparametric framework, where no parametric form of $F_{\epsilon|\alpha}$ is
specified, identification is often achieved by normalizing the magnitude of
the regression coefficients. Assumption \hyperref[Assumption:HK]{A}(e)
assumes that $\beta$ is on the unit circle and has a nonzero first element
$\beta_{1}$.\footnote{Our procedure identifies
$\beta$ and $\gamma$ sequentially, so it is more convenient to normalize the
scale of $\beta$, rather than that of $\theta$, as in HK.}

HK demonstrate that, if $T\geq3$ and $x_{t}$ has time-varying overlap support, $\theta$ can be identified under
Assumption \hyperref[Assumption:HK]{A}.\footnote{As is stated in HK, Assumption \hyperref[Assumption:HK]{A} is not sufficient
for point identifying $\theta$ if $T<3$.} Their proposed approach requires
matching all exogenous covariates over time, and results in an estimator with
a rate that declines as the number of exogenous covariates increases. The main
contribution of this paper is the provision of a set of supplementary
conditions, under which the identification of $\theta$ can escape from the
necessity of element-by-element matching. Specifically, our approach is based
on the following monotonic relationship between a conditional choice
probability and an index of the exogenous covariates: For some $s,t\in
\mathcal{T}$ such that $t-s\geq2$,
\begin{align}
P(y_{t}=1|x_{s},x_{t},y_{s-1}=y_{t-1},y_{s+1}=y_{t+1},\alpha)  &  \geq
P(y_{s}=1|x_{s},x_{t},y_{s-1}=y_{t-1},y_{s+1}=y_{t+1},\alpha)\nonumber\\
&  \Leftrightarrow\nonumber\\
x_{t}^{\prime}\beta &  \geq x_{s}^{\prime}\beta.\label{idinequality}%
\end{align}
Note that (\ref{idinequality}) requires that there be at least
five ($T\geq4$) observations per individual observed by the econometrician  (i.e.,
$s=1$, $t=3$, and $s+1=t-1=2$). In the simplest case with $T=4$,  (\ref{idinequality}) reduces to $P(y_{3}=1|x_{1},x_{3},y_{0}=y_{2}=y_{4},\alpha)    \geq
P(y_{1}=1|x_{1},x_{3},y_{0}=y_{2}=y_{4},\alpha)$ if and only if $x_{3}^{\prime}\beta   \geq x_{1}^{\prime}\beta$.

Result (\ref{idinequality}) states that the indices $x_{s}'\beta$ and $x_{t}'\beta$
rank order the (conditional) probabilities of choosing 1 in periods
$s$ and $t$. To ensure this, conditioning on $y_{s-1}=y_{t-1}$
is obviously necessary. However, as $t-1>s$, $y_{s}$ affects $y_{t-1}$ through the dynamics of the model (specifically, through
the chain $y_{s}\rightarrow y_{s+1}\rightarrow\cdots\rightarrow y_{t-1}$).
Accordingly, we need to include $y_{s+1}$ in the conditioning set
 to cut off such state dependence, and further
impose the restriction $y_{s+1}=y_{t+1}$ to make the events $\{y_{s}=1\}$
and $\{y_{t}=1\}$ have symmetric conditioning sets. Only with
this symmetry can we invoke time stationarity restrictions
on $x_{t}$ to establish the equivalence in (\ref{idinequality}).

In particular, to reach (\ref{idinequality}), we also need to address the following two concerns. First, $x_{s}$ ($x_{t}$) may affect the value of
$y_{t+1}$ ($y_{s+1}$) via its serial dependence on $x_{t+1}$ ($x_{s+1}$).
Second, the dependence between $x_{t}$ and $y_{t+1}$ (via $x_{t+1}$) may
change dramatically over time. Both require additional restrictions to be
placed on the serial dependence of the stochastic process of $x_{t}$. Otherwise, as shown in Appendix \ref{Sec_identification}, $x_t'\beta$ is not the unique factor that can rank order the choice probabilities in (\ref{idinequality}).

The following condition, together with Assumption \hyperref[Assumption:HK]{A}, is sufficient to establish (\ref{idinequality}), as shown in Appendix \ref{Sec_identification}.
\begin{assumptionSI}\label{Assumption:SI}
For all $s,t\in\mathcal{T}$ such that $s\neq t$, (a) $x_{s}\perp x_{t}| \alpha$, and
(b) $x_{s}\overset{d}{=}x_{t}|\alpha$.
\end{assumptionSI}

In Appendix \ref{Sec_identification}, we first prove (\ref{idinequality}) under Assumptions \hyperref[Assumption:HK]{A} and \hyperref[Assumption:SI]{SI} for
a special case of model (\ref{model}) with $T=4$ and $\gamma<0$. This serves
as a roadmap to help readers understand the main ideas. The same
arguments can be applied analogously to prove the most general
case; see Lemma \ref{Lemma:A4}.

Assumption \hyperref[Assumption:SI]{SI} imposes a strong
restriction on the dynamic process of the covariate sequence, which requires
the process $\{x_{t}\}$ to be serially independent and strictly stationary,
conditional on the individual-specific effects $\alpha$. In a dynamic fixed effects model,
$\alpha$ collects all time-invariant covariates, as well as
unobserved individual preferences, abilities, or character traits. In such
models, if $x_{t}$ includes only observed individual
characteristics naturally correlated with $\alpha$, it may be reasonable
to further assume that the serial dependence in the process $\{x_{t}%
\}$ is also derived from $\alpha$. Assumption \hyperref[Assumption:SI]{SI} implies that we cannot accommodate time trends. We do allow
random time effects $\lambda_t$ that satisfy $(\lambda_{s},x_{s})\perp(\lambda
_{t},x_{t})|\alpha$, and $(\lambda_{s},x_{s})\overset{d}{=}(\lambda_{t}%
,x_{t})|\alpha$.\footnote{We can apply essentially the same arguments used in Appendix \ref{Sec_identification} to establish a monotonic relationship analogous to (\ref{idinequality}) for index $\lambda+x_t'\beta $, and identify $\beta$ and $\gamma$ similarly.}

If $x_{t}$
contains covariates related to some institutional factors that lead to
exogenous variation in, for example, costs of participation, across individuals,
Assumption \hyperref[Assumption:SI]{SI} may be approximately satisfied by
using the differencing, demeaning, or de-trending transformation of these
variables. This applies to cases where $\{x_{t}\}$ exhibits some
long-run equilibrium (trend, deterministic or stochastic). The transformed
regressor then measures the deviation of $x_{t}$ from its long-run
equilibrium (trend), which, in some cases, might be assumed
to be a white noise process affecting the short-run dynamics of the
model.\footnote{For example, consider a case where $\{x_{t}\}$ is a
random-walk-plus-drift process (i.e.,
$x_{t}=x_{0}+a_{0}t+\sum_{\tau=1}^{t}e_{\tau}$). Although $\{x_{t}\}$ violates
Assumption SI, its first differencing $\Delta x_{t}=x_{t}-x_{t-1}=a_{0}+e_{t}$
is i.i.d. over time.} Note that such variable transformation may involve model reparameterization.  For example, when model (\ref{model}) is the reduced form derived from some structural model, one would need to first reparameterize the structural model accordingly, so that the estimates of the coefficient vector in model (\ref{model}) with transformed covariates can be interpreted in a meaningful way.

HK assume that the support of $x_{t}$ is
overlapping over time, so the differences in the regressors across different time
periods have a positive density in a neighborhood of zero. However, evidence presented in
\cite*{HonoreTamer2006} implies that some additional assumption is needed to achieve point identification without performing an element-by-element match, as in HK. Indeed, Assumption \hyperref[Assumption:SI]{SI} is the extra condition needed for our approach, compared with the semiparametric estimator in HK.

Under Assumptions \hyperref[Assumption:HK]{A} and \hyperref[Assumption:SI]{SI},  the identification of
$\theta$ proceeds in two steps. Proposition \ref{Lemma:beta} demonstrates that
$\beta$ can be identified based on moment inequality (\ref{idinequality}), and
Proposition \ref{Lemma:gamma} establishes the identification of $\gamma$ by
matching the value of the index function $x_{t}^{\prime}\beta$ in different periods.

\begin{proposition}
[Identification of $\beta$]\label{Lemma:beta} Assume $T \geq 4$. For all $s,t\in\mathcal{T}$ such
that $t\geq s+2$, define
\begin{align*}
Q_{1}(b) =  &  \mathbb{E}\left\{  [P(y_{t}=1|x_{s},x_{t},y_{s-1}%
=y_{t-1},y_{s+1}=y_{t+1})- P(y_{s}=1|x_{s},x_{t},y_{s-1}=y_{t-1}%
,y_{s+1}=y_{t+1})]\right. \\
&  \left.  \times sgn(x_{ts}^{\prime}b)\right\}  . \label{pobjI}%
\end{align*}
Suppose Assumptions \hyperref[Assumption:HK]{A} and \hyperref[Assumption:SI]{SI} hold. Then, $Q_{1}(\beta)>Q_{1}(b)$, for all $b\in\mathcal{B}%
\setminus\{\beta\}$.
\end{proposition}

The proof of Proposition \ref{Lemma:beta} can be found in Appendix
\ref{Sec_identification}. Note that our
identification strategy for $\beta$ requires $T=4$, as a minimum. In this
case, $t=s+2$ must hold with $s=1$ and $t=3$, and thus $Q_{1}(b)$ is
\begin{equation}
Q_{1}(b)=\mathbb{E}\{[P(y_{3}=1|x_{1},x_{3},y_{0}=y_{2}=y_{4})-P(y_{1}%
=1|x_{1},x_{3},y_{0}=y_{2}=y_{4})]\cdot sgn(x_{31}^{\prime}b)\}.
\label{popobj:betaT4}%
\end{equation}

Proposition \ref{Lemma:beta} establishes the identification of $\beta$, which
enables us to identify $\gamma$, with $\beta$ being treated as a
known, constant vector. Then, the following proposition shows that $\gamma$
can be identified by matching the deterministic utility $w_{t}\equiv x_{t}^{\prime
}\beta$ in different periods; the proof is presented in Appendix
\ref{Sec_identification}. Because the key idea for identifying $\gamma$ uses the
insight of Section 4.1 in HK, in what follows, we keep the notation as close to that of
HK as possible.

We define the event
\[
A=\{y_{0}=d_{0},...,y_{s-1}=d_{s-1},y_{s}=0,y_{s+1}=d_{s+1},...,y_{t-1}%
=d_{t-1},y_{t}=1,y_{t+1}=d_{t+1},...,y_{T}=d_{T}\},
\]
and its counterpart
\[
B=\{y_{0}=d_{0},...,y_{s-1}=d_{s-1},y_{s}=1,y_{s+1}=d_{s+1},...,y_{t-1}%
=d_{t-1},y_{t}=0,y_{t+1}=d_{t+1},...,y_{T}=d_{T}\},
\]
where $d_{\tau}\in\{0,1\}$, for $0\leq\tau\leq T$. Note that $y$ takes the
same values other than at time periods $(s,t)$ for $A$ and $B$: $y$ switches
from $0$ to $1$ at time $s$ and $t$, respectively, for $A$, and $y$ switches from $1$ to
$0$ at time $s$ and $t$, respectively, for $B$.

We have two cases, based on whether $s$ and $t$ are adjacent. When $s$ and $t$ are adjacent ($t=s+1$), we define the objective function
\[
Q_{2}(r;\beta)=\mathbb{E}\left\{  \left[  P(A|x^{T},w_{t}=w_{t+1}%
)-P(B|x^{T},w_{t}=w_{t+1})\right]  sgn\left(  (w_{t}-w_{t-1})+r(d_{t+1}%
-d_{t-2})\right)  \right\}  .
\]
For the case where $s$ and $t$ are not adjacent ($t>s+1$), we define the objective function%
\begin{align*}
\tilde{Q}_{2}(r;\beta)= &  \mathbb{E}\left\{  \left[  P(A|x^{T},w_{s+1}%
=w_{t+1},y_{s+1}=y_{t+1})-P(B|x^{T},w_{s+1}=w_{t+1},y_{s+1}=y_{t+1})\right]
\right.  \\
&  \left.  \times sgn\left(  (w_{t}-w_{s})+r(d_{t-1}-d_{s-1})\right)
\right\}  .
\end{align*}
In the following proposition, we establish the identification of $\gamma$ by showing that $\gamma$ uniquely
maximizes both $Q_{2}(r;\beta)$ and $\tilde{Q}_{2}(r;\beta)$.

From the definitions of $A,B,$ and $Q_{2}(r;\beta),$ we require $T\geq3$. In
the simplest case when $T=3$, we have
\[
A=\{y_{0}=d_{0},y_{1}=0,y_{2}=1,y_{3}=d_{3}\}\text{ and }B=\{y_{0}=d_{0}%
,y_{1}=1,y_{2}=0,y_{3}=d_{3}\},
\]
and
\[
Q_{2}(r;\beta)=\mathbb{E}\left\{  \left[  P(A|x^{T},w_{2}=w_{3})-P(B|x^{T}%
,w_{2}=w_{3})\right]  sgn\left(  (w_{2}-w_{1})+r(d_{3}-d_{0})\right)
\right\}  .
\]
$\tilde{Q}_{2}(r;\beta)$ is not applicable for this case.

\begin{proposition}
[Identification of $\gamma$]\label{Lemma:gamma} Suppose Assumption
\hyperref[Assumption:HK]{A} holds. We have
\begin{enumerate}
\item[(i)] $Q_{2}(\gamma;\beta)>Q_{2}(r;\beta)$ for all $r\in\mathcal{R}%
\setminus\{\gamma\}$, and
\item[(ii)] $\tilde{Q}_{2}(\gamma;\beta)>\tilde{Q}_{2}(r;\beta)$ for all
$r\in\mathcal{R}\setminus\{\gamma\}$.
\end{enumerate}
\end{proposition}

\begin{remark}
When $T\geq4$, any combination $(s,t)$ of the elements of $\{1,...,T-1\}$
taken two at a time can be used to construct the population objective function
to identify $\gamma$. For example, in the simplest case $T=4$, feasible
choices of $(s,t)$ include $(1,2)$, $(1,3)$, and $(2,3)$. One can use any of these pairs to define the population objective function, either $Q_{2}%
(\cdot;\beta)$ or $\tilde{Q}_{2}(\cdot;\beta)$. Clearly, any one (or combination, e.g., by simply summing them up) of these objective functions
can be used to identify $\gamma$.
\end{remark}

Propositions \ref{Lemma:beta} and \ref{Lemma:gamma} outline a two-step
procedure for identifying the preference parameters $\beta$ and $\gamma$, of
which Proposition \ref{Lemma:gamma} uses HK's insight. Note that, as
Proposition \ref{Lemma:beta} suggests, an additional assumption,
\hyperref[Assumption:SI]{SI},  enables us to
establish the identification of $\beta$ independently to that of $\gamma$ in
the first step. As a result, it suffices to match the index $x_{t}^{\prime
}\beta$, rather than each component of $x_{t}$ over time, as in HK, when
identifying $\gamma$ in the second step. The benefit of doing so is that the two-step procedure avoids the curse of dimensionality caused by
matching many explanatory variables (see Theorem \ref{TH:betahat} in Section
\ref{sec_asymptotics}). Our method is particularly competitive
when handling high-dimensional models.

The following theorem is an immediate result of Propositions \ref{Lemma:beta}
and \ref{Lemma:gamma}.

\begin{theorem}
[Identification of $\theta$]\label{Thm1} Suppose $T\geq 4$ and Assumptions
\hyperref[Assumption:HK]{A} and
\hyperref[Assumption:SI]{SI} hold. Then, $\beta$ is identified based on population objective function
$Q_{1}(\cdot)$, and $\gamma$ is identified based on either population
objective function $Q_{2}(\cdot;\beta)$ or $\tilde{Q}_{2}(\cdot;\beta)$.
\end{theorem}

\subsection{Alternative Sufficient Conditions for Identification}
Here, we provide an alternative sufficient condition that permits limited
dependence of the covariates for the identification. We show in Lemma \ref{Lemma:A3} that Assumptions \hyperref[Assumption:HK]{A} and \hyperref[Assumption:SD]{SD} (in below) are sufficient for the inequality in (\ref{idinequality}), which, in turn, shows the identification. In this section, we present and discuss this assumption.

\begin{assumptionSD}\label{Assumption:SD}
For all $\alpha$ and $s,t\in\mathcal{T}$,
\begin{enumerate}
\item[(a)] $f_{\epsilon|\alpha}(\cdot)/F_{\epsilon|\alpha}(\cdot)$
is a non-increasing function, or equivalently, $f_{\epsilon|\alpha}(\cdot)/[1-F_{\epsilon|\alpha}(\cdot)]$
is a non-decreasing function.
\item[(b)] Let $w_t\equiv x_t'\beta$.  The joint PDF of $w^{T}$ conditional on $\alpha$ is exchangeable, i.e.,
\[
f_{w^{T}|\alpha}(\omega_{1},...,\omega_{T})=f_{w^{T}|\alpha}(\omega_{\pi(1)},...,\omega_{\pi(T)})
\]
for all permutations $\{\pi(1),...,\pi(T)\}$ defined on the set $\mathcal{T}$.
\end{enumerate}
\end{assumptionSD}

Assumption \hyperref[Assumption:SD]{SD}(a) states that $F_{\epsilon|\alpha}$ has a
decreasing inverse Mills ratio, which, together with Assumption
\hyperref[Assumption:SD]{SD}(b), guarantees the monotonic relation in
(\ref{idinequality}), as proved in Appendix \ref{Sec_identification}. Assumption \hyperref[Assumption:SD]{SD}(a) is satisfied
by many common continuous distributions, such as the Gaussian, logistic, Laplace,
uniform, gamma, log-normal, Gumbel, and Weibull distributions.\footnote{In a mixture model,
e.g.,
\[
f_{\epsilon|\alpha}(e)=\sum_{m=1}^{M}\pi_{m}f_{\epsilon|\alpha}(e;\vartheta
_{m})
\]
with mixing proportions $\pi_{m}$, $\sum_{m=1}^{M}\pi_{m}=1$, where each
component density has a different parameter vector $\vartheta_{m}$, Assumption
\hyperref[Assumption:SD]{SD}(a) holds for $F_{\epsilon_{t}|\alpha}(\cdot)$
if it is satisfied by all component distributions $F_{\epsilon|\alpha}%
(\cdot;\vartheta_{m})$.} However, this property fails if $F_{\epsilon
|\alpha}$ has heavy tails (e.g., Student's $t$-distribution and Cauchy
distribution).\footnote{More precisely, Assumption \hyperref[Assumption:SD]%
{SD}(a) does not hold globally for these
distributions. For example, it is not difficult to find that this assumption holds
for Student's $t$ and Cauchy on $\lbrack-L,\infty)$ for some positive $L$.}
Note that Assumption \hyperref[Assumption:SD]{SD}(a) is a key
condition imposed in \cite*{McFadden1976} and \cite*{Silvapulle1981} for both
$-\log F_{\epsilon|\alpha}(\cdot)$ and $-\log(1-F_{\epsilon|\alpha}(\cdot)
)$ being convex, which guarantees a unique solution for the MLE in
cross-sectional models with errors that follow a general distribution.

In model (\ref{model}), the exogenous utility $w_{t}$ affects the value of
$y_{t+1}$ via $y_{t}$ and its serial dependence with $w_{t+1}$, conditioning
on $\alpha$. The former is explicitly captured by the coefficient $\gamma$.
For the latter, Assumption \hyperref[Assumption:SD]{SD}(b) restricts the serial
dependence of $\{w_{t}\}$. Assumption \hyperref[Assumption:SD]{SD}(b) is weaker than Assumption \hyperref[Assumption:SI]{SI}. Under Assumption \hyperref[Assumption:SI]{SI}, $w_{t}$'s are i.i.d. over time, conditional on $\alpha$. Then, $w_{t}$'s have an exchangeable joint PDF, as defined in Assumption \hyperref[Assumption:SD]{SD}(b). However, the other direction is not always true. As noted in \cite{Fox2007}, a common example of an exchangeable PDF for non-independent $w_{t}$'s is a multivariate normal density with $\mathbb{E}[w_{t}]=\mu$, $\textrm{Var}(w_{t})=\sigma^{2}$, and $\textrm{Corr}(w_{s},w_{t})=\rho$, for all $s,t\in\{1,...,T\}$. More generally, if each $w_{t}$ can be expressed as $w_{t}=\varphi(u_{t},v)$, where $u_t$'s are i.i.d. random variables, $v$ is a random variable independent of all $u_t$'s, and $\varphi(\cdot,\cdot)$ is some measurable function, then $w^{T}$ satisfies Assumption \hyperref[Assumption:SD]{SD}(b), but not Assumption \hyperref[Assumption:SI]{SI}. Similar exchangeability assumptions are imposed in
\cite*{AltonjiMatzkin2005} and \cite*{ChenEtal2018}.

A few remarks are in order about how our identification conditions are related
to the existing literature. 

\begin{remark}
Compared with HK, our approach relies on
additional assumptions restricting the serial dependence of strictly exogenous
regressors $x_{t}$ and requires $T\geq4$. These conditions make identification
without element-by-element matching of $x_{t}$ possible. HK construct identifying inequalities similar to our (\ref{idinequality}). To obtain point identification, HK use probabilities of specific sequences
of $y^{T}$ conditional on event $\{x_{s}=x_{t}\}$, for some $s,t\in\{1,...,T\}$. Instead, our approach matches $y_{t}$ in different time periods to construct identifying inequalities, allowing for point identification without element-by-element matching of $x_{t}$. However, as discussed after the introduction of our key identifying condition (\ref{idinequality}), $x_{t}$ can affect the choice probabilities in (\ref{idinequality}) through the utility index $x_{t}'\beta$ and its serial dependence with $x$ in other time periods. Assumptions \hyperref[Assumption:SI]{SI} and \hyperref[Assumption:SD]{SD} restrict this dependence, ensuring that choice probabilities are solely rank-ordered by $x_{t}'\beta$. Additionally, our approach requires comparing $x_{t}'\beta$ in non-adjacent time periods (i.e., $t>s+1$ in (\ref{idinequality})). As a result, we need observation of one more  period for each individual, compared with HK's method, which can achieve point identification using $x_{t}'\beta$ in adjacent time periods ($t=s+1$).
\end{remark}

\begin{remark}
Second, our identification conditions are non-nested with those in the literature, assuming
exclusion restrictions, such as \cite*{HonoreLewbel2002}, \cite*{ChenEtal2018, ChenEtal2019},
and \cite*{Williams2019}. \cite*{ChenEtal2019} show that
\cite*{HonoreLewbel2002} essentially require the serial independence of the
excluded regressor. 
\cite*{Williams2019} requires that the other strictly exogenous
regressors are conditionally independent of the past values of the excluded
regressor. In addition to specific restrictions on the dynamic process for the
covariates, the identification results of these studies rely on the existence
of at least one \textquotedblleft excluded regressor\textquotedblright%
\ conditionally independent of the individual fixed effects $\alpha$. Conversely, our approach allows for arbitrary
correlation between $x_{t}$ and $\alpha$.
\end{remark}

\section{Estimation}\label{sec_estimation} 
Applying the analogy principle, the identification
results presented in Section \ref{sec_identification} can be translated
into a two-step estimation procedure. In the first step, we obtain an MS estimator (with
binary weights) $\hat{\beta}$ of $\beta$. In the second step,
$\gamma$ is estimated by a localized MS procedure matching the estimated index
$x_{t}^{\prime}\hat{\beta}$ over time. Each of the two steps is described, in
turn, below.

In Sections \ref{Section:betahat} and \ref{Section:gammahat}, we restrict our discussion to the model with $T=4$
to streamline exposition in subsequent sections. The same method can be applied, with straightforward
modification, to models with longer panels. We provide objective functions for
general cases with $T\geq4$ in Section \ref{SEC:generalobjs}.

\subsection{Estimation of \texorpdfstring{$\beta$}{Lg} with \texorpdfstring{$T=4$}{Lg}}\label{Section:betahat}
Assuming a random sample of $n$ individuals, we propose the following weighted
MS estimator $\hat{\beta}$ of $\beta$, defined as the maximizer over the
parameter space $\mathcal{B}$:
\begin{equation}
\hat{\beta}=\arg\max_{b\in\mathcal{B}}Q_{1n}(b), \label{eq:3.1}%
\end{equation}
where
\begin{equation}
Q_{1n}(b)=\frac{1}{n}\sum_{i=1}^{n}1[y_{i0}=y_{i2}=y_{i4}](y_{i3}-y_{i1})\cdot
sgn(x_{i31}^{\prime}b). \label{eq:3.2}%
\end{equation}
Because we restrict the search within a compact set $\mathcal{B}$ and the objective function (\ref{eq:3.2}) is bounded and continuous, the maximizer $\hat{\beta}$ of the maximization problem (\ref{eq:3.1}) does exist. However, $\hat{\beta}$ may not be unique because the objective function (\ref{eq:3.2}) is essentially a step function for any finite samples. Nonetheless, as guaranteed by Theorem \ref{TH:betahat} in Section \ref{sec_asymptotics}, $\hat{\beta}$ is in a small neighborhood of the true parameter $\beta$ with a high probability for a sufficiently large sample size.

It is clear from expression (\ref{eq:3.2}) that only observations that
satisfy $y_{i1}\neq y_{i3}$, $y_{i0}=y_{i2}$, and $y_{i2}=y_{i4}$ are used in
the estimation. That is, the objective function uses only ``switchers'' with choice changes in periods $1$ and $3$, with the same choices in their previous and subsequent periods, respectively. This feature reduces the ``effective'' sample size for the estimator of $\beta$. HK's estimator has a similar problem, because it also uses only switchers, and needs to match $x_t$ over time. Our estimator (\ref{eq:3.1}) is more applicable when the model has many regressors, especially discrete regressors that must be matched exactly over time when applying HK's procedure.

\subsection{Estimation of \texorpdfstring{$\gamma$}{Lg} with \texorpdfstring{$T=4$}{Lg}}\label{Section:gammahat}
Proposition \ref{Lemma:gamma} motivates a localized MS estimator $\hat{\gamma
}$ of $\gamma$, defined here as the maximizer over the parameter space
$\mathcal{R}$ of the objective function\footnote{If one knew $\beta$, the estimation of $\gamma$ requires only $T=3$, that is, using the first line of (\ref{eq:3.3}) as the objective function.}
\begin{align}
Q_{2n}(r;\beta)= &  \frac{1}{n}\sum_{i=1}^{n}\left\{  1[x_{i2}^{\prime}%
\beta=x_{i3}^{\prime}\beta](y_{i2}-y_{i1})\cdot sgn(x_{i21}^{\prime}%
\beta+r(y_{i3}-y_{i0}))\right.  \nonumber\\
&  \left.  +1[x_{i3}^{\prime}\beta=x_{i4}^{\prime}\beta](y_{i3}-y_{i2})\cdot
sgn(x_{i32}^{\prime}\beta+r(y_{i4}-y_{i1}))\right\}.\label{eq:3.3}%
\end{align}
Expression (\ref{eq:3.3}) is the sample analogue of $Q_{2}(r;\beta)$ in
Proposition \ref{Lemma:gamma} after taking the union of events $A$ and $B$ for
all possible values of $d_0,d_{1},...,d_{4}$. As with objective function
(\ref{eq:3.2}), (\ref{eq:3.3}) also uses only data on switchers (i.e., satisfying $A\cup B$) who make different
choices in the two periods compared. In addition, (\ref{eq:3.3}) also
requires a match in $x_{t}^{\prime}\beta$.

Note that this estimator is not feasible, because $\beta$ is unknown, and it is
of probability zero to have exactly matched indices ($x_{is}^{\prime}%
\beta=x_{it}^{\prime}\beta$) in the presence of continuous regressors. To
resolve the first concern, we propose replacing the unknown parameter $\beta$
in expression (\ref{eq:3.3}) with the $\hat{\beta}$ obtained from
(\ref{eq:3.1}), which is shown to be (cube-root $n$) consistent in
Section \ref{sec_asymptotics}.

For the second concern, we use kernel weights
\[
\mathcal{K}_{h_{n}}((x_{it}-x_{is})^{\prime}b),\text{ for all }s,t\in
\mathcal{T}\text{ and }b\in\mathcal{B},
\]
instead of $1[x_{is}^{\prime}b=x_{it}^{\prime}b].$ $\mathcal{K}_{h_{n}}%
(\cdot)$ is defined as $h_{n}^{-1}\mathcal{K}(\cdot/h_{n})$, where
$\mathcal{K}(\cdot)$ is a kernel density function and $h_{n}$ is a bandwidth sequence
that converges to zero as $n\rightarrow\infty$. The idea is to replace the binary
weights for $x_{is}^{\prime}\hat{\beta}=x_{it}^{\prime}\hat{\beta}$ with weights that depend inversely on the magnitude
of $(x_{it}-x_{is})^{\prime}\hat{\beta}$, giving more weight to observations
with $(x_{it}-x_{is})^{\prime}\hat{\beta}$ closer to zero. We discuss the choice of the tuning parameter $h_n$ with illustrating examples in Section \ref{SEC:tuning}.

Then, we propose the following kernel weighted MS estimator $\hat{\gamma}$ of
$\gamma$:
\begin{equation}
\hat{\gamma}=\arg\max_{r\in\mathcal{R}}Q_{2n}^{K}(r;\hat{\beta}),
\label{eq:3.4}%
\end{equation}
where
\begin{align}
Q_{2n}^{K}(r;\hat{\beta})=  &  \frac{1}{n}\sum_{i=1}^{n}\left\{
\mathcal{K}_{h_{n}}(x_{i32}^{\prime}\hat{\beta})(y_{i2}-y_{i1})\cdot
sgn(x_{i21}^{\prime}\hat{\beta}+r(y_{i3}-y_{i0}))\right. \nonumber\\
&  \left.  +\mathcal{K}_{h_{n}}(x_{i43}^{\prime}\hat{\beta})(y_{i3}%
-y_{i2})\cdot sgn(x_{i32}^{\prime}\hat{\beta}+r(y_{i4}-y_{i1}))\right\}  .
\label{eq:3.5}%
\end{align}

\begin{remark}
Note that objective function (\ref{eq:3.5}) is associated with the population
objective function $Q_{2}(r;\beta)$ in Proposition \ref{Lemma:gamma},
which uses only observations of adjacent time periods. Applying the same idea
to {the} population objective function $\tilde{Q}_{2}(r;\beta)$ yields the
following objective function, using observations which are not adjacent:
\[
\tilde{Q}_{2n}^{K}(r;\hat{\beta})=\frac{1}{n}\sum_{i=1}^{n}1[y_{i2}%
=y_{i4}]\mathcal{K}_{h_{n}}(x_{i42}^{\prime}\hat{\beta})(y_{i3}-y_{i1})\cdot
sgn(x_{i31}^{\prime}\hat{\beta}+r(y_{i2}-y_{i0})).
\]
In practice, to make full use of all observations, one can consider using
$Q_{2n}^{K}(r;\hat{\beta})+\tilde{Q}_{2n}^{K}(r;\hat{\beta})$ as {the} objective
function for the estimation of $\gamma$.
\end{remark}

\begin{remark}\label{DEoptim}
Calculating the MS-type of estimators (\ref{eq:3.1}) and (\ref{eq:3.4}) is challenging, as it is for the semiparametric estimator of HK. Following \cite{Fox2007} and \cite{YanYoo2019}, we suggest using a global optimization method called the \textit{differential evolution} (DE) algorithm. The DE algorithm, introduced by \cite{StornPrice1997}, is specifically designed to search the global optimum of a real-valued function with real-valued parameters. Notably, it does not require the objective function to be continuous or differentiable. The DE algorithm has been widely used in engineering applications, and its performance as a global optimization algorithm has been studied extensively (See, e.g., \cite{StornPrice2006}). To implement the DE algorithm, one can use the ``\texttt{DEoptim}'' package in R.\footnote{\url{https://cran.r-project.org/web/packages/DEoptim/index.html}.} The following statement is quoted from the R documentation for the ``\texttt{DEoptim}'' package, which provides a brief introduction. Interested readers are advised to consult this documentation for more information on the implementation and usage of this algorithm.

\textquotedblleft \textit{Differential Evolution (DE) is a search heuristic
introduced by \cite{StornPrice1997}. Its remarkable performance as a global
optimization algorithm on continuous numerical minimization problems has
been extensively explored; see \cite{StornPrice2006}. DE belongs to the
class of genetic algorithms which use biology-inspired operations of
crossover, mutation, and selection on a population in order to minimize an
objective function over the course of successive generations. As with other
evolutionary algorithms, DE solves optimization problems by evolving a
population of candidate solutions using alteration and selection operators.
DE uses floating-point instead of bit-string encoding of population members,
and arithmetic operations instead of logical operations in mutation. DE is
particularly well-suited to find the global optimum of a real-valued
function of real-valued parameters, and does not require that the function
be either continuous or differentiable.}\textquotedblright
\end{remark}

Note that the 2SMS procedure described in (\ref{eq:3.1})--(\ref{eq:3.2}%
) and (\ref{eq:3.4})--(\ref{eq:3.5}) does not require matching each
covariate in $x_{it}$ over time, as it does in HK. As a result, the
rates of convergence of $\hat{\beta}$ and $\hat{\gamma}$ are independent of
the number of continuous covariates in $x_{it}$, in contrast to the procedure of HK. In view of existing results on the MS estimators (e.g.,
\cite*{Manski1985, Manski1987}, \cite*{KimPollard1990}, and \cite*{SeoOtsu2018}),
we expect the limiting distributions of $\hat{\beta}$ and $\hat{\gamma}$ to be
non-Gaussian and their rates of convergence to be $O_{P}(n^{-1/3})$ and
$O_{P}((nh_{n})^{-1/3})$, respectively. Section \ref{sec_asymptotics} states
sufficient conditions under which these asymptotic properties can be derived.

\subsection{Estimation with \texorpdfstring{$T\geq4$}{Lg}}\label{SEC:generalobjs}
A longer panel allows for more objective functions of
similar form. Collectively, these objective functions (by, for example,
summing them) can be used to obtain more accurate estimates of $\theta$ for finite samples. For the case with $T\geq 4$, estimators for $\beta$ and $\gamma$ that best use
the data can be obtained as follows. For $\beta,$ we find $\hat{\beta}$ by
maximizing%
\[
Q_{1n}(b)=\frac{1}{n}\sum_{i=1}^{n}\sum_{t>s+1}1[y_{is-1}=y_{it-1}%
]1[y_{is+1}=y_{it+1}](y_{it}-y_{is})sgn\left(  (x_{it}-x_{is})^{\prime
}b\right)  .
\]
Once $\hat{\beta}$ is obtained, we estimate $\gamma$ by
maximizing
\[
Q_{2n}^{K}(r;\hat{\beta})+\tilde{Q}_{2n}^{K}(r;\hat{\beta})
\]
with respect to $r$, where
\[
Q_{2n}^{K}(r;\hat{\beta})=\frac{1}{n}\sum_{i=1}^{n}\sum_{t=2}^{T-1}
\mathcal{K}_{h_{n}}((x_{it+1}-x_{it})^{\prime}\hat{\beta}) (y_{it}%
-y_{it-1})sgn((x_{it}-x_{it-1})^{\prime}\hat{\beta}+r(y_{it+1}-y_{it-2}))
\]
is for the case with $t=s+1$, and
\begin{align*}
\tilde{Q}_{2n}^{K}(r;\hat{\beta})=\frac{1}{n}\sum_{i=1}^{n}\sum_{s=1}%
^{T-3}\sum_{t=s+2}^{T-1}  &  \left\{  1[y_{is+1}=y_{it+1}]\mathcal{K}_{h_{n}%
}\left(  (x_{it+1}-x_{is+1})^{\prime}\hat{\beta}\right)  \right. \\
&  \left.  \times(y_{it}-y_{is})sgn\left(  (x_{it}-x_{is})^{\prime}\hat{\beta
}+r(y_{it -1}-y_{is-1})\right)  \right\}
\end{align*}
is for the case with $t>s+1$.

\section{Asymptotic Properties}\label{sec_asymptotics}

The estimators proposed in Section \ref{sec_estimation} are of the same
structure and differ only in that they each use a different fraction of
observations in the sample. We expect that they have similar asymptotic
properties. Therefore, it suffices to show the asymptotics for the
estimators in Sections \ref{Section:betahat} and \ref{Section:gammahat}, for
the case $T=4$. The asymptotic properties of the estimators in Section \ref{SEC:generalobjs} can be derived in
a similar way.

As is standard in the literature, such as \cite*{KimPollard1990}, we start
the analysis by introducing modified objective functions for $\hat{\beta}$
and $\hat{\gamma}$. The new objective functions are
monotone (linear) transformations of (\ref{eq:3.2}) and (\ref{eq:3.5}),
respectively. As a result, working with them does not change the values of
$\hat{\beta}$ and $\hat{\gamma}$, but can facilitate the derivation process.

Because adding terms not related to $b$ will not affect the optimization over $b,$ and  $1[a>0]=(sgn(a)+1)/2$ for all $a\in\mathbb{R}$,
$\hat{\beta}$ obtained from the following objective function is identical to
that from (\ref{eq:3.2}),%
\[
\hat{\beta}=\arg\max_{b\in\mathcal{B}}n^{-1}\sum_{i=1}^{n}\xi_{i}\left(
b\right)  ,
\]
where
\begin{equation}
\xi_{i}\left(  b\right)  \equiv1\left[  y_{i0}=y_{i2}=y_{i4}\right]  \left(
y_{i3}-y_{i1}\right)  \left(  1\left[  x_{i31}^{\prime}b>0\right]  -1\left[
x_{i31}^{\prime}\beta>0\right]  \right)  . \label{EQ:xi_ib}%
\end{equation}%
For the same reason,\ $\hat{\gamma}$ can be obtained equivalently from
\[
\hat{\gamma}=\arg\max_{r\in\mathcal{R}}n^{-1}\sum_{i=1}^{n}\varsigma
_{ni}(  r,\hat{\beta})  ,
\]
where
\begin{align}
\varsigma_{ni}( r,\hat{\beta})  \equiv
&\mathcal{K}_{h_{n}}   \left(  x_{i32}^{\prime}\hat{\beta}\right)  \left(
y_{i2}-y_{i1}\right)  1\left[  x_{i21}^{\prime}\hat{\beta}+r\left(
y_{i3}-y_{i0}\right)  >0\right]   \label{EQ:gammahat}\\
&    +\mathcal{K}_{h_{n}}\left(  x_{i43}^{\prime}\hat{\beta}\right)
\left(  y_{i3}-y_{i2}\right)  1\left[  x_{i32}^{\prime}\hat{\beta}+r\left(
y_{i4}-y_{i1}\right)  >0\right]    .\nonumber
\end{align}

The following technical assumptions are needed for the asymptotics of
$\hat{\beta}$ and $\hat{\gamma}$.

\begin{assumption}
\label{A:iid} The vectors $\left(  x_{i}^{T},y_{i}^{T},y_{i0}\right)
^{\prime}$ with $T\geq4$ are i.i.d. across individuals.
\end{assumption}

\begin{assumption}
\label{A:maximiser}$n^{-1}\sum_{i=1}^{n}\xi_{i}( \hat{\beta}) \geq\max
_{b\in\mathcal{B}}n^{-1}\sum_{i=1}^{n}\xi_{i}\left(  b\right)  -o_{P}\left(
n^{-2/3}\right)  $ and $n^{-1}\sum_{i=1}^{n}\varsigma_{ni}( \hat{\gamma}%
,\hat{\beta}) \geq\max_{r\in\mathcal{R}}n^{-1}\sum_{i=1}^{n}\varsigma_{ni}(
r,\hat{\beta}) -o_{P}\left(  ( nh_{n}\right)  ^{-2/3})$.
\end{assumption}

\begin{assumption}
\label{A:boundedDensity}The joint density function for $\alpha$, covariates
$x^{T}$, and $\epsilon^{T}$ are continuously differentiable. The density
function and its first-order derivatives are uniformly bounded. Further,
\[ f\left(
\epsilon_{t}|\alpha,x^{T},\epsilon_{1},...,\epsilon_{t-1},\epsilon
_{t+1},...,\epsilon_{T}\right)   \text{and }  f\left(  x_{t}|\alpha,x_{1}%
,...,x_{t-1},x_{t+1},...,x_{T},\epsilon^{T}\right)
\]
are continuous
differentiable with respect to all arguments. The conditional densities and their derivatives are
uniformly bounded.
\end{assumption}

\begin{assumption}
\label{A:kernel}The kernel function $\mathcal{K}\left(  u\right)  $ is
nonnegative, symmetric about zero, continuous differentiable, has compact
support, and satisfies $\int_{\mathbb{R}}\mathcal{K}\left(  u\right)  du=1$.
\end{assumption}

\begin{assumption}
\label{A:hn}$h_{n}\rightarrow0,$ $nh_{n}\rightarrow\infty,$ and $nh_{n}%
^{4}\rightarrow0$ as $n\rightarrow\infty.$
\end{assumption}

Assumption \ref{A:maximiser} is standard in the literature and precisely
defines our estimator. Assumption \ref{A:boundedDensity} is for technical
convenience; it ensures certain functions defined in the proof of Theorem \ref{TH:betahat} are differentiable, so that $V_1$ and $V_2$ (defined in Theorem \ref{TH:betahat}) have simple representations. In the case of discrete explanatory variables which violates Assumption \ref{A:boundedDensity}, $V_1$ and $V_2$ can be shown to be well defined, but with more tedious calculations and notation. Assumption \ref{A:kernel} collects some standard
restrictions on kernel functions. The symmetry of $\mathcal{K}\left(
u\right)$ ensures that the bias term is of the order $h_{n}^{2}$. In Assumption \ref{A:hn}, $nh_{n}\rightarrow\infty$ is
standard, and $nh_{n}^{4}\rightarrow0$ ensures the bias term from
the kernel estimation is asymptotically negligible.

\begin{theorem}\label{TH:betahat}
Suppose $T\geq 4$ and Assumptions \hyperref[Assumption:HK]{A},
\hyperref[Assumption:SI]{SI} (or \hyperref[Assumption:SD]{SD}), and
\ref{A:iid}--\ref{A:hn}\ hold. Then,

\begin{enumerate}
\item $\hat{\beta}-\beta=O_{P}\left(  n^{-1/3}\right)  ,$ and%
\[
n^{1/3}(  \hat{\beta}-\beta)  \overset{d}{\rightarrow}\arg
\max_{\boldsymbol{s}\in\mathbb{R}^{K}}Z_{1}\left(  \boldsymbol{s}\right)  ,
\]
where $Z_{1}\left(  \boldsymbol{s}\right)  $ is a Gaussian process with
continuous sample paths, expected value $\frac{1}{2}\boldsymbol{s}^{\prime
}V_{1}\boldsymbol{s}$, and covariance kernel $H_{1}\left(  \boldsymbol{s}%
,\boldsymbol{t}\right)  .$ $V_{1}$ and $H_{1}$ are defined in expressions
(\ref{EQ:V1}) and (\ref{EQ:H1_f}), respectively$.$

\item $\hat{\gamma}-\gamma=O_{P}(  \left(  nh_{n}\right)  ^{-1/3})
,$ and%
\[
\left(  nh_{n}\right)  ^{1/3}\left(  \hat{\gamma}-\gamma\right)
\overset{d}{\rightarrow}\arg\max_{s\in\mathbb{R}}Z_{2}\left(  s\right)  .
\]
where $Z_{2}\left(  s\right)  $ is a Gaussian process with continuous path,
expected value $\frac{1}{2}V_{2}s^{2}$, and covariance kernel $H_{2}\left(
s,t\right)  .$ $V_{2}$ and $H_{2}$ are defined in expressions (\ref{EQ:V2_11})
and (\ref{EQ:H2}), respectively.
\end{enumerate}
\end{theorem}

\cite*{KimPollard1990} and \cite*{SeoOtsu2018} derive the cube-root
asymptotics for a class of estimators by means of empirical processes. For a
comprehensive treatment on this technique, see \cite*{vdVaartWellner2000}. Our
estimators fall into this category. In particular, they are more closely related
to \cite*{SeoOtsu2018}. The main body of the proof for Theorem \ref{TH:betahat}
verifies the technical conditions in \cite*{SeoOtsu2018}, applies their
asymptotics results to our estimators, and calculates the technical terms needed for
the asymptotics such as $V_{1},H_{1},V_{2},$ and $H_{2}.$

Note that the asymptotics of $\hat{\gamma}$ are the same
as in the case where the true value of $\beta$ is used. Intuitively, $\hat{\beta}$ converges to $\beta$ faster than $\hat{\gamma}$ does to
$\gamma,$ and the objective function (\ref{EQ:gammahat}), after proper normalization, uniformly converges to the limit over a compact set
of $\left(b',r\right)'$ around $\left(\beta',\gamma\right)'$. The details
can be found in the proof of Theorem \ref{TH:betahat}, which is presented in Appendix
\ref{SEC:main_proof}.

\section{Inference\label{SEC:inferences}}

The asymptotic distributions of $\hat{\beta}$ and $\hat{\gamma}$\ are
complicated and do not have an analytical form. As a result, inference using the asymptotic distribution directly is difficult to implement. One
may consider smoothing the objective functions, in the spirit of
\cite*{Horowitz1992}, to attain faster rates of convergence and asymptotic
normality.\footnote{See also \cite*{Kyriazidou1997} and \cite*{charlier1997limited}.} However, this requires selecting additional kernel functions and tuning
parameters, and then computing consistent estimates for asymptotic variances. As an alternative, we seek to use more direct sampling methods (e.g., bootstrap). Unfortunately,
\cite*{AbrevayaHuang2005} have proved the inconsistency of the classic
bootstrap for the MS score estimators. We expect that the classic
bootstrap does not work for our estimators either.

For the ordinary MS estimator, valid inference can be conducted
using subsampling (\cite*{DelgadoEtal2001}), $m$-out-of-$n$ bootstrap
(\cite*{LeePun2006}), the numerical bootstrap (\cite*{HongLi2020}), and a model-based bootstrap procedure that analytically modifies the criterion function
(\cite*{CattaneoEtal2020}), among other procedures.\footnote{The case-specific, smooth bootstrap method proposed by \cite*{PatraEtal2018} is also valid for the MS estimator of \cite*{Manski1975, Manski1985}. However, this method is difficult to generalize to
our case.} These methods, with certain modifications, can be justified to be valid for our estimators.

Monte Carlo evidence demonstrated in \cite*{HongLi2020} suggests that their proposed approach outperforms the subsampling and the $m$-out-of-$n$ bootstrap in finite samples.
Based on these results, we focus on the numerical bootstrap. We provide a brief discussion on the classic bootstrap and the $m$-out-of-$n$ bootstrap in Appendix E.\footnote{We show in Appendix E that the classic bootstrap is not consistent for our estimators (Appendix E.2), while the  $m$-out-of-$n$ bootstrap is still valid (Appendix E.3). Note that we re-use some notation in this appendix for notational convenience. To avoid confusion, all notation in each subsection is specific to the procedure discussed in that subsection.}

We next introduce some additional notation. Let $(  y_{j}^{T\ast\prime},x_{j}%
^{T\ast\prime})  ^{\prime},$ $j=1,...,n,$ be a random sample drawn with
replacement from the collection of the sample values $\left(  y_{1}^{T\prime}%
,x_{1}^{T\prime}\right)  ^{\prime},$ $\left(  y_{2}^{T\prime},x_{2}^{T\prime
}\right)  ^{\prime},$ $...,$ $\left(  y_{n}^{T\prime},x_{n}^{T\prime}\right)
^{\prime}.$ Let $\xi_{j}^{\ast}\left(  b\right)  $ denote $\xi\left(
b\right)  $ evaluated at $(  y_{j}^{T\ast\prime},x_{j}^{T\ast\prime})
^{\prime}$, specifically,%
\[
\xi_{j}^{\ast}\left(  b\right)  \equiv1\left[  y_{j0}^{\ast}=y_{j2}^{\ast
}=y_{j4}^{\ast}\right]  \left(  y_{j3}^{\ast}-y_{j1}^{\ast}\right)  \left(
1\left[  x_{j31}^{\ast\prime}b>0\right]  -1\left[  x_{j31}^{\ast\prime}%
\beta>0\right]  \right)  .
\]
Similarly, we define $\varsigma_{nj}^{\ast}\left(  r,b\right)  $ as%
\begin{align*}
\varsigma_{nj}^{\ast}\left(  r,b\right)   &  \equiv\mathcal{K}_{h_{n}}\left(
x_{j32}^{\ast\prime}b\right)  \left(  y_{j2}^{\ast}-y_{j1}^{\ast}\right)
\left(  1\left[  x_{j21}^{\ast\prime}b+r\left(  y_{j3}^{\ast}-y_{j0}^{\ast
}\right)  >0\right]  -1\left[  x_{j21}^{\ast\prime}\beta+\gamma\left(
y_{j3}^{\ast}-y_{j0}^{\ast}\right)  >0\right]  \right) \\
&  +\mathcal{K}_{h_{n}}\left(  x_{j43}^{\ast\prime}b\right)  \left(
y_{j3}^{\ast}-y_{j2}^{\ast}\right)  \left(  1\left[  x_{j32}^{\ast\prime
}b+r\left(  y_{j4}^{\ast}-y_{j1}^{\ast}\right)  >0\right]  -1\left[
x_{j32}^{\ast\prime}\beta+\gamma\left(  y_{j4}^{\ast}-y_{j1}^{\ast}\right)
>0\right]  \right)  .
\end{align*}


\subsection{Numerical Bootstrap\label{SEC:bootnumerical}}

\cite*{HongLi2020} develop a numerical bootstrap procedure for cases in which
the classic bootstrap does not work. They demonstrate that their method works for a class of M-estimators that converge at rate $n^{a}$ for some $a\in(1/4,1].$ The estimator $\hat{\beta}$ proposed in Section \ref{sec_estimation} fits their framework directly, but $\hat{\gamma}$ does not.
With a slight modification of their proof, we show that the numerical bootstrap also works for $\hat{\gamma}$.

The numerically bootstrapped $\hat{\beta}^{\ast}$ and $\hat{\gamma}^{\ast}$ are
constructed from
\begin{equation}
\hat{\beta}^{\ast}=\arg\max_{b\in\mathcal{B}}\left\{  n^{-1}\sum_{i=1}^{n}%
\xi_{i}\left(  b\right)  +\left(  n\varepsilon_{n}\right)  ^{1/2}\cdot
n^{-1}\sum_{j=1}^{n}\left(  \xi_{j}^{\ast}\left(  b\right)  -n^{-1}\sum
_{i=1}^{n}\xi_{i}\left(  b\right)  \right)  \right\}  \label{EQ:beta*num}%
\end{equation}
and
\begin{equation}
\hat{\gamma}^{\ast}=\arg\max_{r\in\mathcal{R}}\left\{  n^{-1}\sum_{i=1}%
^{n}\varsigma_{ni}\left(  r,\hat{\beta}\right)  +\left(  n\varepsilon
_{n}\right)  ^{1/2}\cdot n^{-1}\sum_{j=1}^{n}\left(  \varsigma_{nj}^{\ast
}\left(  r,\hat{\beta}\right)  -n^{-1}\sum_{i=1}^{n}\varsigma_{ni}\left(
r,\hat{\beta}\right)  \right)  \right\}  , \label{EQ:gamma*num}%
\end{equation}
where $\varepsilon_{n}\rightarrow0,$ $n\varepsilon_{n}\rightarrow\infty$%
,$\ $and $(  y_{j}^{T\ast\prime},x_{j}^{T\ast\prime})  ^{\prime},$
$j=1,...,n,$ are drawn independently from the collection of the sample values
$\left(  y_{1}^{T\prime},x_{1}^{T\prime}\right)  ^{\prime},\left(  y_{2}^{T\prime}%
,x_{2}^{T\prime}\right)  ^{\prime},...,\left(  y_{n}^{T\prime},x_{n}^{T\prime
}\right)  ^{\prime}$, with replacement$.$ $\varepsilon_{n}^{-1}$ plays a
similar role as the $m$ in the $m$-out-of-$n$ bootstrap procedure. For
$\hat{\gamma}^{\ast}$, we additionally require $\varepsilon_{n}^{-1}%
h_{n}\rightarrow\infty$ and $\varepsilon_{n}^{-1}h_{n}^{4}\rightarrow0$. Following the same arguments as in the discussion below (\ref{eq:3.2}) for $\hat{\beta}$, the maximizer $\hat{\beta}^{\ast}$ exists, but its uniqueness is not guaranteed due to the non-smoothness of its step objective function. The second term in (\ref{EQ:beta*num}) can be shown to be $o_P(1)$. The first term, sharing a similar structure to the objective function (\ref{eq:3.2}), dominates in (\ref{EQ:beta*num}). As a result, we anticipate $\hat{\beta}^{\ast}$ to be close to $\beta$ as $n\rightarrow\infty$. Similar arguments apply to $\hat{\gamma}^{\ast}$.

We claim that
\[
\varepsilon_{n}^{-1/3}\left(  \hat{\beta}^{\ast}-\hat{\beta}\right)
\overset{d}{\rightarrow}\arg\max_{\boldsymbol{s}\in\mathbb{R}^{K}}\left(
\frac{1}{2}\boldsymbol{s}^{\prime}V_{1}\boldsymbol{s}+W_{1}\left(
\boldsymbol{s}\right)  \right)
\]
and%
\[
\left(  \varepsilon_{n}^{-1}h_{n}\right)  ^{1/3}\left(  \hat{\gamma}^{\ast
}-\hat{\gamma}\right)  \overset{d}{\rightarrow}\arg\max_{s\in\mathbb{R}%
}\left(  \frac{1}{2}V_{2}s^{2}+W_{2}\left(  s\right)  \right)  ,
\]
where $W_1$ and $W_2$ are mean zero Gaussian processes with covariance kernels $H_1$ and $H_2$, respectively.
An outline of the proof of why the numerical bootstrap works and the way to modify the proof in \cite*{HongLi2020} to accommodate $\hat{\gamma}$ is provided in Appendix E.1.

\subsection{Procedures in Details\label{SEC:inferenceDetails}}

We investigate the finite-sample properties of the  bootstrap method discussed in Section
\ref{SEC:bootnumerical} using Monte Carlo experiments in Section \ref{sec_simulation}, and defer the discussion on the choices of
their tuning parameters to Section \ref{SEC:tuning}. Here, we provide the algorithm for constructing the 95\% confidence intervals (CIs) for $\beta$ and
$\gamma$. 

\noindent The numerical bootstrap proceeds as follows.

\begin{enumerate}
\item Draw $(  y_{j}^{T\ast\prime},x_{j}^{T\ast\prime})  ^{\prime},$
$j=1,...,n,$ independently, with replacement, from the original sample.

\item Obtain $\hat{\beta}^{\ast}$ and $\hat{\gamma}^{\ast}$ from equations
(\ref{EQ:beta*num}) and (\ref{EQ:gamma*num}).

\item Repeat Steps 1 and 2 for $B$ times independently and obtain a sequence
of $(\hat{\beta}^{\ast},\hat{\gamma}^{\ast})$, say, $\{ (\hat{\beta}_{{}%
}^{\ast\left(  b\right)  },\hat{\gamma}_{{}}^{\ast\left(  b\right)  })\}
_{b=1}^{B}.$

\item Let $Q_{\hat{\beta}^{\ast}}\left(  \tau\right)  $ denote the $\tau$-th
quantile of $\{  \hat{\beta}_{{}}^{\ast\left(  b\right)  }\}
_{b=1}^{B},$ $0\leq\tau\leq1$. Define $Q_{\hat{\gamma}^{\ast}}\left(
\tau\right)  $ analogously. The 95\% CIs for $\beta$ and
$\gamma$ are constructed, respectively, as
\[
\left[  \hat{\beta}-n^{-1/3}\cdot\varepsilon_{n}^{-1/3}(  Q_{\hat{\beta
}^{\ast}}\left(  0.975\right)  -\hat{\beta})  ,\text{ }\hat{\beta
}-n^{-1/3}\cdot\varepsilon_{n}^{-1/3}(  Q_{\hat{\beta}^{\ast}}\left(
0.025\right)  -\hat{\beta})  \right]
\]
and
\[
\left[  \hat{\gamma}-n^{-1/3}\cdot\varepsilon_{n}^{-1/3}\left(  Q_{\hat
{\gamma}^{\ast}}\left(  0.975\right)  -\hat{\gamma}\right)  ,\text{ }%
\hat{\gamma}-n^{-1/3}\cdot\varepsilon_{n}^{-1/3}\left(  Q_{\hat{\gamma}^{\ast
}}\left(  0.025\right)  -\hat{\gamma}\right)  \right]  .
\]

\end{enumerate}

\section{Monte Carlo Experiments}\label{sec_simulation}

\subsection{Simulation Setup}

In this section, we investigate the finite-sample performance of the
proposed estimators by means of Monte Carlo experiments. We start by
considering a benchmark design similar to that used in HK. Specifically,
this design (referred to as Design 1 hereafter) is specified as follows: 
\begin{align*}
y_{i0} & =1\left[ \beta_{1}x_{i0,1}+\beta_{2}x_{i0,2}+\alpha_{i}-%
\epsilon_{i0}>0\right], \\
y_{it} & =1\left[ \beta_{1}x_{it,1}+\beta_{2}x_{it,2}+\gamma
y_{it-1}+\alpha_{i}-\epsilon_{it}>0\right] ,\text{ \ }t\in\left\{
1,2,3,4\right\} ,
\end{align*}
where
\begin{itemize}
\item[-] $\beta\equiv(\beta_{1},\beta_{2})^{\prime}=(1,1)^{\prime}$ and $%
\gamma=-1$,

\item[-] $x_{it,j}=\frac{\sqrt{15}}{4}u_{it,j}+\frac{1}{4}u_{it,3},j=1,2$, $%
\left( u_{it,1},u_{it,2},u_{it,3}\right)\overset{d}{\sim }N\left( 0_{3\times
1},I_{3\times 3}\right)$, and $\left( u_{it,1},u_{it,2},u_{it,3}\right) $
are i.i.d. across $i$ and $t$,\footnote{Note that all covariates are correlated with each other and have a standard deviation of one.}

\item[-] $\alpha _{i}=\left(
x_{i0,2}+x_{i1,2}+x_{i2,2}+x_{i3,2}+x_{i4,2}\right) /5$,

\item[-] $\epsilon _{it}\overset{d}{\sim }\left( \pi ^{2}/3\right)
^{-1/2}\cdot \text{Logistic}\left( 0,1\right) $ and are i.i.d. across $i$
and $t$, and

\item[-] $\left( u_{\cdot ,1},u_{\cdot ,2},u_{\cdot ,3}\right) $ and $%
\epsilon _{\cdot }$\ are independent of each other.
\end{itemize}

In the second design (hereafter, Design 2), the model and the
coefficients are the same as in Design 1, but $x_{\cdot ,1}$ and $x_{\cdot
,2}$ are autocorrelated over time. Specifically, we have

\begin{itemize}
\item[-] $x_{i0,j}=\frac{\sqrt{15}}{4}u_{i0,j}+\frac{1}{4}u_{i0,3},$ $j=1,2,$
and $x_{it,j}=\frac{1}{2}x_{it-1,j}+\frac{\sqrt{3}}{2}\left( \frac{\sqrt{15}}{4}u_{it,j}+%
\frac{1}{4}u_{it,3}\right) ,$ $j=1,2$ for all $t\geq 1$, where  $\left(
u_{it,1},u_{it,2},u_{it,3}\right) \overset{d}{\sim }N\left( 0_{3\times 1},I_{3\times
3}\right) $ and $\left( u_{it,1},u_{it,2},u_{it,3}\right) $ are i.i.d.
across $i$ and $t$,  

\item[-] $\left( u_{\cdot ,1},u_{\cdot ,2},u_{\cdot ,3}\right) $ and $%
\epsilon _{\cdot }$\ are independent of each other.
\end{itemize}

Note that the setup of Design 2 violates both Assumption \hyperref[Assumption:SI]%
{SI} and the exchangeability condition stated in Assumption \hyperref[Assumption:SD]%
{SD}. We conduct this Monte Carlo study to develop insight into the
practical consequences of the failure of these sufficient (but not
necessary) conditions. That is, we examine the extent to which serial dependence in
exogenous covariates may affect the identification.

In the third to fifth designs (Designs 3, 4, and 5,
respectively), the setup is the same as that in Design 1, except that we add
one, two, and three more covariates (in Designs 3, 4, and 5, respectively) to
examine how our estimators perform in higher-dimensional designs.
Specifically, in Design $k,$ $k=3,4,$ and $5,$%
\begin{align*}
y_{i0}& =1\left[ \beta _{1}x_{i0,1}+\beta _{2}x_{i0,2}+...+\beta
_{k}x_{i0,k}+\alpha _{i}-\epsilon _{i0}>0\right] , \\
y_{it}& =1\left[ \beta _{1}x_{it,1}+\beta _{2}x_{it,2}+...+\beta
_{k}x_{it,k}+\gamma y_{it-1}+\alpha _{i}-\epsilon _{it}>0\right] ,\text{ \ }%
t\in \left\{ 1,2,3,4\right\} ,
\end{align*}%
where

\begin{itemize}
\item[-] $\beta \equiv (\beta _{1},\beta _{2},...,\beta _{k})^{\prime
}=(1,1,...,1)^{\prime }$ and $\gamma =-1$,

\item[-] $x_{it,j}=\frac{\sqrt{15}}{4}u_{it,j}+\frac{1}{4}%
u_{it,k+1},j=1,2,...,k$, $\left( u_{it,1},u_{it,2},...,u_{it,k+1}\right) \overset{d}{\sim }N\left( 0_{(k+1)\times 1},I_{(k+1)\times (k+1)}\right) $, and $\left(
u_{it,1},u_{it,2},...,u_{it,k+1}\right) $ are i.i.d. across $i$ and $t$,

\item[-] $\alpha _{i}=\left(
x_{i0,2}+x_{i1,2}+x_{i2,2}+x_{i3,2}+x_{i4,2}\right) /5$,

\item[-] $\epsilon _{it}\overset{d}{\sim }\left( \pi ^{2}/3\right)
^{-1/2}\cdot \text{Logistic}\left( 0,1\right) $ and are i.i.d. across $i$
and $t$, and

\item[-] $\left( u_{\cdot ,1},u_{\cdot ,2},...,u_{\cdot ,k+1}\right) $ and $%
\epsilon _{\cdot }$\ are independent of each other.
\end{itemize}

We also explore the impact of the serial dependence of $x_{it}$ on the estimation and inference for the models in Designs 3 through 5 in Appendix F. We adopt a similar method to Design 2 for this analysis, which offers additional insights into the robustness and performance of our proposed method.

For the estimation of $\beta$, we adopt the objective function (\ref{eq:3.2}). To
estimate $\gamma$, we use the objective function (\ref{eq:3.5}) with the
Epanechnikov kernel function. That is, 
\begin{equation*}
\mathcal{K}\left( u\right) =\frac{3}{4}\left( 1-u^{2}\right) 1\left[
\left\vert u\right\vert \leq1\right],
\end{equation*}
which satisfies Assumption \ref{A:kernel} with a compact support. We discuss the choice
of bandwidth sequence $h_{n}$ in Section \ref{SEC:tuning}. 

For inference, we investigate the finite-sample performance of the numerical
bootstrap (Section \ref{SEC:bootnumerical}). The 95\% CIs
are obtained from $B=199$ independent draws and estimations. See Section \ref{SEC:inferenceDetails} for the details of the
implementation.

Recall that only the observations with $\left\{ y_{i0}=y_{i2}=y_{i4}\text{
and }y_{i1}\neq y_{i3}\right\} $ are used to estimate $\beta$. In
all designs, the effective observations, which are useful for estimating $\beta$, comprise about $14\%$
of the whole sample. Similarly, for $\gamma$, only observations with
either $\left\{ y_{i1}\neq y_{i2}\text{ and }y_{i0}\neq y_{i3}\right\} $ or $%
\left\{ y_{i2}\neq y_{i3}\text{ and }y_{i1}\neq y_{i4}\right\} $ are useful.
In all designs, about $31\%$ to $39\%$ of the observations are effective for $\gamma$. For each
design, we consider sample sizes of 2,500, 5,000, 10,000, and 20,000. All
the estimation and inference results (based on 199 draws and estimation)
presented in this section are based on 1,000 replications of each design and
each sample size.

Furthermore, we compare our method with the parametric (Logit) and semiparametric (distribution-free) estimators of HK. We use the objective functions linked to these two estimators, as defined in Section 4.1 of HK (for panel data where $T>3$), to implement their methods. To facilitate the comparison, we apply the same scale normalization, specified in Assumption \hyperref[Assumption:HK]{A}(e), to both HK's two estimators and our own. Specifically, we normalize the vector of $\beta$'s to have a Euclidean norm of one. Note that HK's Logit estimator does not require scale normalization for the preference coefficients, because it assumes that the error terms follow a standard logistic distribution. Therefore, if we choose to apply scale normalization to $\beta$, we should also estimate a scale parameter in the Logit model to regain one degree of freedom in the parameter space. For example, for two adjacent time periods $t$ and $t+1$ in Designs 1 and 2, the log-likelihood function is written as
\begin{align*}
 &\sum_{i=1}^{n}1\left[ y_{it}+y_{it+1}=1\right]
\sigma _{n}^{-2}\mathcal{K}\left( \frac{x_{it+1,1}-x_{it+2,1}}{\sigma _{n}}%
\right) \mathcal{K}\left( \frac{x_{it+1,2}-x_{it+2,2}}{\sigma _{n}}\right) \\
&\times \log \left( \frac{\exp \left( \left[ \left(
x_{it,1}-x_{it+1,1}\right) b_{1}+\left( x_{it,2}-x_{it+1,2}\right)
b_{2}+r\left( y_{it-1}-y_{it+2}\right) \right] /s\right) ^{y_{it}}}{1+\exp
\left( \left[ \left( x_{it,1}-x_{it+1,1}\right) b_{1}+\left(
x_{it,2}-x_{it+1,2}\right) b_{2}+r\left( y_{it-1}-y_{it+2}\right) \right]
/s\right) }\right),
\end{align*}
where we impose restriction $\sqrt{b _{1}^{2}+b _{2}^{2}}=1$ and add the
scale parameter of the logistic distribution, $s>0$, to estimate.\footnote{HK normalize $s$ to one, but do not require $(b_1, b_2)$ to be on the unit circle. Both methods of scale normalization are equivalent.}

\subsection{Tuning Parameters and Computation\label{SEC:tuning}}

There is only one tuning parameter used for estimation, namely, $h_{n},$ in the
objective function (\ref{eq:3.5}). In Assumption \ref{A:hn}, we restrict $%
nh_{n}^{4}\rightarrow0$, so that the bias term (of order $h^{2}_{n}$) is a
small order term of $\left( nh_{n}\right) ^{-2/3}$. Because the convergence
rate of $\hat{\gamma}$ is $\left( nh_{n}\right) ^{-1/3}$, the condition, $%
nh_{n}^{4}\rightarrow0,$ makes the bias term much smaller than the
convergence rate. To attain a faster convergence rate, we set $h_{n}$
as large as possible, and thus set $h_{n}=n^{-1/4}\left( \log n\right) ^{-1}$.

For the numerical bootstrap, we have one additional tuning parameter $\varepsilon
_{n}.$ As recommended in \cite{HongLi2020}, we set $\varepsilon_{n}$
proportional to $n^{-2/3}\log n$ for the inferences of $\hat{\beta}$ and $%
\hat{\gamma}.$ Apparently, $\varepsilon_{n}$ of this order satisfies the
additional requirements for $\hat{\gamma}_{{}}^{\ast}$\ that $%
\varepsilon_{n}^{-1}h_{n}\rightarrow\infty$ and $%
\varepsilon_{n}^{-1}h_{n}^{4}\rightarrow0.$ To check how sensitive the
procedure is to the choice of $\varepsilon_{n}$, we conduct the procedure
with $\varepsilon_{n}=c\cdot n^{-2/3}\log n$ and $c=0.8,0.9,1.0,1.1,$ and $%
1.2.$

Following the recommendation in HK, we adopt the bandwidth $\sigma_n = c \cdot n^{-1/(4+k)}$ for HK's estimators, where $k$ is the dimension of $x_{it}$. We conduct experiments with $c=1,2,3,4$, and report the simulation results corresponding to $c=3$. For this value, the HK estimators of $\gamma $ exhibit the smallest bias and have relatively smaller root mean squared errors among all tested values.

For all simulation designs, we employ the DE algorithm (using the \texttt{DEoptim} package in R, see Remark \ref{DEoptim}) to compute our proposed estimators and those of HK. To ensure efficient convergence and robustness, we set the lower and upper bounds for searching each parameter to $[-3, 3]$, the maximum number of iterations to 500, and the relative convergence tolerance to $10^{-8}$. The DE algorithm does not require explicit initial values; instead, it randomly assigns $NP \times (\text{the number of parameters})$ initial values, where we adopt the default value of $NP=10$ in \texttt{DEoptim}. Additionally, we use the default settings for the other algorithm controls. In all simulation runs for each estimator, we consistently observe successful convergence of the algorithm. All estimators can be computed very quickly for all sample sizes considered. Using an Intel\textsuperscript{\tiny\textregistered} Core\textsuperscript{\tiny TM} i7-4790 processor, each replication takes only seconds to complete. 

\subsection{Simulation Results}

We normalize the preference coefficients $\beta$ on exogenous covariates to $%
1$ in Euclidean norm. Because of this normalization, we lose one degree of
freedom in the parameter space. As a result, we report only the results for $\left(
\beta_{2},\gamma\right) $ in Designs 1 and 2 and the results for $\left(
\beta_{2}, ..., \beta_{k},\gamma\right) $ in Design $k$, for $k=3,4,$ and $5$.

We report the mean bias (BIAS), the standard deviation (STD), the median
absolute deviation (MAD), and the root mean squared error (RMSE) for $\hat{%
\beta}$ and $\hat{\gamma}.$ All results are expressed as percentages of the true
values of the parameters, so that the results are independent of how we
normalize the parameters.\footnote{We thank the co-editor for this suggestion.} For inference, we
report the coverage rates (COV) of the true values and lengths (LEN) of the
95\% CIs for the numerical inference procedure for our
method only. Furthermore, we report the computation time (TIME) for one replication of each estimator for Design 1, and omit this for other designs, owing to their similarity.

Results for Design 1 are reported in tables numbered
``1'', and so on for other designs. We report the performance of the estimators and the numerical
bootstrap procedure in tables labeled ``A'' and ``B'',
respectively. For example, Table 1A reports the performance of the
estimators for Design 1. The results for our estimator are denoted as ``OY''
in the tables. The parametric and semiparametric estimators in HK are
denoted as ``HK1'' and ``HK2'', respectively. Due to space limitations, only the results of Designs 1 and 2 are presented in Appendix \ref{SEC:tables}. The results of Designs 3--5, along with additional simulation studies, are reported in Appendix F of the Online Supplement. In what follows, we briefly summarize our findings.

The RMSEs of $\hat{\beta}$ and $\hat{\gamma}$ become smaller as the sample
size increases in all designs, with the RMSE of $\hat{\gamma}$ slightly greater
than that of $\hat{\beta}$. This shows the consistency of our estimators,
though the rates of convergence are clearly slower than $\sqrt{n}$. The
numerical bootstrap inference procedures perform reasonably well in all
designs. In general, they yield shrinking CIs with coverage rates
approaching 95\% as the sample size grows. The coverage rates of these CIs are
greater than 90\%, but are slightly lower than 95\% in most cases. The
coverage rates of the CIs for $\gamma$ do not perform as well as those for $%
\beta, $ which is not surprising, considering the complication of using two
tuning parameters. The inference procedure is not very sensitive to the
choice of tuning parameters.

Despite Design 2 not satisfying Assumption \href{assumptionSI}{SI} or \href{assumptionSD}{SD}, our proposed estimators still perform reasonably well in this setting.\footnote{These assumptions guarantee that the utility index $x_t'\beta$ rank orders the (conditional) probabilities in (\ref{idinequality}). Relaxing them could potentially invalidate the ``if and only if'' result in equation (\ref{idinequality}) and cause bias.} Surprisingly, its performance is similar to that of Design 1, where these assumptions are completely fulfilled. This finding indicates that our method exhibits certain robustness, and can function effectively even when the two sufficient identifying assumptions are not met. Furthermore, the results from Designs 3--5 provide evidence supporting our asymptotic analysis. In these designs, we observe that the convergence rates of our estimators remain relatively stable as the dimension of the model increases. 

The HK1 estimator (HK's Logit estimator) performs the best for Designs 1 and 2, which is not surprising, because the error terms are scaled logistic. Our proposed estimators exhibit higher RMSEs compared with those of HK1, approximately more than twice for these designs. For Designs 1 and 2, the HK2 estimator demonstrates finite-sample performance similar to our proposed method.\footnote{In particular, our method exhibits smaller RMSEs for $\gamma$, and larger RMSEs for $\beta$ compared to HK2.} However, when the number of regressors increases in Design 3, our estimator outperforms HK2, particularly in estimating the parameter $\gamma$. In this setting, the RMSEs of our estimators become about 50\% greater than those of HK1. In Design 4, where there are two more regressors than in Design 1, our estimators' RMSEs are comparable with those of HK1 for all sample sizes considered. Notably, for sample sizes of $n=10,000$ and $20,000$, our RMSEs are about 40\% lower than those of HK2, demonstrating the advantage of our method in high-dimensional settings. In Design 5, with three more regressors than Design 1, our RMSEs are slightly lower than those of HK1 and only half of HK2's RMSEs for sample sizes of $n=10,000$ and $20,000$. These findings highlight the favorable properties of our method, particularly its resilience to the curse of dimensionality. 

We also conduct additional simulations to assess how our estimators perform on a relatively small sample size of $n=1,000$ for Design 1. The results presented in Table 1C suggest that our estimators do not perform poorly in terms of parameter estimation. The RMSEs are 30\%-35\% of the true parameter values, suggesting reasonable accuracy in estimation, despite the smaller sample size. For inference, we report only the results for $c = 1$. We can see the CIs have lower coverage of approximately 85\%. In conclusion, our estimators perform reasonably well for this sample size, but the CIs may be too short.

A final note is that the serial dependence of $x_{it}$ has limited impact on the estimation and inference, as demonstrated by the results associated with Design 2 and additional simulation studies for higher-dimensional designs presented in Appendix F of the Online Supplement.

\section{Conclusions}\label{sec_conclusion} 
This paper presents new identification
results for preference parameters in panel data binary choice models that
allow for both fixed effects (Heckman's ``spurious'' state dependence) and lagged dependent variables
(``true'' state dependence). The same semiparametric random utility framework as in \cite*{HonoreKyriazidou2000} is
considered. A key innovation in this paper is the assertion that, given additional
restrictions on the dynamic process of observed covariates and the tail behavior of the error distribution, the point
identification no longer needs element-by-element matching of regressors over
time, in contrast to the method proposed in \cite*{HonoreKyriazidou2000}. Our approach requires a minimum panel length of five ($T\geq 4$), which fits in most empirical settings. Our
identification arguments motivate a two-step estimation procedure, adapting
Manski's MS estimator. The proposed estimators are consistent with
rates of convergence independent of the model dimension, unlike the
estimator proposed in \cite*{HonoreKyriazidou2000}. We further derive the limiting
distributions of the proposed estimators, which are non-Gaussian, aligning
with existing literature. We justify the use of several bootstrap
procedures for conducting statistical inference. A Monte Carlo study
indicates that our estimators and inference procedures perform well in finite
samples.

This paper leaves some open questions for future research.  For example, it might be worthwhile extending the framework in this paper to
study the identification with more than one lag of the dependent variable or
the identification in panel data multinomial response models.

\appendix

\part*{{\protect\LARGE Appendices}}\label{appendix}
These appendices are organized as follows. In Appendix \ref{Sec_identification}, we present proofs for identification, including those for Propositions \ref{Lemma:beta} and \ref{Lemma:gamma}, along with the necessary lemmas. Additionally, we provide a roadmap for a key step in these proofs. In Appendix \ref{SEC:main_proof}, we establish the asymptotic theory of our estimators, as summarized in Theorem \ref{TH:betahat}, and include the technical lemmas required for this proof in the same section. Appendix \ref{SEC:tables} compiles tables summarizing simulation results for Designs 1 and 2.

The Online Supplement contains Appendices D--F. In Appendix D, we prove all technical lemmas used in Appendices \ref{Sec_identification} and \ref{SEC:main_proof}. In Appendix E, we provide technical details for Section \ref{SEC:inferences}. Appendix F presents simulation results for Designs 3-5 and supplementary simulation studies (Designs 6--8).

\section{Technical Lemmas and Main Proofs for Identification}\label{Sec_identification}
Building on the results of Lemmas \ref{Lemma:A1}
and \ref{Lemma:A2}, Lemma \ref{Lemma:A3} establishes the identification
inequality (\ref{idinequality}) under Assumptions \hyperref[Assumption:HK]{A} and
\href{assumptionSD}{SD}. Lemma \ref{Lemma:A4} shows that (\ref{idinequality})
also holds under Assumptions \hyperref[Assumption:HK]{A} and
\href{assumptionSI}{SI}. We present these lemmas below and leave their proofs
to Appendix D. Based on these results, we prove
Propositions \ref{Lemma:beta} and \ref{Lemma:gamma}. Throughout this appendix, we assume
$\gamma<0$. The proofs for the case with $\gamma\geq0$ are symmetric. We omit
them for conciseness.

\subsection*{Roadmap for Establishing Moment Inequality (\ref{idinequality})}\label{roadmap}
As indicated in the main text of the paper, the key is to establish the
identifying inequality (\ref{idinequality}). Here, we use the simplest case
with $T=4$ to illustrate how Assumption \hyperref[Assumption:SI]{SI}, together
with Assumption \hyperref[Assumption:HK]{A}, can ensure this inequality. The
proof for the most general case is lengthy but uses all analogous arguments.
We defer it to subsequent sections.

Our proof will repeatedly use the following identities: For any events $A,B$,
and $C$,
\begin{align}
P\left(  A|B\cap C\right)   &  =\frac{P\left(  A\cap B\cap C\right)
}{P\left(  B\cap C\right)  }=\frac{\left[  P\left(  A\cap B\cap C\right)
/P\left(  A\cap C\right)  \right]  \left[  P\left(  A\cap C\right)  /P\left(
C\right)  \right]  }{P\left(  B\cap C\right)  /P\left(  C\right)  }\nonumber\\
&  =\frac{P\left(  B|A\cap C\right)  P\left(  A|C\right)  }{P\left(
B|C\right)  }; \label{explain_pre1}%
\end{align}
if event $A$ implies event $B,$%
\[
A\subseteq B\text{ and }A\cap B=A;
\]
if $A\perp B|C,$%
\begin{equation}
P\left(  A|B\cap C\right)  =P\left(  A|C\right)  ; \label{explain_pre2}%
\end{equation}
if $A\subseteq B|C,$
\begin{equation}
P\left(  A\cap B|C\right)  =P\left(  A|C\right)  ; \label{explain_pre3}%
\end{equation}
and for any partition $\{B_{1},...,B_{m}\}$ of the sample space,
\begin{equation}
P(A|C)=\sum_{k=1}^{m}P(A\cap B_{j}|C)=\sum_{k=1}^{m}P(A|B_{j}\cap
C)P(B_{j}|C). \label{explain_pre4}%
\end{equation}
The following result is useful:%
\[
\left(  x_{t+j},\epsilon_{t+j}\right)  \perp y_{t}|\alpha,
\]
holds for any $j>0$, due to $\left\{  x_{t},\epsilon_{t}\right\}
\perp\left\{  x_{s},\epsilon_{s}\right\}  |\alpha$ for $s\neq t$ and the fact
that $y_{t}$ is a function of current and previous periods of $\left\{
x_{s},\epsilon_{s}\right\}  .$

We present the proof first and defer the explanations of $\overset{(\cdot
)}{=}$ to the end$.$

Consider the conditional probability $P(y_{1}=1|x_{1},x_{3},y_{0}=y_{2}%
=y_{4}=1,\alpha)$. We can write
\begin{align}
&  P(y_{1}=1|x_{1},x_{3},y_{0}=y_{2}=y_{4}=1,\alpha)\nonumber\\
\overset{(\text{i})}{=}  &  \frac{P(y_{4}=1|x_{1},x_{3},y_{0}=y_{1}%
=y_{2}=1,\alpha)P(y_{1}=1|x_{1},x_{3},y_{0}=y_{2}=1,\alpha)}{P(y_{4}%
=1|x_{1},x_{3},y_{0}=y_{2}=1,\alpha)}\nonumber\\
\overset{\left(  \text{ii}\right)  }{=}  &  P(y_{1}=1|x_{1},x_{3},y_{0}%
=y_{2}=1,\alpha)\nonumber\\
\overset{\left(  \text{iii}\right)  }{=}  &  P(y_{1}=1|x_{1},y_{0}%
=y_{2}=1,\alpha). \label{explain_1}%
\end{align}

In what follows, we assume w.l.o.g. that $\gamma<0$. The proof for the case
$\gamma\geq0$ is symmetric. We define a partition of the sample space:
\[
E_{2,1}=\{\epsilon_{2}<x_{2}^{\prime}\beta+\gamma+\alpha\},E_{2,2}%
=\{x_{2}^{\prime}\beta+\gamma+\alpha\leq\epsilon_{2}<x_{2}^{\prime}%
\beta+\alpha\},\text{and }E_{2,3}=\{\epsilon_{2}\geq x_{2}^{\prime}%
\beta+\alpha\}.
\]
From the model, $E_{2,1}$ implies $\left\{  y_{2}=1\right\}  $, so
$E_{2,1}\subseteq\left\{  y_{2}=1\right\}  $ and $E_{2,1}\cap\left\{
y_{2}=1\right\}  =E_{2,1}$. Similarly, $E_{2,3}$ implies $\left\{
y_{2}=0\right\}  $, so $E_{2,3}\subseteq\left\{  y_{2}=0\right\}  $ and
$E_{2,3}\cap\left\{  y_{2}=0\right\}  =E_{2,3}$.

Then, we use this partition and (\ref{explain_pre4}) to write
\begin{align}
P(y_{1} &  =1|x_{1},y_{0}=y_{2}=1,\alpha)\nonumber\\
&  =\sum_{k=1}^{3}P(y_{1}=1|x_{1},y_{0}=y_{2}=1,E_{2,k},\alpha)P(E_{2,k}%
|x_{1},y_{0}=y_{2}=1,\alpha)\nonumber\\
&  \overset{(\text{iv})}{=}P(y_{1}=1|x_{1},y_{0}=y_{2}=1,E_{2,1}%
,\alpha)P(E_{2,1}|x_{1},y_{0}=y_{2}=1,\alpha)\nonumber\\
&  \overset{(\text{v})}{=}P(y_{1}=1|x_{1},y_{0}=1,E_{2,1},\alpha
)P(E_{2,1}|x_{1},y_{0}=y_{2}=1,\alpha)\nonumber\\
&  \overset{(\text{vi})}{=}P(y_{1}=1|x_{1},y_{0}=1,\alpha)P(E_{2,1}%
|x_{1},y_{0}=y_{2}=1,\alpha)\nonumber\\
&  \overset{(\text{vii})}{=}F_{\epsilon|\alpha}(x_{1}^{\prime}\beta
+\gamma+\alpha)P(E_{2,1}|x_{1},y_{0}=y_{2}=1,\alpha).\label{explain_2}%
\end{align}

Applying (\ref{explain_pre4}) to the term $P(E_{2,1}|x_{1},y_{0}%
=y_{2}=1,\alpha)$ gives%
\begin{align}
&  P(E_{2,1}|x_{1},y_{0}=y_{2}=1,\alpha)\nonumber\\
=  &  P(E_{2,1}\cap\left\{  y_{1}=1\right\}  |x_{1},y_{0}=y_{2}=1,\alpha
)+P(E_{2,1}\cap\left\{  y_{1}=0\right\}  |x_{1},y_{0}=y_{2}=1,\alpha
)\nonumber\\
\overset{(\text{viii})}{=}  &  P(y_{1}=1|x_{1},y_{0}=y_{2}=1,\alpha
)+P(E_{2,1}|x_{1},y_{0}=y_{2}=1,y_{1}=0,\alpha)P(y_{1}=0|x_{1},y_{0}%
=y_{2}=1,\alpha)\nonumber\\
\overset{\left(  \text{ix}\right)  }{=}  &  P(y_{1}=1|x_{1},y_{0}%
=y_{2}=1,\alpha)\nonumber\\
&  +\frac{P(E_{2,1}\cap\left\{  y_{2}=1\right\}  |x_{1},y_{0}=1,y_{1}%
=0,\alpha)}{P(y_{2}=1|x_{1},y_{0}=1,y_{1}=0,\alpha)}\left[  1-P(y_{1}%
=1|x_{1},y_{0}=y_{2}=1,\alpha)\right] \nonumber\\
\overset{\left(  \text{x}\right)  }{=}  &  P(y_{1}=1|x_{1},y_{0}%
=y_{2}=1,\alpha)+\frac{P(E_{2,1}|x_{1},y_{0}=1,y_{1}=0,\alpha)}{P(y_{2}%
=1|x_{1},y_{0}=1,y_{1}=0,\alpha)}\left[  1-P(y_{1}=1|x_{1},y_{0}%
=y_{2}=1,\alpha)\right] \nonumber\\
\overset{\left(  \text{xi}\right)  }{=}  &  P(y_{1}=1|x_{1},y_{0}%
=y_{2}=1,\alpha)+\frac{P(E_{2,1}|\alpha)}{P(y_{2}=1|y_{1}=0,\alpha)}\left[
1-P(y_{1}=1|x_{1},y_{0}=y_{2}=1,\alpha)\right]  . \label{explain_3}%
\end{align}

If we set $\Delta=P(y_{1}=1|x_{1},y_{0}=y_{2}=1,\alpha)$ as an unknown,
(\ref{explain_2})--(\ref{explain_3}) imply%
\[
\Delta=F_{\epsilon|\alpha}(x_{1}^{\prime}\beta+\gamma+\alpha)\left[
\Delta+\frac{P(E_{2,1}|\alpha)}{P(y_{2}=1|y_{1}=0,\alpha)}\left(
1-\Delta\right)  \right]  .
\]
Solve $\Delta$ out and apply (\ref{explain_1}), we obtain
\begin{align}
&  P(y_{1}=1|x_{1},x_{3},y_{0}=y_{2}=y_{4}=1,\alpha)\nonumber\\
= &  P(y_{1}=1|x_{1},y_{0}=y_{2}=1,\alpha)=\Delta\nonumber\\
= &  \frac{P(E_{2,1}|\alpha)/P(y_{2}=1|y_{1}=0,\alpha)}{1/F_{\epsilon|\alpha
}(x_{1}^{\prime}\beta+\gamma+\alpha)+1-P(E_{2,1}|\alpha)/P(y_{2}%
=1|y_{1}=0,\alpha)}.\label{explain_4}%
\end{align}
Furthermore, applying arguments for (\ref{explain_1})--(\ref{explain_3}) to
$P(y_{3}=1|x_{1},x_{3},y_{0}=y_{2}=y_{4}=1,\alpha)$ yields
\begin{align}
&  P(y_{3}=1|x_{1},x_{3},y_{0}=y_{2}=y_{4}=1,\alpha)\nonumber\\
= &  \frac{P(E_{4,1}|\alpha)/P(y_{4}=1|y_{3}=0,\alpha)}{1/F_{\epsilon|\alpha
}(x_{3}^{\prime}\beta+\gamma+\alpha)+1-P(E_{4,1}|\alpha)/P(y_{4}%
=1|y_{3}=0,\alpha)},\label{explain_5}%
\end{align}
with $E_{4,1}=\left\{  \epsilon_{4}<x_{4}^{\prime}\beta+\gamma+\alpha\right\}
$. Note that Assumptions \hyperref[Assumption:HK]{A}(a) and
\hyperref[Assumption:SI]{SI}(b) imply $P(E_{2,1}|\alpha)/P(y_{2}%
=1|y_{1}=0,\alpha)=P(E_{4,1}|\alpha)/P(y_{4}=1|y_{3}=0,\alpha)$, and
Assumption \hyperref[Assumption:HK]{A}(b) guarantees the monotonicity of
$F_{\epsilon|\alpha}(\cdot)$. Then identifying inequality (\ref{idinequality})
follows by putting all these results together and comparing (\ref{explain_4})
and (\ref{explain_5}).

To sum up, Assumption \hyperref[Assumption:SI]{SI}(a) eliminates the effects
of $x_{s}$ on $y_{t}$ through its dependence on $x_{t}$ for all $t\neq s$, and
Assumption \hyperref[Assumption:SI]{SI}(b) is placed to ensure that the
probabilities in (\ref{idinequality}) do not have time-varying
representations. Using similar arguments, inequality (\ref{idinequality}) can
be established for general cases.

We provide the explanations for $\overset{(\cdot)}{=}$ in the following.

Equality $($i$)$ in (\ref{explain_1}) follows from (\ref{explain_pre1}) by
setting $A=\left\{  y_{1}=1\right\}  ,$ $B=\left\{  y_{4}=1\right\}  $, and
$C=\left\{  x_{1},x_{3},y_{0}=y_{2}=1,\alpha\right\}  .$

Equality $($ii$)$ holds due to (\ref{explain_pre2}) and $\left\{
y_{4}=1\right\}  \perp\left\{  y_{1}=1\right\}  |\left\{  x_{1},x_{3}%
,y_{0}=y_{2}=1,\alpha\right\}  .$ To see why this conditional independence
holds, recall that by model (\ref{model})
\begin{align*}
y_{4}  &  =1\left[  x_{4}^{\prime}\beta+\gamma y_{3}+\alpha-\epsilon_{4}%
\geq0\right] \\
&  =1\left[  x_{4}^{\prime}\beta+\gamma1\left[  x_{3}^{\prime}\beta+\gamma
y_{2}+\alpha-\epsilon_{3}\geq0\right]  +\alpha-\epsilon_{4}\geq0\right]  ,
\end{align*}
and so conditioning on $\left\{  x_{1},x_{3},y_{0}=y_{2}=1,\alpha\right\}  ,$
the random terms remained in $y_{4}$ are $(x_{4},\epsilon_{3},\epsilon_{4}).$
Then Assumptions \hyperref[Assumption:HK]{A}(a) and \hyperref[Assumption:SI]%
{SI}(a), where we assume $\epsilon^{T}\perp\left(  x^{T},y_{0}\right)
|\alpha,$ and $\left\{  x_{t},\epsilon_{t}\right\}  \perp\left\{
x_{s},\epsilon_{s}\right\}  |\alpha$ for $s\neq t$, imply%
\[
\left(  x_{4},\epsilon_{3},\epsilon_{4}\right)  \perp\left\{  x_{1}%
,x_{3},y_{0}=y_{2}=1\right\}  ,\text{ }\left\{  y_{1}=1\right\}  |\alpha,
\]
and thus%
\[
\left(  x_{4},\epsilon_{3},\epsilon_{4}\right)  \perp\left\{  y_{1}=1\right\}
|\left\{  x_{1},x_{3},y_{0}=y_{2}=1,\alpha\right\}  ,
\]
implying
\[
\left\{  y_{4}=1\right\}  \perp\left\{  y_{1}=1\right\}  |\left\{  x_{1}%
,x_{3},y_{0}=y_{2}=1,\alpha\right\}  .
\]

Equality $($iii$)$ follows by $\left\{  y_{1}=1\right\}  \perp x_{3}|\left\{
x_{1},y_{0}=y_{2}=1,\alpha\right\}  $ which holds for the same reason as above.

Equality (iv) is due to $P(E_{2,3}|x_{1},y_{0}=y_{2}=1,\alpha)=0$ (Since
$E_{2,3}$ implies $\left\{  y_{2}=0\right\}  $ and thus $E_{2,3}%
\subseteq\left\{  y_{2}=0\right\}  $),$\ $and $P(y_{1}=1|x_{1},y_{0}%
=y_{2}=1,E_{2,2},\alpha)=0$ (Since $\left\{  y_{2}=1\right\}  \cap E_{2,2}$
implies $\left\{  y_{1}=0\right\}  $) .

Equality (v) follows by $E_{2,1}\subseteq\{y_{2}=1\}$.

Equality (vi) holds because $P(y_{1}=1|x_{1},y_{0}=1,E_{2,1},\alpha
)=P(y_{1}=1|x_{1},y_{0}=1,\alpha)$ implied by (\ref{explain_pre2}) (letting
$A=\{y_{1}=1\}$, $B=E_{2,1}$, and $C=\{x_{1},y_{0}=1,\alpha\}$).

Equality (vii) holds because of Assumption \hyperref[Assumption:HK]{A}(a) that
$\epsilon^{T}\perp\left(  x^{T},y_{0}\right)  |\alpha.$

Equality (viii) holds because conditional on $\left\{  y_{2}=1\right\}  $ and
$\gamma<0,$ $\left\{  y_{1}=1\right\}  $ implies $E_{2,1},$ and thus $\left\{
y_{1}=1\right\}  \subseteq E_{2,1}$ conditional on $\left\{  x_{1},y_{0}%
=y_{2}=1,\alpha\right\}  .$

Equality (ix) holds due to fact that $P\left(  A|B\cap C\right)  =P\left(
A\cap B|C\right)  /P\left(  B|C\right)  $.

Equality (x) holds since $E_{2,1}$ implies $\{y_{2}=1\}$ and thus
$E_{2,1}\subseteq\{y_{2}=1\}$.

Equality (xi) follows by model (\ref{model}), Assumption
\hyperref[Assumption:HK]{A}(a), Assumption \hyperref[Assumption:SI]{SI}(a),
and applying (\ref{explain_pre2}) (letting $A=E_{2,1}$, $B=\{x_{1}%
,y_{0}=1,y_{1}=0\}$, and $C=\{\alpha\}$ for $P(E_{2,1}|x_{1},y_{0}%
=1,y_{1}=0,\alpha)$, and $A=\{y_{2}=1\}$, $B=\{x_{1},y_{0}=1\}$, and
$C=\{y_{1}=0,\alpha\}$ for $P(y_{2}=1|x_{1},y_{0}=1,y_{1}=0,\alpha)$).

\subsection*{Technical Lemmas and Main Proofs for Identification}
Now we rigorously prove our identification results (Propositions \ref{Lemma:beta}--\ref{Lemma:gamma}). Before that, we first list necessary technical Lemmas whose proofs are presented in Appendix D. For each $t\in\mathcal{T}$, define the following partition of the sample
space:\footnote{For the case with $\gamma\geq0$, the proofs of Lemmas
\ref{Lemma:A1}--\ref{Lemma:A3} work through with the partition $E_{t,1}%
=\{\epsilon_{t}<w_{t}+\alpha\},E_{t,2}=\{w_{t}+\alpha\leq\epsilon_{t}%
<w_{t}+\gamma+\alpha\}$, and $E_{t,3}=\{\epsilon_{t}\geq w_{t}+\gamma
+\alpha\}$.}
\[
E_{t,1}=\{\epsilon_{t}<w_{t}+\gamma+\alpha\},E_{t,2}=\{w_{t}+\gamma+\alpha
\leq\epsilon_{t}<w_{t}+\alpha\},E_{t,3}=\{\epsilon_{t}\geq w_{t}+\alpha\}.
\]

\begin{lemma}
\label{Lemma:A1} Let $s,t\in\mathcal{T}$ such that $t\geq s+2$. Under
Assumption \hyperref[Assumption:HK]{A}, the following equalities hold for
both $\tau=s$ and $\tau=t$.
\begin{align}
&  P(y_{\tau}=1|w^{T},y_{s-1}=y_{t-1},y_{s+1}=y_{t+1}=1,\alpha)\nonumber\\
=  &  F_{\epsilon|\alpha}(w_{\tau}+\gamma y_{\tau-1} +\alpha)P(E_{\tau
+1,1}|w^{T},y_{s-1}=y_{t-1},y_{s+1}=y_{t+1}=1,\alpha), \label{eq:A1}%
\end{align}
and
\begin{align}
&  P(y_{\tau}=1|w^{T},y_{s-1}=y_{t-1},y_{s+1}=y_{t+1}=0,\alpha)\nonumber\\
=  &  P(E_{\tau+1,2}|w^{T},y_{s-1}=y_{t-1},y_{s+1}=y_{t+1}=0,\alpha
)\nonumber\\
&  +F_{\epsilon|\alpha}(w_{\tau}+\gamma y_{\tau-1}+\alpha)P(E_{\tau+1,3}%
|w^{T},y_{s-1}=y_{t-1},y_{s+1}=y_{t+1}=0,\alpha). \label{eq:A2}%
\end{align}

\end{lemma}

\begin{lemma}
\label{Lemma:A2} Let $s,t\in\mathcal{T}$ such that $t\geq s+2$. Under
Assumption \hyperref[Assumption:HK]{A}, the following equalities hold for
both $\tau=s$ and $\tau=t$.
\begin{align}
&  P(E_{\tau+1,1}|w^{T},y_{s-1}=y_{t-1},y_{s+1}=y_{t+1}=1,\alpha)\nonumber\\
=  &  P(y_{\tau}=1|w^{T},y_{s-1}=y_{t-1},y_{s+1}=y_{t+1}=1,\alpha)\nonumber\\
&  +\frac{F_{\epsilon|\alpha}(w_{\tau+1}+\gamma+\alpha)}{F_{\epsilon|\alpha
}(w_{\tau+1}+\alpha)}[1-P(y_{\tau}=1|w^{T},y_{s-1}=y_{t-1},y_{s+1}%
=y_{t+1}=1,\alpha)], \label{eq:A9}%
\end{align}
\begin{align}
&  P(E_{\tau+1,2}|w^{T},y_{s-1}=y_{t-1},y_{s+1}=y_{t+1}=0,\alpha)\nonumber\\
=  &  \frac{F_{\epsilon|\alpha}(w_{s+1}+\alpha)-F_{\epsilon|\alpha}%
(w_{s+1}+\gamma+\alpha)}{1-F_{\epsilon|\alpha}(w_{s+1}+\gamma+\alpha
)}P(y_{\tau}=1|w^{T},y_{s-1}=y_{t-1},y_{s+1}=y_{t+1}=0,\alpha), \label{eq:A10}%
\end{align}
and
\begin{align}
&  P(E_{\tau+1,3}|w^{T},y_{s-1}=y_{t-1},y_{s+1}=y_{t+1}=0,\alpha)\nonumber\\
=  &  \frac{1-F_{\epsilon|\alpha}(w_{\tau+1}+\alpha)}{1-F_{\epsilon|\alpha
}(w_{\tau+1}+\gamma+\alpha)}P(y_{\tau}=1|w^{T},y_{s-1}=y_{t-1},y_{s+1}%
=y_{t+1}=0,\alpha)\nonumber\\
&  +1-P(y_{\tau}=1|w^{T},y_{s-1}=y_{t-1},y_{s+1}=y_{t+1}=0,\alpha).
\label{eq:A11}%
\end{align}

\end{lemma}

\begin{lemma}
\label{Lemma:A3} If Assumptions \hyperref[Assumption:HK]{A} and
\href{assumptionSD}{SD} hold, then for all $s,t\in\mathcal{T}$,
\[
P(y_{t}=1|w_{s},w_{t},y_{s-1}=y_{t-1},y_{s+1}=y_{t+1},\alpha)\geq
P(y_{s}=1|w_{s},w_{t},y_{s-1}=y_{t-1},y_{s+1}=y_{t+1},\alpha)
\]
if and only if $w_{t}\geq w_{s}$.
\end{lemma}

\begin{lemma}
\label{Lemma:A4} If Assumptions \hyperref[Assumption:HK]{A} and
\href{assumptionSI}{SI} hold, then for all $s,t\in\mathcal{T}$,
\[
P(y_{t}=1|w_{s},w_{t},y_{s-1}=y_{t-1},y_{s+1}=y_{t+1},\alpha)\geq
P(y_{s}=1|w_{s},w_{t},y_{s-1}=y_{t-1},y_{s+1}=y_{t+1},\alpha)
\]
if and only if $w_{t}\geq w_{s}$.
\end{lemma}

We next prove Propositions \ref{Lemma:beta}--\ref{Lemma:gamma} in order.

\begin{proof}
[Proof of Proposition \ref{Lemma:beta}]The monotonic relation established in
either Lemma \ref{Lemma:A3} or Lemma \ref{Lemma:A4} implies that $\beta$
maximizes $Q_{1}(\cdot;\alpha)$. The remaining task is to show the uniqueness
of $\beta$ in $\mathcal{B}$, i.e., $Q_{1}(b;\alpha)=Q_{1}(\beta;\alpha)$
implies $b=\beta$. Here we assume $\beta_{1}>0$ w.l.o.g. as the case
$\beta_{1}<0$ is symmetric.

First note that for any $b\in\mathcal{B}$ such that $Q_{1}(b;\alpha
)=Q_{1}(\beta;\alpha)$, if
\[
P([x_{ts,1}b_{1}+\tilde{x}_{ts}^{\prime}\tilde{b}<0<x_{ts,1}\beta_{1}%
+\tilde{x}_{ts}^{\prime}\tilde{\beta}]\cup[x_{ts,1}\beta_{1}+\tilde{x}%
_{ts}^{\prime}\tilde{\beta}<0<x_{ts,1}b_{1}+\tilde{x}_{ts}^{\prime}\tilde
{b}])>0,
\]
then $\beta$ and $b$ will yield different realized values of the sign function
in $Q_{1}(\cdot;\alpha)$ with strictly positive probability, and thus
$Q_{1}(\beta;\alpha)>Q_{1}(b;\alpha)$. It then follows that $b_{1}>0$ must
hold, for otherwise by Assumption \hyperref[Assumption:HK]{A}(c) we have
\[
P(x_{ts,1}b_{1}+\tilde{x}_{ts}^{\prime}\tilde{b}<0<x_{ts,1}\beta_{1}+\tilde
{x}_{ts}^{\prime}\tilde{\beta}) = P(x_{ts,1}>-\tilde{x}_{ts}^{\prime}\tilde
{b}/b_{1},x_{ts,1}>-\tilde{x}_{ts}^{\prime}\tilde{\beta}/\beta_{1})>0.
\]
Then focusing on the case with $b_{1}>0$, we can write
\begin{align}
&  P([x_{ts,1}b_{1}+\tilde{x}_{ts}^{\prime}\tilde{b}<0<x_{ts,1}\beta
_{1}+\tilde{x}_{ts}^{\prime}\tilde{\beta}]\cup[x_{ts,1}\beta_{1}+\tilde
{x}_{ts}^{\prime}\tilde{\beta}<0<x_{ts,1}b_{1}+\tilde{x}_{ts}^{\prime}%
\tilde{b}])\nonumber\\
=  &  P([-\tilde{x}_{ts}^{\prime}\tilde{\beta}/\beta_{1}<x_{ts,1}<-\tilde
{x}_{ts}^{\prime}\tilde{b}/b_{1}]\cup[-\tilde{x}_{ts}^{\prime}\tilde{b}%
/b_{1}<x_{ts,1}<-\tilde{x}_{ts}^{\prime}\tilde{\beta}/\beta_{1}]),\nonumber
\end{align}
which implies that to make $Q_{1}(b;\alpha)=Q_{1}(\beta;\alpha)$ hold we must
have $P(\tilde{x}_{ts}^{\prime}\tilde{\beta}/\beta_{1}=\tilde{x}_{ts}^{\prime
}\tilde{b}/b_{1})=1$ by Assumption \hyperref[Assumption:HK]{A}(c).

However, whenever $b$ is not a scalar multiple of $\beta$, $P(\tilde{x}%
_{ts}^{\prime}\tilde{\beta}/\beta_{1}=\tilde{x}_{ts}^{\prime}\tilde{b}%
/b_{1})=1$ implies that $\tilde{\mathcal{X}}_{ts}$ is contained in a proper
linear subspace of $\mathbb{R}^{K-1}$ a.e., violating Assumption
\hyperref[Assumption:HK]{A}(d). As a result, we must have $b$ as a scalar
multiple of $\beta$, which leads to the desired result $b=\beta$ as $\Vert
b\Vert_{2}=\Vert\beta\Vert_{2}=1$ by the construction of the parameter space
$\mathcal{B}$ in Assumption \hyperref[Assumption:HK]{A}(e).
\end{proof}

\begin{proof}
[Proof of Proposition \ref{Lemma:gamma}]The proof uses the insight of HK. Here
we only prove case (ii) of Proposition \ref{Lemma:gamma} for $t>s+1$ as the
same method can be applied to case (i) where $s$ and $t$ are adjacent. Note
that it also suffices to prove that $\gamma$ uniquely maximizes the following
population objective function conditional on $\alpha$:
\begin{align*}
Q_{2,2}(\gamma;\beta,\alpha)\equiv &  \mathbb{E}\left\{  \left[
P(A|x^{T},w_{s+1}=w_{t+1},y_{s+1}=y_{t+1},\alpha)-P(B|x^{T},w_{s+1}%
=w_{t+1},y_{s+1}=y_{t+1},\alpha)\right]  \right. \\
&  \left.  \times sgn\left(  (w_{t}-w_{s})+r(d_{t-1}-d_{s-1})\right)
|\alpha\right\}  .
\end{align*}

First, note that under Assumptions \hyperref[Assumption:HK]{A}(a) and
\hyperref[Assumption:HK]{A}(b), we can write
\begin{align*}
&  P(A|x^{T},w_{s+1}=w_{t+1}=w,y_{s+1}=y_{t+1}=d,\alpha)\\
=  &  p_{0}(x^{T},\alpha)^{d_{0}}(1-p_{0}(x^{T},\alpha))^{1-d_{0}}\times
F_{\epsilon|\alpha}(w_{1}+\gamma d_{0}+\alpha)^{d_{1}}(1-F_{\epsilon|\alpha
}(w_{1}+\gamma d_{0}+\alpha))^{1-d_{1}}\\
&  \times\cdots\times(1-F_{\epsilon|\alpha}(w_{s}+\gamma d_{s-1}%
+\alpha))\times F_{\epsilon|\alpha}(w+\alpha)^{d}(1-F_{\epsilon|\alpha
}(w+\alpha))^{1-d}\\
&  \times\cdots\times F_{\epsilon|\alpha}(w_{t}+\gamma d_{t-1}+\alpha)\times
F_{\epsilon|\alpha}(w+\gamma+\alpha)^{d}(1-F_{\epsilon|\alpha}(w+\gamma
+\alpha))^{1-d}\\
&  \times\cdots\times F_{\epsilon|\alpha}(w_{T}+\gamma d_{T-1}+\alpha)^{d_{T}%
}(1-F_{\epsilon|\alpha}(w_{T}+\gamma d_{T-1}+\alpha))^{1-d_{T}}%
\end{align*}
for all $w\in\mathbb{R}$ and $d\in\{0,1\}$, and similarly,
\begin{align*}
&  P(B|x^{T},w_{s+1}=w_{t+1}=w,y_{s+1}=y_{t+1}=d,\alpha)\\
=  &  p_{0}(x^{T},\alpha)^{d_{0}}(1-p_{0}(x^{T},\alpha))^{1-d_{0}}\times
F_{\epsilon|\alpha}(w_{1}+\gamma d_{0}+\alpha)^{d_{1}}(1-F_{\epsilon|\alpha
}(w_{1}+\gamma d_{0}+\alpha))^{1-d_{1}}\\
&  \times\cdots\times F_{\epsilon|\alpha}(w_{s}+\gamma d_{s-1}+\alpha)\times
F_{\epsilon|\alpha}(w+\gamma+\alpha)^{d}(1-F_{\epsilon|\alpha}(w+\gamma
+\alpha))^{1-d}\\
&  \times\cdots\times(1-F_{\epsilon|\alpha}(w_{t}+\gamma d_{t-1}%
+\alpha))\times F_{\epsilon|\alpha}(w+\alpha)^{d}(1-F_{\epsilon|\alpha
}(w+\alpha))^{1-d}\\
&  \times\cdots\times F_{\epsilon|\alpha}(w_{T}+\gamma d_{T-1}+\alpha)^{d_{T}%
}(1-F_{\epsilon|\alpha}(w_{T}+\gamma d_{T-1}+\alpha))^{1-d_{T}}.
\end{align*}
Then, we obtain
\begin{align*}
&  \frac{P(A|x^{T},w_{s+1}=w_{t+1}=w,y_{s+1}=y_{t+1}=d,\alpha)}{P(B|x^{T}%
,w_{s+1}=w_{t+1}=w,y_{s+1}=y_{t+1}=d,\alpha)}\\
=  &  \frac{(1-F_{\epsilon|\alpha}(w_{s}+\gamma d_{s-1}+\alpha))\times
F_{\epsilon|\alpha}(w_{t}+\gamma d_{t-1}+\alpha)}{F_{\epsilon|\alpha}%
(w_{s}+\gamma d_{s-1}+\alpha)\times(1-F_{\epsilon|\alpha}(w_{t}+\gamma
d_{t-1}+\alpha))}\\
&  \times\frac{F_{\epsilon|\alpha}(w+\alpha)^{d}(1-F_{\epsilon|\alpha
}(w+\alpha))^{1-d}\times F_{\epsilon|\alpha}(w+\gamma+\alpha)^{d}%
(1-F_{\epsilon|\alpha}(w+\gamma+\alpha))^{1-d}}{F_{\epsilon|\alpha}%
(w+\gamma+\alpha)^{d}(1-F_{\epsilon|\alpha}(w+\gamma+\alpha))^{1-d}\times
F_{\epsilon|\alpha}(w+\alpha)^{d}(1-F_{\epsilon|\alpha}(w+\alpha))^{1-d}}\\
=  &  \frac{(1-F_{\epsilon|\alpha}(w_{s}+\gamma d_{s-1}+\alpha))\times
F_{\epsilon|\alpha}(w_{t}+\gamma d_{t-1}+\alpha)}{F_{\epsilon|\alpha}%
(w_{s}+\gamma d_{s-1}+\alpha)\times(1-F_{\epsilon|\alpha}(w_{t}+\gamma
d_{t-1}+\alpha))}%
\end{align*}
and therefore,
\[
P(A|x^{T},w_{s+1}=w_{t+1}=w,y_{s+1}=y_{t+1}=d,\alpha)\geq P(B|x^{T}%
,w_{s+1}=w_{t+1}=w,y_{s+1}=y_{t+1}=d,\alpha)
\]
if and only if $w_{t}+\gamma d_{t-1}\ge w_{s}+\gamma d_{s-1}$, which implies
that $\gamma$ maximizes $Q_{2,2}(\gamma;\beta,\alpha)$.

The remaining task is to show that $\gamma$ is unique in $\mathcal{R}$.
Suppose that there exists an $r\in\mathcal{R}\setminus\{ \gamma\}$ such that
$Q_{2,2}\left(  r;\beta,\alpha\right)  =Q_{2,2}\left(  \gamma;\beta
,\alpha\right)  $. Note that the value of $r$ (and $\gamma$) affects
$Q_{2,2}\left(  \cdot;\beta,\alpha\right)  $ only when $d_{s-1}\neq d_{t-1}$.
Here we assume that $d_{t-1}=1$ and $d_{s-1}=0$ (the case with $d_{t-1}=0$ and
$d_{s-1}=1$ is symmetric). Then by Assumption \hyperref[Assumption:HK]{A}(c),
the following probability is non-zero:
\[
P\left(  \left[  -\gamma<w_{t}-w_{s}<-r\right]  \cup\left[  -r<w_{t}%
-w_{s}<-\gamma\right]  \right)  .
\]
Consequently, $\gamma$ and $r$ yield different realized values of the sign
function in objective function $Q_{2,2}\left(  \cdot;\beta,\alpha\right)  $
with strictly positive probability, and hence $Q_{2,2}\left(  r;\beta
,\alpha\right)  <Q_{2,2}\left(  \gamma;\beta,\alpha\right)  $, a
contradiction. Then we can conclude that $Q_{2,2}\left(  r;\beta
,\alpha\right)  =Q_{2,2}\left(  \gamma;\beta,\alpha\right)  $ if and only if
$r=\gamma$, or equivalently $\gamma$ uniquely maximizes $Q_{2,2}\left(
\cdot;\beta,\alpha\right)  $ in $\mathcal{R}$.
\end{proof}

\section{Technical Lemmas and Main Proofs for Asymptotics}\label{SEC:main_proof}

In this section, we define a few technical terms and a few more technical
notations, present some technical lemmas, and prove our main
asymptotic theory, Theorem \ref{TH:betahat}. The proofs for the technical
lemmas are relegated to Appendix D.

The outline of the proof of Theorem \ref{TH:betahat} is as follows. Lemmas
\ref{LE:xi_ib} and \ref{LE:ita_ib} verify the technical conditions as required
in \cite*{SeoOtsu2018}. Those conditions can ensure the class of functions is
\textit{manageable }as in \cite*{KimPollard1990}. After that, the maximal
inequalities and asymptotics in \cite*{SeoOtsu2018} can be readily applied to
our estimator. Lemma \ref{LE:equi} deals with the impact of using $\hat{\beta
}$ on estimating $\hat{\gamma}$, using maximal inequalities established  in
\cite*{SeoOtsu2018}.\ Lemmas \ref{LE:asym_xi} and \ref{LE:asym_ita} obtain the
technical terms for the final asymptotics for $\hat{\beta}$ and $\hat{\gamma
}.$

Let $c$ and $C$\ denote some constants that may vary from line to line.
$\mathbb{E}_{n}$ denotes the expectation conditional on observations being
fixed. $\rightsquigarrow$ denotes weakly convergence in the sense of
\cite*{vdVaartWellner2000}. Let%
\[
\mathbb{G}_{n}\left(  f_{ni}\right)  \equiv n^{1/2}\sum_{i=1}^{n}\left[
f_{ni}-\mathbb{E}_{n}\left(  f_{n}\right)  \right]  ,
\]
for any $f_{ni}.$ To facilitate calculation, occasionally we may decompose
covariate $x$ into $\varpi\beta+x_{\beta}$ with a scalar $\varpi$ and
$x_{\beta}$ orthogonal to $\beta.$

We define the following technical terms used in lemmas:

\[
Z_{n,1}\left(  \boldsymbol{s}\right)  \equiv n^{2/3}\cdot n^{-1}\sum_{i=1}%
^{n}\xi_{i}\left(  \beta+\boldsymbol{s}n^{-1/3}\right)  ,
\]%
\[
Z_{n,2}\left(  s\right)  \equiv\left(  nh_{n}\right)  ^{2/3}\cdot n^{-1}%
\sum_{i=1}^{n}\varsigma_{ni}\left(  \gamma+s\left(  nh_{n}\right)
^{-1/3},\beta\right)  ,
\]
and%
\[
\hat{Z}_{n,2}\left(  s\right)  \equiv\left(  nh_{n}\right)  ^{2/3}\cdot
n^{-1}\sum_{i=1}^{n}\varsigma_{ni}\left(  \gamma+s\left(  nh_{n}\right)
^{-1/3},\hat{\beta}\right)  .
\]
Note that the $\boldsymbol{s}$ in $Z_{n,1}\left(  \boldsymbol{s}\right)  $ is
a $K\times1$ vector, and the $s$ in $Z_{n,2}\left(  s\right)  $ and $\hat{Z}_{n,2}\left(  s\right)  $ is a scalar.

\begin{lemma}
\label{LE:xi_ib}Suppose Assumptions \hyperref[Assumption:HK]{A},
\hyperref[Assumption:SI]{SI} (or \hyperref[Assumption:SD]{SD}),
and \ref{A:boundedDensity} hold. Then $\xi_{i}\left(  b\right)  $
satisfies Assumption M in \cite*{SeoOtsu2018}\textbf{.}
\end{lemma}

\begin{lemma}
\label{LE:ita_ib}Suppose Assumptions \hyperref[Assumption:HK]{A},
\hyperref[Assumption:SI]{SI} (or \hyperref[Assumption:SD]{SD}) and
\ref{A:boundedDensity}--\ref{A:hn}\ hold. Then $\varsigma_{ni}\left(
r,b\right)  $ satisfies Assumption M in \cite*{SeoOtsu2018}\textbf{.}
\end{lemma}

\begin{lemma}
\label{LE:asym_xi}Suppose Assumptions \hyperref[Assumption:HK]{A} and
\ref{A:boundedDensity}\ hold. Then
\[
\lim_{n\rightarrow\infty}n^{2/3}\mathbb{E}\left(  \xi_{i}\left(
\beta+\boldsymbol{s}n^{-1/3}\right)  \right)  =\frac{1}{2}\boldsymbol{s}%
^{\prime}V_{1}\boldsymbol{s},
\]
and
\[
\lim_{n\rightarrow\infty}n^{1/3}\mathbb{E}\left[  \xi_{i}\left(
\beta+\boldsymbol{s}n^{-1/3}\right)  \xi_{i}\left(  \beta+\boldsymbol{t}%
n^{-1/3}\right)  \right]  =H_{1}\left(  \boldsymbol{s},\boldsymbol{t}\right)
.
\]
$V_{1}$ is defined as%
\begin{equation}
V_{1}=-\int1\left[  x_{31}^{\prime}\beta=0\right]  \left(  \frac
{\partial\kappa\left(  x_{31}\right)  }{\partial x_{31}}^{\prime}\beta\right)
f_{x_{31}}\left(  x_{31}\right)  x_{31}x_{31}^{\prime}d\sigma_{0},
\label{EQ:V1}%
\end{equation}
\textbf{\medskip}with $\sigma_{0}$\ being the surface measure on $\left\{
x_{31}:x_{31}^{\prime}\beta=0\right\}  $ and%
\[
\kappa\left(  x\right)  =\mathbb{E}\left\{  P\left(  y_{i0}=y_{i2}%
=y_{i4}|x_{i1},x_{i3}\right)  \left\{  \mathbb{E}\left[  y_{i3}|y_{i2}%
=y_{i4},x_{i3}\right]  -\mathbb{E}\left[  y_{i1}|y_{i0}=y_{i2},x_{i1}\right]
\right\}  |x_{i31}=x\right\}  .
\]
$H_{1}\left(  \boldsymbol{s},\boldsymbol{t}\right)  $ is defined as%
\begin{equation}
H_{1}\left(  \boldsymbol{s},\boldsymbol{t}\right)  =\frac{1}{2}\int%
_{\mathbb{R}^{K-1}}\psi\left(  x_{\beta}\right)  \left[  \left\vert x_{\beta
}^{\prime}\boldsymbol{s}\right\vert +\left\vert x_{\beta}^{\prime
}\boldsymbol{t}\right\vert -\left\vert x_{\beta}^{\prime}\left(
\boldsymbol{s}-\boldsymbol{t}\right)  \right\vert \right]  f_{x_{31}}\left(
x_{\beta}\right)  dx_{\beta}, \label{EQ:H1_f}%
\end{equation}
where $\boldsymbol{s},\boldsymbol{t}$ are $K\times1$ vectors,
\[
\psi\left(  x\right)  =\mathbb{E}\left\{  P\left(  y_{i0}=y_{i2}=y_{i4}%
|x_{i1},x_{i3}\right)  \left\vert \mathbb{E}\left[  y_{i3}|y_{i2}%
=y_{i4},x_{i3}\right]  -\mathbb{E}\left[  y_{i1}|y_{i0}=y_{i2},x_{i1}\right]
\right\vert \text{ }|x_{i31}=x\right\}  ,
\]
and $x_{\beta}$ is orthogonal to $\beta.$
\end{lemma}

\begin{lemma}
\label{LE:equi}Suppose Assumptions \hyperref[Assumption:HK]{A} and
\ref{A:boundedDensity}--\ref{A:hn}\ hold. Then
\[
\hat{Z}_{n,2}\left(  s\right)  -Z_{n,2}\left(  s\right)  =o_{P}\left(
1\right)  ,
\]
where the small order term holds uniformly over $\left\vert s\right\vert \leq
C$ for any positive $C.$
\end{lemma}

\begin{lemma}
\label{LE:asym_ita}Suppose Assumptions \hyperref[Assumption:HK]{A},
\ref{A:boundedDensity}, \ref{A:kernel}, and \ref{A:hn}\ hold. Then
\[
\lim_{n\rightarrow\infty}\left(  nh_{n}\right)  ^{2/3}\mathbb{E}\left(
\varsigma_{ni}\left(  \gamma+s\left(  nh_{n}\right)  ^{-1/3},\beta\right)
\right)  =\frac{1}{2}V_{2}s^{2},
\]
and
\[
\lim_{n\rightarrow\infty}\left(  nh_{n}\right)  ^{1/3}\mathbb{E}\left(
h_{n}\varsigma_{ni}\left(  \gamma+s\left(  nh_{n}\right)  ^{-1/3}%
,\beta\right)  \varsigma_{ni}\left(  \gamma+t\left(  nh_{n}\right)
^{-1/3},\beta\right)  \right)  =H_{2}\left(  s,t\right)  .
\]
$V_{2}$ is defined as
\begin{align}
V_{2}  &  =-\int_{\mathbb{R}^{K-1}}\int1\left[  x_{21}^{\prime}\beta+\gamma
y_{30}=0\right]  \left(  \frac{\partial\mathbb{E}\left(  y_{21}|x_{21}%
,y_{30},x_{32}=x_{\beta}\right)  }{\partial\left(  y_{30},x_{21}^{\prime
}\right)  ^{\prime}}^{\prime}\left(
\begin{array}
[c]{c}%
\gamma\\
\beta
\end{array}
\right)  \right) \label{EQ:V2_11}\\
&  f\left(  x_{21},y_{30}|x_{32}=x_{\beta}\right)  \left\vert y_{30}%
\right\vert d\sigma_{0}f_{x_{32}}\left(  x_{\beta}\right)  dx_{\beta
}\nonumber\\
&  -\int_{\mathbb{R}^{K-1}}\int1\left[  x_{32}^{\prime}\beta+\gamma
y_{41}=0\right]  \left(  \frac{\partial\mathbb{E}\left(  y_{32}|x_{32}%
,y_{41},x_{43}=x_{\beta}\right)  }{\partial\left(  y_{41},x_{32}^{\prime
}\right)  ^{\prime}}^{\prime}\left(
\begin{array}
[c]{c}%
\gamma\\
\beta
\end{array}
\right)  \right) \nonumber\\
&  f\left(  x_{32},y_{41}|x_{43}=x_{\beta}\right)  \left\vert y_{41}%
\right\vert d\sigma_{0}f_{x_{43}}\left(  x_{\beta}\right)  dx_{\beta}\nonumber
\end{align}
with $\sigma_{0}$ denoting the surface measure of $\left\{  \left.  \left(
x_{21}^{\prime},y_{30}\right)  ^{\prime}\right\vert x_{21}^{\prime}%
\beta+\gamma y_{30}=0\right\}  $ in the first integral and the surface measure
of $\left\{  \left.  \left(  x_{32}^{\prime},y_{41}\right)  ^{\prime
}\right\vert x_{32}^{\prime}\beta+\gamma y_{41}=0\right\}  $ in the second
integral$.$\ $H_{2}\left(  s,t\right)  $ is defined as%
\begin{align}
&  H_{2}\left(  s,t\right) \label{EQ:H2}\\
&  =\frac{1}{2}\left(  \left\vert s\right\vert +\left\vert t\right\vert
-\left\vert s-t\right\vert \right)  \mathcal{\bar{K}}_{2}\int_{\mathbb{R}%
^{K-1}}\left\{  \mathbb{E}\left[  \left\vert y_{21}\right\vert |x_{21}%
^{\prime}\beta=-\gamma,y_{30}=1,x_{32}=x_{\beta}\right]  f\left(
y_{30}=1,x_{21}^{\prime}\beta=-\gamma|x_{32}=x_{\beta}\right)  \right.
\nonumber\\
&  +\left.  \mathbb{E}\left[  \left\vert y_{21}\right\vert |x_{21}^{\prime
}\beta=\gamma,y_{30}=-1,x_{32}=x_{\beta}\right]  f\left(  y_{30}%
=-1,x_{21}^{\prime}\beta=\gamma|x_{32}=x_{\beta}\right)  \right\}  f_{x_{32}%
}\left(  x_{\beta}\right)  dx_{\beta}\nonumber\\
&  +\frac{1}{2}\left(  \left\vert s\right\vert +\left\vert t\right\vert
-\left\vert s-t\right\vert \right)  \mathcal{\bar{K}}_{2}\int_{\mathbb{R}%
^{K-1}}\left\{  \mathbb{E}\left[  \left\vert y_{32}\right\vert |x_{32}%
^{\prime}\beta=-\gamma,y_{41}=1,x_{43}=x_{\beta}\right]  f\left(
y_{41}=1,x_{32}^{\prime}\beta=-\gamma|x_{43}=x_{\beta}\right)  \right.
\nonumber\\
&  \left.  +\mathbb{E}\left[  \left\vert y_{32}\right\vert |x_{32}^{\prime
}\beta=\gamma,y_{41}=-1,x_{43}=x_{\beta}\right]  f\left(  y_{41}%
=-1,x_{32}^{\prime}\beta=\gamma|x_{43}=x_{\beta}\right)  \right\}  f_{x_{43}%
}\left(  x_{\beta}\right)  dx_{\beta},\nonumber
\end{align}
where $s,t$ are scalars, $\mathcal{\bar{K}}_{2}=\int_{\mathbb{R}}%
\mathcal{K}\left(  u\right)  ^{2}du,$ and $x_{\beta}$ is orthogonal to
$\beta.$
\end{lemma}

Note $y_{30}$ can only take values $-1,0,$ or $1.$ We
assume that $y_{30}$ takes hypothetical values in $\left(
-1-\varepsilon,-1+\varepsilon\right)  $, $\left(  -\varepsilon,\varepsilon
\right)  ,$ and ($1-\varepsilon,1+\varepsilon$) for a small $\varepsilon>0$
when calculating derivatives with respect to $y_{30}.$ For example, $\left.
\frac{d\Phi\left(  y_{30}\right)  }{dy_{30}}\right\vert _{y_{30}=1}=\left.
\Phi^{\prime}\left(  y_{30}\right)  \right\vert _{y_{30}=1}=\Phi^{\prime
}\left(  1\right)$ for any continuous differentiable $\Phi$.

\begin{proof}
[Proof of Theorem \ref{TH:betahat}]We prove the first part of this theorem first.

Lemma \ref{LE:xi_ib} verifies the key technical conditions needed for applying
Theorem 1 in\textbf{ }\cite*{SeoOtsu2018}$.$ $\hat{\beta}-\beta=O_{P}\left(
n^{-1/3}\right)  $ by Assumption \ref{A:maximiser} and Lemma 1 in
\cite*{SeoOtsu2018}.

Notice that $\hat{\beta}$ can be equivalently obtained from%
\[
\arg\max_{b\in\mathcal{B}}n^{2/3}\cdot n^{-1}\sum_{i=1}^{n}\xi_{i}\left(
\beta+n^{-1/3}\cdot n^{1/3}\left(  b-\beta\right)  \right)  .
\]
Intuitively, we get the asymptotics of $n^{1/3}\left(  \hat{\beta}%
-\beta\right)  $ if we can get the asymptotics of
\[
Z_{n,1}\left(  \boldsymbol{s}\right)  =n^{2/3}\cdot n^{-1}\sum_{i=1}^{n}%
\xi_{i}\left(  \beta+\boldsymbol{s}n^{-1/3}\right)  .
\]

Lemma \ref{LE:asym_xi} calculates the mean and covariance kernel of
$Z_{n,1}\left(  \boldsymbol{s}\right)  .$ $\xi_{i}\left(  b\right)  $ is
uniformly bounded, so the Lindeberg condition for $Z_{n,1}\left(
\boldsymbol{s}\right)  $ is satisfied. Therefore, $Z_{n,1}\left(
\boldsymbol{s}\right)  $ is pointwise asymptotically normal. With Assumption
\ref{A:maximiser}, Theorem 1 in \cite*{SeoOtsu2018} implies the equicontinuity
of $Z_{n,1}\left(  \boldsymbol{s}\right)  ,$ and it yields $Z_{n,1}\left(
\boldsymbol{s}\right)  \rightsquigarrow Z_{1}\left(  \boldsymbol{s}\right)  ,$
where $Z_{1}\left(  \boldsymbol{s}\right)  $ is a Gaussian Process with
continuous sample paths, expected value $-\frac{1}{2}\boldsymbol{s}^{\prime
}V_{1}\boldsymbol{s},$ and covariance kernel $H_{1}\left(  \boldsymbol{s}%
,\boldsymbol{t}\right)  $ that can be calculated as in equation (\ref{EQ:H1_f}%
)$.$ As a result,
\[
n^{1/3}\left(  \hat{\beta}-\beta\right)  \overset{d}{\rightarrow}\arg
\max_{\boldsymbol{s}\in\mathbb{R}^{K}}Z_{1}\left(  \boldsymbol{s}\right)  ,
\]
by applying Theorem 1 in \cite*{SeoOtsu2018}.

We now prove the second part. The calculation of equation (D.31) in the proof of Lemma
\ref{LE:asym_ita} shows,%
\begin{align}
\mathbb{E}_{n}\left(  \varsigma_{ni}\left(  r,\hat{\beta}\right)
-\varsigma_{ni}\left(  \gamma,\beta\right)  \right)   &  =\frac{1}{2}\left(
r-\gamma,\left(  \hat{\beta}-\beta\right)  ^{\prime}\right)  \tilde{V}%
_{2}\left(
\begin{array}
[c]{c}%
r-\gamma\\
\hat{\beta}-\beta
\end{array}
\right) \label{EQ:EExp}\\
&  +o\left(  \left\Vert \left(  r-\gamma,\left(  \hat{\beta}-\beta\right)
^{\prime}\right)  ^{\prime}\right\Vert _{2}\right)  +o\left(  \left(
nh_{n}\right)  ^{-2/3}\right)  ,\nonumber
\end{align}
where $\tilde{V}_{2}$ is defined in equation (D.30).

The convergence rate of $\hat{\gamma}$ is $\left(  nh_{n}\right)  ^{-1/3},$
which can be seen from%
\begin{align*}
o_{P}\left(  \left(  nh_{n}\right)  ^{-2/3}\right)   &  \leq n^{-1}\sum
_{i=1}^{n}\varsigma_{ni}\left(  \hat{\gamma},\hat{\beta}\right)  -n^{-1}%
\sum_{i=1}^{n}\varsigma_{ni}\left(  \gamma,\hat{\beta}\right) \\
&  =n^{-1}\sum_{i=1}^{n}\varsigma_{ni}\left(  \hat{\gamma},\hat{\beta}\right)
-n^{-1}\sum_{i=1}^{n}\varsigma_{ni}\left(  \gamma,\beta\right)  +n^{-1}%
\sum_{i=1}^{n}\varsigma_{ni}\left(  \gamma,\beta\right)  -n^{-1}\sum_{i=1}%
^{n}\varsigma_{ni}\left(  \gamma,\hat{\beta}\right) \\
&  \leq\mathbb{E}_{n}\left(  \varsigma_{ni}\left(  \hat{\gamma},\hat{\beta
}\right)  -\varsigma_{ni}\left(  \gamma,\beta\right)  \right)  +\varepsilon
\left(  \left(  \hat{\gamma}-\gamma\right)  ^{2}+\left\Vert \hat{\beta}%
-\beta\right\Vert _{2}^{2}\right)  +O_{P}\left(  \left(  nh_{n}\right)
^{-2/3}\right) \\
&  +\mathbb{E}_{n}\left(  \varsigma_{ni}\left(  \gamma,\hat{\beta}\right)
-\varsigma_{ni}\left(  \gamma,\beta\right)  \right)  +\varepsilon\left\Vert
\hat{\beta}-\beta\right\Vert _{2}^{2}+O_{P}\left(  \left(  nh_{n}\right)
^{-2/3}\right) \\
&  \leq\left(  -c+\varepsilon\right)  \left(  \left(  \hat{\gamma}%
-\gamma\right)  ^{2}+2\left\Vert \hat{\beta}-\beta\right\Vert _{2}^{2}\right)
+o\left(  \left(  \hat{\gamma}-\gamma\right)  ^{2}+\left\Vert \hat{\beta
}-\beta\right\Vert _{2}^{2}\right)  +O_{P}\left(  \left(  nh_{n}\right)
^{-2/3}\right)  ,
\end{align*}
for each $\varepsilon>0$, where the first line holds by Assumption
\ref{A:maximiser}, the third to fourth lines hold by applying Lemma 1
in\textbf{ }\cite*{SeoOtsu2018}, the fifth line holds by equation (\ref{EQ:EExp}). By noting $\left\Vert \hat{\beta
}-\beta\right\Vert _{2}=O_{P}\left(  n^{-1/3}\right)  =o_{P}\left(  \left(
nh_{n}\right)  ^{-1/3}\right)  ,$ the inequality above implies%
\[
0\leq\left(  -c+\varepsilon\right)  \left(  \hat{\gamma}-\gamma\right)
^{2}+o\left(  \left(  \hat{\gamma}-\gamma\right)  ^{2}\right)  +O_{P}\left(
\left(  nh_{n}\right)  ^{-2/3}\right)  .
\]
Taking an $\varepsilon$ to satisfy $\varepsilon<<c$ yields that $\hat{\gamma
}-\gamma=O_{P}\left(  \left(  nh_{n}\right)  ^{-1/3}\right)  .$ So we only
need to get the limiting distribution of $\hat{Z}_{n,2}\left(  s\right)  $.

The analysis of $\hat{Z}_{n,2}\left(  s\right)  $ is complicated by including
the first-stage estimator $\hat{\beta}.$ Lemma \ref{LE:equi} shows that
$\hat{\beta}$ has no impact on the asymptotics of $\hat{\gamma}$. More
specifically,%
\begin{align}
\hat{Z}_{n,2}\left(  s\right)   &  =Z_{n,2}\left(  s\right)  +o_{P}\left(
1\right) \nonumber\\
&  =n^{1/6}h_{n}^{2/3}\mathbb{G}_{n}\left(  \varsigma_{ni}\left(
\gamma+s\left(  nh_{n}\right)  ^{-1/3},\beta\right)  \right)  +\left(
nh_{n}\right)  ^{2/3}\mathbb{E}\left(  \varsigma_{ni}\left(  \gamma+s\left(
nh_{n}\right)  ^{-1/3},\beta\right)  \right)  +o_{P}\left(  1\right)  ,
\label{EQ:Z2_pro}%
\end{align}
where $\mathbb{G}_{n}\left(  \varsigma_{ni}\left(  r,b\right)  \right)
=n^{-1/2}\sum_{i=1}^{n}\left(  \varsigma_{ni}\left(  r,b\right)
-\mathbb{E}_{n}\left(  \varsigma_{ni}\left(  r,b\right)  \right)  \right)  $.
As a result, the asymptotics is established if the weak convergence of the
leading term in equation (\ref{EQ:Z2_pro}) is proved.

Lemma \ref{LE:asym_ita} calculates the the mean of $\left(  nh_{n}\right)
^{2/3}\mathbb{E}\left(  \varsigma_{ni}\left(  \gamma+s\left(  nh_{n}\right)
^{-1/3},\beta\right)  \right)  $ and covariance kernel $n^{1/6}h_{n}%
^{2/3}\mathbb{G}_{n}\left(  \varsigma_{ni}\left(  \gamma+s\left(
nh_{n}\right)  ^{-1/3},\beta\right)  \right)  .$

Note%
\begin{align*}
&  \sum_{i=1}^{n}\left(  \left(  nh_{n}\right)  ^{2/3}\cdot n^{-1}\right)
^{2+\delta}\mathbb{E}\left[  \left\vert \varsigma_{ni}\left(  \gamma+s\left(
nh_{n}\right)  ^{-1/3},\beta\right)  \right\vert ^{2+\delta}\right] \\
&  =\left(  nh_{n}\right)  ^{-\delta/3}\cdot\left(  nh_{n}\right)
^{1/3}\mathbb{E}\left[  h_{n}^{1+\delta}\left\vert \varsigma_{ni}\left(
\gamma+s\left(  nh_{n}\right)  ^{-1/3},\beta\right)  \right\vert ^{2+\delta
}\right]  \rightarrow0
\end{align*}
for a small $\delta>0,$ because $\left(  nh_{n}\right)  ^{-\delta
/3}\rightarrow0$ and $\left(  nh_{n}\right)  ^{1/3}\mathbb{E}\left[
h_{n}^{1+\delta}\left\vert \varsigma_{ni}\left(  \gamma+s\left(
nh_{n}\right)  ^{-1/3},\beta\right)  \right\vert ^{2+\delta}\right]
\rightarrow c$ for a finite $c$. This verifies the Lyapunov condition for
$n^{1/6}h_{n}^{2/3}\mathbb{G}_{n}\left(  \varsigma_{ni}\left(  \gamma+s\left(
nh_{n}\right)  ^{-1/3},\beta\right)  \right)  .$ Therefore, it converges to
normal in distribution for each $s$. Lemma \ref{LE:ita_ib} verifies the key
technical conditions for applying Theorem 1 in\textbf{ }\cite*{SeoOtsu2018} to
$Z_{n,2}\left(  s\right)  $. Together with Assumption \ref{A:maximiser}, all
technical conditions in Theorem 1 of\textbf{ }\cite*{SeoOtsu2018} are satisfied
for $Z_{n,2}\left(  s\right)  $. That\textbf{ }implies the stochastic
equicontinuity of $Z_{n,2}\left(  s\right)  $ in $s$ and
\[
Z_{n,2}\left(  s\right)  \rightsquigarrow Z_{2}\left(  s\right)  ,
\]
where$\ Z_{2}\left(  s\right)  $ is a Gaussian process with continuous path,
expected value $\frac{1}{2}V_{2}s^{2}$ and covariance kernel $H_{2}\left(
s,t\right)$. Then, the following result follows by the continuous mapping theorem:
\[
\left(  nh_{n}\right)  ^{1/3}\left(  \hat{\gamma}-\gamma\right)
\overset{d}{\rightarrow}\arg\max_{s\in\mathbb{R}}Z_{2}\left(  s\right).
\]
\end{proof}


\section{Simulation Results of Designs 1 and 2\label{SEC:tables}}
\begin{center}%
\begin{tabular}
[c]{rcccccccc}%
\multicolumn{9}{c}{Table 1A: Design 1, Performance of $\hat{\beta}$ and
$\hat{\gamma}$}\\\hline\hline
& \multicolumn{2}{c}{$n=2500$} & \multicolumn{2}{c}{$n=5000$} & \multicolumn{2}{c}{$n=10000$} &
\multicolumn{2}{c}{$n=20000$}\\\cline{2-9}
& $\hat{\beta}_{2}$ & $\hat{\gamma}$ & $\hat{\beta}_{2}$ & $\hat{\gamma}$ &
$\hat{\beta}_{2}$ & $\hat{\gamma}$ & $\hat{\beta}_{2}$ & $\hat{\gamma}$ \\\hline
 OY \ BIAS & $1.8\%$ & $-0.9\%$ & $1.7\%$ & $-0.3\%$ & $0.8\%$ & $1.5\%$ & $0.7\%$ & $0.5\%$ \\
     STD& $20.2\%$ & $25.0\%$ & $14.7\%$ & $18.9\%$ & $11.4\%$ & $14.7\%$ & $9.1\%$ & $12.0\%$ \\
     MAD& $15.9\%$ & $19.4\%$ & $11.7\%$ & $15.1\%$ & $9.0\%$ & $11.8\%$ & $7.3\%$ & $9.7\%$ \\
    RMSE& $20.2\%$ & $25.0\%$ & $14.8\%$ & $18.9\%$ & $11.4\%$ & $14.8\%$ & $9.1\%$ & $12.0\%$ \\
    TIME & \multicolumn{2}{c}{0.33 seconds} & \multicolumn{2}{c}{0.62 seconds} & \multicolumn{2}{c}{1.16 seconds}  & \multicolumn{2}{c}{2.50 seconds} \\\hline

HK1 BIAS & $-0.4\%$ & $2.4\%$ & $-0.0\%$ & $1.8\%$ & $-0.2\%$ & $1.4\%$ & $-0.1\%$ & $1.2\%$ \\
     STD& $5.7\%$ & $15.5\%$ & $4.6\%$ & $11.6\%$ & $3.6\%$ & $8.8\%$ & $2.8\%$ & $6.8\%$ \\
     MAD& $4.6\%$ & $12.3\%$ & $3.7\%$ & $9.2\%$ & $2.9\%$ & $7.2\%$ & $2.2\%$ & $5.5\%$ \\
    RMSE& $5.7\%$ & $15.6\%$ & $4.6\%$ & $11.7\%$ & $3.6\%$ & $9.0\%$ & $2.8\%$ & $6.9\%$ \\
    TIME & \multicolumn{2}{c}{4.27 seconds} & \multicolumn{2}{c}{7.68 seconds} & \multicolumn{2}{c}{14.01 seconds}  & \multicolumn{2}{c}{30.02 seconds} \\\hline

 HK2 BIAS & $-0.0\%$ & $2.0\%$ & $0.2\%$ & $1.5\%$ & $-0.1\%$ & $2.8\%$ & $0.2\%$ & $1.2\%$ \\
     STD& $12.1\%$ & $29.8\%$ & $10.0\%$ & $24.8\%$ & $8.8\%$ & $21.0\%$ & $7.4\%$ & $18.1\%$ \\
     MAD& $9.6\%$ & $24.1\%$ & $8.2\%$ & $19.8\%$ & $7.0\%$ & $17.1\%$ & $5.9\%$ & $14.6\%$ \\
    RMSE& $12.1\%$ & $29.9\%$ & $10.0\%$ & $24.8\%$ & $8.8\%$ & $21.2\%$ & $7.4\%$ & $18.1\%$ \\
        TIME & \multicolumn{2}{c}{0.58 seconds} & \multicolumn{2}{c}{1.05 seconds} & \multicolumn{2}{c}{3.00 seconds}  & \multicolumn{2}{c}{3.96 seconds} \\\hline\hline
\end{tabular}

\begin{tabular}
[c]{rcccccccc}%
\multicolumn{9}{c}{Table 1B: Design 1, Numerical Bootstrap}\\\hline\hline
& \multicolumn{2}{c}{$n=2500$} & \multicolumn{2}{c}{$n=5000$} & \multicolumn{2}{c}{$n=10000$} &
\multicolumn{2}{c}{$n=20000$}\\\cline{2-9}
 & $\hat{\beta}_{2}$ & $\hat{\gamma}$ & $\hat{\beta}_{2}$ & $\hat{\gamma}$ &
$\hat{\beta}_{2}$ & $\hat{\gamma}$ & $\hat{\beta}_{2}$ & $\hat{\gamma}$\\\hline
 $c = 0.8$ COV & $87.8\%$ & $89.8\%$ & $91.0\%$ & $90.7\%$ & $91.6\%$ & $93.3\%$ & $91.1\%$ & $92.5\%$ \\
 LEN & $96.7\%$ & $99.1\%$ & $81.6\%$ & $83.2\%$ & $68.7\%$ & $68.4\%$ & $57.8\%$ & $54.7\%$ \\\hline
 $c = 0.9$ COV & $87.6\%$ & $89.8\%$ & $91.7\%$ & $91.5\%$ & $91.7\%$ & $94.1\%$ & $90.8\%$ & $93.5\%$ \\
 LEN & $94.5\%$ & $97.3\%$ & $80.2\%$ & $82.5\%$ & $67.4\%$ & $68.2\%$ & $56.8\%$ & $55.1\%$ \\\hline
 $c = 1.0$ COV & $88.4\%$ & $90.1\%$ & $91.6\%$ & $91.1\%$ & $92.0\%$ & $93.5\%$ & $90.9\%$ & $93.2\%$ \\
 LEN & $93.1\%$ & $96.3\%$ & $78.7\%$ & $81.8\%$ & $66.2\%$ & $68.0\%$ & $55.8\%$ & $55.5\%$ \\\hline
 $c = 1.1$ COV & $88.1\%$ & $89.9\%$ & $92.4\%$ & $90.9\%$ & $91.9\%$ & $93.9\%$ & $91.3\%$ & $93.2\%$ \\
 LEN & $91.5\%$ & $94.7\%$ & $77.8\%$ & $81.1\%$ & $65.4\%$ & $67.7\%$ & $55.1\%$ & $55.7\%$ \\\hline
 $c = 1.2$ COV & $88.5\%$ & $89.8\%$ & $91.4\%$ & $90.6\%$ & $91.6\%$ & $93.5\%$ & $91.6\%$ & $94.1\%$ \\
 LEN & $90.0\%$ & $93.0\%$ & $76.8\%$ & $80.0\%$ & $64.5\%$ & $67.5\%$ & $54.3\%$ & $55.7\%$ \\\hline\hline
\end{tabular}

\begin{tabular}
[c]{rcccccc}
\multicolumn{7}{c}{Table 1C: Design 1, $n=1000$} \label{T:n1000}\\\hline\hline
 & BIAS & STD & MAD  & RMSE & COV & LEN \\\hline
OY \  $\hat{\beta}_{2}$ & $6.4\%$ & $28.7\%$ & $22.5\%$ & $29.4\%$ & $83.5\%$ & $114.1\%$ \\
$\hat{\gamma}$ & $-4.2\%$ & $35.0\%$ & $27.0\%$ & $35.2\%$ & $85.5\%$ & $116.3\%$ \\\hline
HK1 $\hat{\beta}_{2}$  & $-0.2\%$ & $7.6\%$ & $6.1\%$ & $7.6\%$ & -- & -- \\
$\hat{\gamma}$ & $2.5\%$ & $21.7\%$ & $17.2\%$ & $21.8\%$ & -- & -- \\\hline
HK2 $\hat{\beta}_{2}$  & $0.2\%$ & $14.4\%$ & $11.7\%$ & $14.4\%$ & -- & -- \\
$\hat{\gamma}$ & $1.0\%$ & $36.9\%$ & $29.4\%$ & $36.9\%$ & -- & -- \\
\hline\hline
\end{tabular}

\begin{tabular}
[c]{rcccccccc}%
\multicolumn{9}{c}{Table 2A: Design 2, Performance of $\hat{\beta}$ and
$\hat{\gamma}$}\\\hline\hline
& \multicolumn{2}{c}{$n=2500$} & \multicolumn{2}{c}{$n=5000$} & \multicolumn{2}{c}{$n=10000$} &
\multicolumn{2}{c}{$n=20000$}\\\cline{2-9}
& $\hat{\beta}_{2}$ & $\hat{\gamma}$ & $\hat{\beta}_{2}$ & $\hat{\gamma}$ &
$\hat{\beta}_{2}$ & $\hat{\gamma}$ & $\hat{\beta}_{2}$ & $\hat{\gamma}$ \\\hline
 OY BIAS & $2.2\%$ & $0.9\%$ & $1.9\%$ & $0.9\%$ & $0.3\%$ & $1.3\%$ & $-0.1\%$ & $1.1\%$ \\
     STD& $19.0\%$ & $22.9\%$ & $15.0\%$ & $17.7\%$ & $11.7\%$ & $14.7\%$ & $9.2\%$ & $11.2\%$ \\
     MAD& $15.1\%$ & $18.5\%$ & $11.9\%$ & $14.1\%$ & $9.4\%$ & $11.9\%$ & $7.3\%$ & $9.0\%$ \\
    RMSE& $19.2\%$ & $22.9\%$ & $15.1\%$ & $17.7\%$ & $11.7\%$ & $14.8\%$ & $9.2\%$ & $11.2\%$ \\\hline

 HK1 BIAS & $-0.0\%$ & $2.9\%$ & $0.1\%$ & $2.6\%$ & $-0.1\%$ & $2.0\%$ & $-0.0\%$ & $1.8\%$ \\
     STD& $5.8\%$ & $12.9\%$ & $4.2\%$ & $10.4\%$ & $3.5\%$ & $8.2\%$ & $2.7\%$ & $6.5\%$ \\
     MAD& $4.7\%$ & $10.6\%$ & $3.4\%$ & $8.7\%$ & $2.8\%$ & $6.8\%$ & $2.2\%$ & $5.4\%$ \\
    RMSE& $5.8\%$ & $13.3\%$ & $4.2\%$ & $10.8\%$ & $3.5\%$ & $8.5\%$ & $2.7\%$ & $6.8\%$ \\\hline

 HK2 BIAS & $0.2\%$ & $3.4\%$ & $0.2\%$ & $2.6\%$ & $-0.2\%$ & $2.3\%$ & $0.3\%$ & $2.0\%$ \\
     STD& $11.9\%$ & $26.4\%$ & $10.1\%$ & $23.5\%$ & $8.9\%$ & $19.1\%$ & $6.9\%$ & $16.2\%$ \\
     MAD& $9.6\%$ & $21.3\%$ & $8.1\%$ & $19.0\%$ & $7.2\%$ & $15.4\%$ & $5.5\%$ & $12.9\%$ \\
    RMSE& $11.9\%$ & $26.6\%$ & $10.1\%$ & $23.6\%$ & $8.9\%$ & $19.2\%$ & $6.9\%$ & $16.3\%$ \\\hline\hline
\end{tabular}

\begin{tabular}
[c]{rcccccccc}%
\multicolumn{9}{c}{Table 2B: Design 2, Numerical Bootstrap}\\\hline\hline
& \multicolumn{2}{c}{$n=2500$} & \multicolumn{2}{c}{$n=5000$} & \multicolumn{2}{c}{$n=10000$} &
\multicolumn{2}{c}{$n=20000$}\\\cline{2-9}
 & $\hat{\beta}_{2}$ & $\hat{\gamma}$ & $\hat{\beta}_{2}$ & $\hat{\gamma}$ &
$\hat{\beta}_{2}$ & $\hat{\gamma}$ & $\hat{\beta}_{2}$ & $\hat{\gamma}$\\\hline
 $c = 0.8$ COV & $89.4\%$ & $91.7\%$ & $90.1\%$ & $91.1\%$ & $89.2\%$ & $91.4\%$ & $91.2\%$ & $92.3\%$ \\
 LEN & $97.2\%$ & $99.8\%$ & $81.5\%$ & $82.8\%$ & $68.8\%$ & $67.6\%$ & $57.9\%$ & $54.4\%$ \\\hline
 $c = 0.9$ COV & $89.8\%$ & $91.6\%$ & $90.5\%$ & $91.4\%$ & $89.7\%$ & $91.6\%$ & $91.6\%$ & $93.0\%$ \\
 LEN & $95.2\%$ & $98.1\%$ & $80.3\%$ & $82.4\%$ & $67.5\%$ & $68.0\%$ & $57.0\%$ & $54.8\%$ \\\hline
 $c = 1.0$ COV & $89.6\%$ & $91.4\%$ & $91.7\%$ & $91.4\%$ & $89.0\%$ & $92.3\%$ & $92.2\%$ & $93.7\%$ \\
 LEN & $93.4\%$ & $96.4\%$ & $78.7\%$ & $81.7\%$ & $66.5\%$ & $67.8\%$ & $56.1\%$ & $55.2\%$ \\\hline
 $c = 1.1$ COV & $89.1\%$ & $90.7\%$ & $91.0\%$ & $91.6\%$ & $89.0\%$ & $92.2\%$ & $91.6\%$ & $93.9\%$ \\
 LEN & $92.0\%$ & $95.1\%$ & $77.7\%$ & $80.8\%$ & $65.4\%$ & $67.4\%$ & $55.2\%$ & $55.5\%$ \\\hline
 $c = 1.2$ COV & $88.8\%$ & $90.8\%$ & $91.0\%$ & $91.6\%$ & $89.4\%$ & $91.6\%$ & $91.9\%$ & $93.8\%$ \\
 LEN & $90.5\%$ & $93.4\%$ & $76.6\%$ & $80.0\%$ & $64.6\%$ & $67.3\%$ & $54.5\%$ & $55.5\%$ \\\hline\hline
\end{tabular}
\end{center}

\newpage
\noindent \textbf{\LARGE{Online Supplementary Material}}

\noindent Appendix \ref{SEC:tech_proof} contains proofs for all technical lemmas used in Appendices \ref{Sec_identification} and \ref{SEC:main_proof}. Appendix \ref{SEC:Tinfereneces} offers technical details for Section \ref{SEC:inferences}. Additionally, Appendix \ref{app:moretables} presents tables summarizing simulation results for Designs 3--5. It also includes supplementary Monte Carlo experiments (Designs 6--8) that investigate the impact of serial correlations of $x_{it}$ on our proposed estimation and inference procedure.

\section{Proofs for Technical Lemmas \label{SEC:tech_proof}}
\begin{proof}
[Proof of Lemma \ref{Lemma:A1}]Here we only prove the case $\tau=s$. The
derivation for case $\tau=t$ is analogous.

{First,} note that by {the} law of total probability, we can write for all $d_{1}%
\in\{0,1\}$,
\begin{align}
&  P(y_{s}=1|w^{T},y_{s-1}=y_{t-1},y_{s+1}=y_{t+1}=d_{1},\alpha)\nonumber\\
=  &  \sum_{j=1}^{3}\left\{  P(y_{s}=1|w^{T},y_{s-1}=y_{t-1},y_{s+1}%
=y_{t+1}=d_{1},\alpha,E_{s+1,j})\right. \nonumber\\
&  \left.  \times P(E_{s+1,j}|w^{T},y_{s-1}=y_{t-1},y_{s+1}=y_{t+1}%
=d_{1},\alpha)\right\}  . \label{eq:A3}%
\end{align}

When $d_{1}=1$, (\ref{eq:A3}) reduces to
\begin{align}
&  P(y_{s}=1|w^{T},y_{s-1}=y_{t-1},y_{s+1}=y_{t+1}=1,\alpha)\nonumber\\
=  &  P(y_{s}=1|w^{T},y_{s-1}=y_{t-1},y_{s+1}=y_{t+1}=1,\alpha,E_{s+1,1}%
)\nonumber\\
&  \times P(E_{s+1,1}|w^{T},y_{s-1}=y_{t-1},y_{s+1}=y_{t+1}=1,\alpha)
\label{eq:A4}%
\end{align}
as by definition $E_{s+1,3}\cap\{y_{s+1}=1\}=\emptyset$ and by Bayes'
theorem\footnote{The Bayes' theorem is stated mathematically as the following
equation
\[
P(A|B,C)=P(B|A,C)P(A|C)/P(B|C)
\]
where $A$, $B$ and $C$ are events and $P(B|C)>0$. Here, we apply Bayes' theorem
by letting $A=\{y_{s}=1\}$, $B=\{y_{s+1}=1,E_{s+1,2}\}$, and $C=\{w^{T}%
,y_{s-1}=y_{t-1},y_{t+1}=1,\alpha\}$.}
\begin{align*}
&  P(y_{s}=1|w^{T},y_{s-1}=y_{t-1},y_{s+1}=y_{t+1}=1,\alpha,E_{s+1,2})\\
=  &  \frac{P(y_{s+1}=1,E_{s+1,2}|w^{T},y_{s-1}=y_{t-1},y_{t+1}=1,\alpha
,y_{s}=1)P(y_{s}=1|w^{T},y_{s-1}=y_{t-1},y_{t+1}=1,\alpha)}{P(y_{s+1}%
=1,E_{s+1,2}|w^{T},y_{s-1}=y_{t-1},y_{t+1}=1,\alpha)}\\
=  &  0,
\end{align*}
where the last equality is due to {the fact} that $E_{s+1,2}\cap\{y_{s+1}%
=1\}=E_{s+1,2}\cap E_{s+1,1}=\emptyset$ conditional on $\{y_{s}=1\}$.
Furthermore, under Assumption \hyperref[Assumption:HK]{A}(a), we can write
\begin{align}
&  P(y_{s}=1|w^{T},y_{s-1}=y_{t-1},y_{s+1}=y_{t+1}=1,\alpha,E_{s+1,1}%
)\nonumber\\
=  &  P(y_{s}=1|w^{T},y_{s-1}=y_{t-1},y_{t+1}=1,\alpha,E_{s+1,1}%
)=P(y_{s}=1|w^{T},y_{s-1},\alpha,E_{s+1,1})\nonumber\\
=  &  P(y_{s}=1|w^{T},y_{s-1},\alpha)=F_{\epsilon|\alpha}(w_{s}+\gamma
y_{s-1}+\alpha), \label{eq:A5}%
\end{align}
where the first equality uses the fact that $E_{s+1,1}\subset\{y_{s+1}=1\}$,
the second equality follows from noticing that $y_{s}$ (depends only on
$\epsilon_{s}$) is independent of $(y_{t-1},y_{t+1})$ (depend only on
$(\epsilon_{s+2},...,\epsilon_{t+1})$) conditional on $(w^{T},y_{s-1},\alpha)$
and event $E_{s+1,1}$, and the third equality is because $y_{s}\perp
E_{s+1,1}$ conditional on $(w^{T},y_{s-1},\alpha)$. Plugging (\ref{eq:A5})
into (\ref{eq:A4}) yields (\ref{eq:A1}).

When $d_{1}=0$, (\ref{eq:A3}) reduces to
\begin{align}
&  P(y_{s}=1|w^{T},y_{s-1}=y_{t-1},y_{s+1}=y_{t+1}=0,\alpha)\nonumber\\
=  &  \sum_{j=2}^{3}\left\{  P(y_{s}=1|w^{T},y_{s-1}=y_{t-1},y_{s+1}%
=y_{t+1}=0,\alpha,E_{s+1,j})\right. \nonumber\\
&  \left.  \times P(E_{s+1,j}|w^{T},y_{s-1}=y_{t-1},y_{s+1}=y_{t+1}%
=0,\alpha)\right\}  \label{eq:A6}%
\end{align}
as by definition $E_{s+1,1}\cap\{y_{s+1}=0\}=\emptyset$.

Using Bayes' theorem and the fact that $E_{s+1,2}\cap\{y_{s+1}=0\}=E_{s+1,2}%
\cap E_{s+1,3}=\emptyset$ conditional on $\{y_{s}=0\}$, we have
\begin{align*}
&  P(y_{s}=0|w^{T},y_{s-1}=y_{t-1},y_{s+1}=y_{t+1}=0,\alpha,E_{s+1,2})\\
=  &  \frac{P(y_{s+1}=0,E_{s+1,2}|w^{T},y_{s-1}=y_{t-1},y_{t+1}=0,\alpha
,y_{s}=0)P(y_{s}=0|w^{T},y_{s-1}=y_{t-1},y_{t+1}=0,\alpha)}{P(y_{s+1}%
=0,E_{s+1,2}|w^{T},y_{s-1}=y_{t-1},y_{t+1}=0,\alpha)}\\
=  &  \frac{P(E_{s+1,2}\cap E_{s+1,3}|w^{T},y_{s-1}=y_{t-1},y_{t+1}%
=0,\alpha,y_{s}=0)P(y_{s}=0|w^{T},y_{s-1}=y_{t-1},y_{t+1}=0,\alpha)}%
{P(y_{s+1}=0,E_{s+1,2}|w^{T},y_{s-1}=y_{t-1},y_{t+1}=0,\alpha)}\\
=  &  0,
\end{align*}
and thus
\begin{equation}
P(y_{s}=1|w^{T},y_{s-1}=y_{t-1},y_{s+1}=y_{t+1}=0,\alpha,E_{s+1,2})=1.
\label{eq:A7}%
\end{equation}

Applying similar arguments for deriving (\ref{eq:A5}) gives
\begin{align}
&  P(y_{s}=1|w^{T},y_{s-1}=y_{t-1},y_{s+1}=y_{t+1}=0,\alpha,E_{s+1,3}%
)\nonumber\\
=  &  P(y_{s}=1|w^{T},y_{s-1}=y_{t-1},y_{t+1}=0,\alpha,E_{s+1,3}%
)=P(y_{s}=1|w^{T},y_{s-1}=d,\alpha,E_{s+1,3})\nonumber\\
=  &  P(y_{s}=1|w^{T},y_{s-1},\alpha)=F_{\epsilon|\alpha}(w_{s}+\gamma
y_{s-1}+\alpha). \label{eq:A8}%
\end{align}
Then plugging (\ref{eq:A7}) and (\ref{eq:A8}) into (\ref{eq:A6}) yields
(\ref{eq:A2}).
\end{proof}

\begin{proof}
[Proof of Lemma \ref{Lemma:A2}]Again we only prove the case $\tau=s$ as the
same arguments can be applied to derive the case $\tau=t$. Note that for all
$j=1,2,3$, we can use {the} law of total probability to write
\begin{align*}
&  P(E_{s+1,j}|w^{T},y_{s-1}=y_{t-1},y_{s+1}=y_{t+1},\alpha)\\
=  &  P(E_{s+1,j}|w^{T},y_{s-1}=y_{t-1},y_{s+1}=y_{t+1},\alpha,y_{s}%
=0)P(y_{s}=0|w^{T},y_{s-1}=y_{t-1},y_{s+1}=y_{t+1},\alpha)\\
&  +P(E_{s+1,j}|w^{T},y_{s-1}=y_{t-1},y_{s+1}=y_{t+1},\alpha,y_{s}%
=1)P(y_{s}=1|w^{T},y_{s-1}=y_{t-1},y_{s+1}=y_{t+1},\alpha)\\
=  &  P(E_{s+1,j}|w^{T},y_{s-1},y_{s+1},\alpha,y_{s}=0)P(y_{s}=0|w^{T}%
,y_{s-1}=y_{t-1},y_{s+1}=y_{t+1},\alpha)\\
&  +P(E_{s+1,j}|w^{T},y_{s-1},y_{s+1},\alpha,y_{s}=1)P(y_{s}=1|w^{T}%
,y_{s-1}=y_{t-1},y_{s+1}=y_{t+1},\alpha),
\end{align*}
where the second equality follows from $E_{s+1,j}\perp\{y_{t-1},y_{t+1}%
\}|(w^{T},y_{s-1},y_{s},y_{s+1},\alpha)$ by Assumption
\hyperref[Assumption:HK]{A}(a). Therefore, to prove (\ref{eq:A9})--(\ref{eq:A11}), it suffices to verify the following equalities:

\begin{enumerate}
\item[(1)] $P(E_{s+1,1}|w^{T},y_{s-1},y_{s+1}=1,\alpha,y_{s}=1)=1$

\item[(2)] $P(E_{s+1,1}|w^{T},y_{s-1},y_{s+1}=1,\alpha,y_{s}=0)=\frac
{F_{\epsilon|\alpha}(w_{s+1}+\gamma+\alpha)}{F_{\epsilon|\alpha}%
(w_{s+1}+\alpha)}$

\item[(3)] $P(E_{s+1,2}|w^{T},y_{s-1},y_{s+1}=0,\alpha,y_{s}=1)=\frac
{F_{\epsilon|\alpha}(w_{s+1}+\alpha)-F_{\epsilon|\alpha}(w_{s+1}+\gamma
+\alpha)}{1-F_{\epsilon|\alpha}(w_{s+1}+\gamma+\alpha)}$

\item[(4)] $P(E_{s+1,2}|w^{T},y_{s-1},y_{s+1}=0,\alpha,y_{s}=0)=0$

\item[(5)] $P(E_{s+1,3}|w^{T},y_{s-1},y_{s+1}=0,\alpha,y_{s}=1)=\frac
{1-F_{\epsilon|\alpha}(w_{s+1}+\alpha)}{1-F_{\epsilon|\alpha}(w_{s+1}%
+\gamma+\alpha)}$

\item[(6)] $P(E_{s+1,3}|w^{T},y_{s-1},y_{s+1}=0,\alpha,y_{s}=0)=1$
\end{enumerate}

Equalities (1), (4), and (6) can be easily verified using the facts that
$E_{s+1,1}=\{y_{s+1}=1\}$ conditional on $\{y_{s}=1\}$, $E_{s+1,2}%
\cap\{y_{s+1}=0\}=\emptyset$ conditional on $\{y_{s}=0\}$, and $E_{s+1,3}%
=\{y_{s+1}=0\}$ conditional on $\{y_{s}=0\}$, respectively.

For equality (2), note that using {the} conditional probability formula, we have
\begin{align*}
&  P(E_{s+1,1}|w^{T},y_{s-1},y_{s+1}=1,\alpha,y_{s}=0)\\
=  &  \frac{P(y_{s+1}=1,E_{s+1,1}|w^{T},y_{s-1},\alpha,y_{s}=0)}%
{P(y_{s+1}=1|w^{T},y_{s-1},\alpha,y_{s}=0)}= \frac{P(E_{s+1,1}|w^{T}%
,y_{s-1},\alpha,y_{s}=0)}{P(E_{s+1,1}\cup E_{s+1,2}|w^{T},y_{s-1},\alpha
,y_{s}=0)}\\
=  &  \frac{P(E_{s+1,1}|w_{s+1},\alpha)}{P(E_{s+1,1}\cup E_{s+1,2}%
|w_{s+1},\alpha)}= \frac{F_{\epsilon|\alpha}(w_{s+1}+\gamma+\alpha
)}{F_{\epsilon|\alpha}(w_{s+1}+\alpha)}%
\end{align*}
where the second equality uses the fact that $\{y_{s+1}=1\}=E_{s+1,1}\cup
E_{s+1,2}$ conditional on $\{y_{s}=0\}$, and the third equality follows by
Assumption \hyperref[Assumption:HK]{A}(a).

Similar arguments, along with the fact that $\{y_{s+1}=0\}=E_{s+1,2}\cup
E_{s+1,3}$ conditional on $\{y_{s}=1\}$, can be used to verify equalities (3)
and (5). Specifically, we can write for equality (3),
\begin{align*}
&  P(E_{s+1,2}|w^{T},y_{s-1},y_{s+1}=0,\alpha,y_{s}=1)\\
=  &  \frac{P(y_{s+1}=0,E_{s+1,2}|w^{T},y_{s-1},\alpha,y_{s}=1)}%
{P(y_{s+1}=0|w^{T},y_{s-1},\alpha,y_{s}=1)} = \frac{P(E_{s+1,2}|w^{T}%
,y_{s-1},\alpha,y_{s}=1)}{P(E_{s+1,2}\cup E_{s+1,3}|w^{T},y_{s-1},\alpha
,y_{s}=1)}\\
=  &  \frac{P(E_{s+1,2}|w_{s+1},\alpha)}{P(E_{s+1,2}\cup E_{s+1,3}%
|w_{s+1},\alpha)}=\frac{F_{\epsilon|\alpha}(w_{s+1}+\alpha)-F_{\epsilon
|\alpha}(w_{s+1}+\gamma+\alpha)}{1-F_{\epsilon|\alpha}(w_{s+1}+\gamma+\alpha)}%
\end{align*}
and analogously for equality (5),
\begin{align*}
&  P(E_{s+1,3}|w^{T},y_{s-1},y_{s+1},\alpha,y_{s}=1)\\
=  &  \frac{P(y_{s+1}=0,E_{s+1,3}|w^{T},y_{s-1},\alpha,y_{s}=1)}%
{P(y_{s+1}=0|w^{T},y_{s-1},\alpha,y_{s}=1)} = \frac{P(E_{s+1,3}|w^{T}%
,y_{s-1},\alpha,y_{s}=1)}{P(E_{s+1,2}\cup E_{s+1,3}|w^{T},y_{s-1},\alpha
,y_{s}=1)}\\
=  &  \frac{P(E_{s+1,3}|w_{s+1},\alpha)}{P(E_{s+1,2}\cup E_{s+1,3}%
|w_{s+1},\alpha)}= \frac{1-F_{\epsilon|\alpha}(w_{s+1}+\alpha)}{1-F_{\epsilon
|\alpha}(w_{s+1}+\gamma+\alpha)}.
\end{align*}
{Then, the proof is completed}.
\end{proof}

We then prove Lemma \ref{Lemma:A3} with a weaker version of Assumption \href{assumptionSD}{SD}.  Particularly,   Assumption \href{assumptionSD}{SD}(b) will be replaced by the following stochastic dominance condition: For all $v\in\mathbb{R}$ and $d_{0},d_1\in\{0,1\}$, if $w_{t}\geq w_{s}$, then
\begin{equation}
F_{w_{s+1}|w_{s},w_{t},y_{s-1}=y_{t-1}=d_{0},y_{s+1}=y_{t+1}=d_1,\alpha}(v)\geq F_{w_{t+1}|w_{s},w_{t},y_{s-1}=y_{t-1}=d_{0},y_{s+1}=y_{t+1}=d_1,\alpha}(v), \label{sd_1}
\end{equation}
and if $w_{t}\leq w_{s}$, then
\begin{equation}
F_{w_{s+1}|w_{s},w_{t},y_{s-1}=y_{t-1}=d_{0},y_{s+1}=y_{t+1}=d_1,\alpha}(v)\leq F_{w_{t+1}|w_{s},w_{t},y_{s-1}=y_{t-1}=d_{0},y_{s+1}=y_{t+1}=d_1,\alpha}(v). \label{sd_2}
\end{equation}
Inequalities (\ref{sd_1}) and (\ref{sd_2})  say that,
conditional on $\alpha$ and the same \textquotedblleft
initial\textquotedblright\ and \textquotedblleft ending\textquotedblright%
\ statuses ($y_{s-1}=y_{t-1}$, $y_{s+1}=y_{t+1}$), if the value of $w_{t}$ is
higher than that of $w_{s}$, then $w_{t+1}$ has a better chance of taking a
large value than $w_{s+1}$. This restriction rules out the case in which high
utility in one period has negative effects on the utility in the next period.
This assumption is more likely to hold in applications where $\{w_{t}\}$ represents a positively autocorrelated stochastic process
of the \textquotedblleft utility\textquotedblright, \textquotedblleft
benefits\textquotedblright, or \textquotedblleft profits\textquotedblright\ of
a decision.  This assumption is of high level, for which a sufficient, but not
necessary, condition is Assumption \hyperref[Assumption:SD]{SD}(b), which is formally stated in Lemma
\ref{Lemma:SD} below.

\begin{lemma}
\label{Lemma:SD} Suppose that Assumption \hyperref[Assumption:HK]{A} is
satisfied. Then inequalities (\ref{sd_1}) and (\ref{sd_2}) hold with equality, if
the joint PDF of $w^{T}$ conditional on $\alpha$ is exchangeable, i.e.,
\[
f_{w^{T}|\alpha}(\omega_{1},...,\omega_{T})=f_{w^{T}|\alpha}(\omega_{\pi(1)},...,\omega_{\pi(T)})
\]
for all permutations $\{\pi(1),...,\pi(T)\}$ defined on the set $\mathcal{T}$.
\end{lemma}
The proof of Lemma \ref{Lemma:SD} can be found at the end of this section.

Note that inequalities (\ref{sd_1}) and (\ref{sd_2}) can be thought of as a conditional
\textquotedblleft first-order stochastic dominance\textquotedblright%
\ condition, which implies that, for any non-decreasing (non-increasing)
function $u(\cdot)$,
\[
\int u(v)dF_{w_{s+1}|w_{s}=w,w_{t}=w^{\prime},y_{s-1}=y_{t-1},y_{s+1}%
=y_{t+1},\alpha}(v)\leq\int u(v)dF_{w_{t+1}|w_{s}=w,w_{t}=w^{\prime}%
,y_{s-1}=y_{t-1},y_{s+1}=y_{t+1},\alpha}(v)
\]
($\int u(v)dF_{w_{s+1}|w_{s}=w,w_{t}=w^{\prime},y_{s-1}=y_{t-1},y_{s+1}%
=y_{t+1},\alpha}(v)\geq\int u(v)dF_{w_{t+1}|w_{s}=w,w_{t}=w^{\prime}%
,y_{s-1}=y_{t-1},y_{s+1}=y_{t+1},\alpha}(v)$) whenever $w^{\prime}\geq w$. This
property is needed for establishing the monotonic relation in
(\ref{idinequality}) as demonstrated in the proof of Lemma \ref{Lemma:A3} below.

\begin{proof}
[Proof of Lemma \ref{Lemma:A3}]Let $\varpi$ denote the sub-vector of $w^{T}$
comprising all its elements other than $w_{s}$ and $w_{t}$. Note that for all
$\tau\in\{s,t\}$,
\begin{align}
&  P(y_{\tau}=1|w_{s},w_{t},y_{s-1}=y_{t-1},y_{s+1}=y_{t+1},\alpha)\nonumber\\
=  &  \int P(y_{\tau}=1|w^{T},y_{s-1}=y_{t-1},y_{s+1}=y_{t+1},\alpha
)dF_{\varpi|w_{s},w_{t},y_{s-1}=y_{t-1},y_{s+1}=y_{t+1},\alpha}.
\label{eq:A12}%
\end{align}
In what follows, we consider two cases, $y_{s+1}=y_{t+1}=1$ and $y_{s+1}%
=y_{t+1}=0$, in turn.

\paragraph*{Case 1 ($y_{s+1}=y_{t+1}=1$)}

Plug (\ref{eq:A9}) into (\ref{eq:A1}) to obtain
\begin{align}
&  P(y_{\tau}=1|w^{T},y_{s-1}=y_{t-1},y_{s+1}=y_{t+1}=1,\alpha)\nonumber\\
=  &  F_{\epsilon|\alpha}(w_{\tau}+\gamma y_{\tau-1}+\alpha)\{P(y_{\tau
}=1|w^{T},y_{s-1}=y_{t-1},y_{s+1}=y_{t+1}=1,\alpha)\nonumber\\
&  +\frac{F_{\epsilon|\alpha}(w_{\tau+1}+\gamma+\alpha)}{F_{\epsilon|\alpha
}(w_{\tau+1}+\alpha)}[1-P(y_{\tau}=1|w^{T},y_{s-1}=y_{t-1},y_{s+1}%
=y_{t+1}=1,\alpha)]\}. \label{eq:A13}%
\end{align}
Let $\psi(w)\equiv F_{\epsilon|\alpha}(w+\gamma y_{\tau-1}+\alpha)$ and
$\phi_{1}(w)\equiv F_{\epsilon|\alpha}(w+\gamma+\alpha)/F_{\epsilon|\alpha
}(w+\alpha)$. Deduce from (\ref{eq:A13}) that
\begin{align*}
&  P(y_{\tau}=1|w^{T},y_{s-1}=y_{t-1},y_{s+1}=y_{t+1}=1,\alpha)\\
=  &  \frac{\psi(w_{\tau})\phi_{1}(w_{\tau+1})}{1-\psi(w_{\tau})+\psi(w_{\tau
})\phi_{1}(w_{\tau+1})}\equiv G_{1}(w_{\tau},w_{\tau+1}).
\end{align*}
Then, (\ref{eq:A12}) reduces to
\[
\int G_{1}(w_{\tau},w)dF_{w_{\tau+1}|w_{s},w_{t},y_{s-1}=y_{t-1}%
,y_{s+1}=y_{t+1}=1,\alpha}(w),
\]
and hence
\begin{align}
&  P(y_{t}=1|w_{s},w_{t},y_{s-1}=y_{t-1},y_{s+1}=y_{t+1}=1,\alpha)\nonumber\\
&  -P(y_{s}=1|w_{s},w_{t},y_{s-1}=y_{t-1},y_{s+1}=y_{t+1}=1,\alpha)\nonumber\\
=  &  \int G_{1}(w_{t},w)dF_{w_{t+1}|w_{s},w_{t},y_{s-1}=y_{t-1}%
,y_{s+1}=y_{t+1}=1,\alpha}(w)\nonumber\\
&  -\int G_{1}(w_{s},w)dF_{w_{s+1}|w_{s},w_{t},y_{s-1}=y_{t-1},y_{s+1}%
=y_{t+1}=1,\alpha}(w)\nonumber\\
=  &  \int\left[  G_{1}(w_{t},w)-G_{1}(w_{s},w)\right]  dF_{w_{t+1}%
|w_{s},w_{t},y_{s-1}=y_{t-1},y_{s+1}=y_{t+1}=1,\alpha}(w)\label{eq:A14}\\
&  +\int G_{1}(w_{s},w)d\left[  F_{w_{t+1}|w_{s},w_{t},y_{s-1}=y_{t-1}%
,y_{s+1}=y_{t+1}=1,\alpha}(w)-F_{w_{s+1}|w_{s},w_{t},y_{s-1}=y_{t-1}%
,y_{s+1}=y_{t+1}=1,\alpha}(w)\right]  .\nonumber
\end{align}

It is easy to verify that $\psi^{\prime}(\cdot)>0$, $\phi_{1}^{\prime}%
(\cdot)>0$ (by Assumption \href{assumptionSD}{SD}(a)). Therefore,
$G_{1}^{\prime}(\cdot,w)>0$ and $G_{1}^{\prime}(w,\cdot)>0$ hold true for all
$w$. The former monotonicity result implies that the first term in
(\ref{eq:A14}) is positive if and only if $w_{t}\geq w_{s}$. The latter,
together with Assumption \href{assumptionSD}{SD}(b), implies that the second
term in (\ref{eq:A14}) is positive if and only if $w_{t}\geq w_{s}$. Put these
results to establish the desired result.

\paragraph*{Case 2 ($y_{s+1}=y_{t+1}=0$)}

Plug (\ref{eq:A10}) and (\ref{eq:A11}) into (\ref{eq:A2}) to obtain
\begin{align}
&  P(y_{\tau}=1|w^{T},y_{s-1}=y_{t-1},y_{s+1}=y_{t+1}=0,\alpha)\nonumber\\
=  &  \frac{F_{\epsilon|\alpha}(w_{\tau+1}+\alpha)-F_{\epsilon|\alpha}%
(w_{\tau+1}+\gamma+\alpha)}{1-F_{\epsilon|\alpha}(w_{\tau+1}+\gamma+\alpha
)}P(y_{\tau}=1|w^{T},y_{s-1}=y_{t-1},y_{s+1}=y_{t+1}=0,\alpha)\nonumber\\
&  +F_{\epsilon|\alpha}(w_{\tau}+\gamma y_{\tau-1}+\alpha)[\frac
{1-F_{\epsilon|\alpha}(w_{\tau+1}+\alpha)}{1-F_{\epsilon|\alpha}(w_{\tau
+1}+\gamma+\alpha)}P(y_{\tau}=1|w^{T},y_{s-1}=y_{t-1},y_{s+1}=y_{t+1}%
=0,\alpha)\nonumber\\
&  +1-P(y_{\tau}=1|w^{T},y_{s-1}=y_{t-1},y_{s+1}=y_{t+1}=0,\alpha)].
\label{eq:A15}%
\end{align}
Let $\phi_{0}(w)\equiv[1-F_{\epsilon|\alpha}(w+\alpha)]/[1-F_{\epsilon|\alpha
}(w+\gamma+\alpha)]$. We deduce from (\ref{eq:A15}) that
\begin{align*}
&  P(y_{\tau}=1|w^{T},y_{s-1}=y_{t-1},y_{s+1}=y_{t+1}=0,\alpha)\\
=  &  \frac{\psi(w_{\tau})}{\psi(w_{\tau})+\phi_{0}(w_{\tau+1})-\psi(w_{\tau
})\phi_{0}(w_{\tau+1})}\equiv G_{0}(w_{\tau},w_{\tau+1}).
\end{align*}
Then, (\ref{eq:A12}) reduces to
\[
\int G_{0}(w_{\tau},w)dF_{w_{\tau+1}|w_{s},w_{t},y_{s-1}=y_{t-1}%
,y_{s+1}=y_{t+1}=0,\alpha}(w),
\]
and hence
\begin{align}
&  P(y_{t}=1|w_{s},w_{t},y_{s-1}=y_{t-1},y_{s+1}=y_{t+1}=0,\alpha)\nonumber\\
&  -P(y_{s}=1|w_{s},w_{t},y_{s-1}=y_{t-1},y_{s+1}=y_{t+1}=0,\alpha)\nonumber\\
=  &  \int G_{0}(w_{t},w)dF_{w_{t+1}|w_{s},w_{t},y_{s-1}=y_{t-1}%
,y_{s+1}=y_{t+1}=0,\alpha}(w)\nonumber\\
&  -\int G_{0}(w_{s},w)dF_{w_{s+1}|w_{s},w_{t},y_{s-1}=y_{t-1},y_{s+1}%
=y_{t+1}=0,\alpha}(w)\nonumber\\
=  &  \int\left[  G_{0}(w_{t},w)-G_{0}(w_{s},w)\right]  dF_{w_{t+1}%
|w_{s},w_{t},y_{s-1}=y_{t-1},y_{s+1}=y_{t+1}=0,\alpha}(w)\label{eq:A16}\\
&  +\int G_{0}(w_{s},w)d\left[  F_{w_{t+1}|w_{s},w_{t},y_{s-1}=y_{t-1}%
,y_{s+1}=y_{t+1}=0,\alpha}(w)-F_{w_{s+1}|w_{s},w_{t},y_{s-1}=y_{t-1}%
,y_{s+1}=y_{t+1}=0,\alpha}(w)\right]  .\nonumber
\end{align}
By Assumption \href{assumptionSD}{SD}(a), $\phi_{0}^{\prime}(\cdot)<0$.
Therefore, $G_{0}^{\prime}(\cdot,w)>0$ and $G_{0}^{\prime}(w,\cdot)>0$ hold
true for all $w$. The former monotonicity result implies that the first term
in (\ref{eq:A16}) is positive if and only if $w_{t}\geq w_{s}$. The latter,
together with Assumption \href{assumptionSD}{SD}(b), implies that the second
term in (\ref{eq:A16}) is positive if and only if $w_{t}\geq w_{s}$. The proof
is complete.
\end{proof}

\begin{proof}
[Proof of Lemma \ref{Lemma:A4}]The proof adopts similar arguments used in the
proofs of Lemmas \ref{Lemma:A1}--\ref{Lemma:A3}. Here, we only outline the
proof procedure and omit repetitive technical details for {brevity.}

First note that, under Assumptions \hyperref[Assumption:HK]{A} and
\href{assumptionSI}{SI}, we can write for both $\tau=s$ and $\tau=t$,
\begin{equation}
P(y_{\tau}=1|w_{s},w_{t},y_{s-1}=y_{t-1},y_{s+1}=y_{t+1},\alpha)=P(y_{\tau
}=1|w_{\tau},y_{\tau-1},y_{\tau+1},\alpha). \label{eq:A17}%
\end{equation}
To see this, note that for $\tau=s$ and all $d_{0},d_{1}\in\{0,1\}$
\begin{align*}
&  P(y_{s}=1|w_{s},w_{t},y_{s-1}=y_{t-1}=d_{0},y_{s+1}=y_{t+1}=d_{1},\alpha)\\
=  &  \frac{P(y_{t-1}=d_{0},y_{t+1}=d_{1}|w_{s},w_{t},y_{s-1}=d_{0}%
,y_{s}=1,y_{s+1}=d_{1},\alpha)}{P(y_{t-1}=d_{0},y_{t+1}=d_{1}|w_{s}%
,w_{t},y_{s-1}=d_{0},y_{s+1}=d_{1},\alpha)}\\
&  \times P(y_{s}=1|w_{s},w_{t},y_{s-1}=d_{0},y_{s+1}=d_{1},\alpha)\\
=  &  \frac{P(y_{t-1}=d_{0},y_{t+1}=d_{1}|w_{t},y_{s+1}=d_{1},\alpha
)P(y_{s}=1|w_{s},w_{t},y_{s-1}=d_{0},y_{s+1}=d_{1},\alpha)}{P(y_{t-1}%
=d_{0},y_{t+1}=d_{1}|w_{t},y_{s+1}=d_{1},\alpha)}\\
=  &  P(y_{s}=1|w_{s},w_{t},y_{s-1}=d_{0},y_{s+1}=d_{1},\alpha)\\
=  &  \frac{P(y_{s+1}=d_{1}|w_{s},w_{t},y_{s-1}=d_{0},y_{s}=1,\alpha
)P(y_{s}=1|w_{s},w_{t},y_{s-1}=d_{0},\alpha)}{P(y_{s+1}=d_{1}|w_{s}%
,w_{t},y_{s-1}=d_{0},\alpha)}\\
=  &  \frac{P(y_{s+1}=d_{1}|w_{s},y_{s-1}=d_{0},y_{s}=1,\alpha)P(y_{s}%
=1|w_{s},y_{s-1}=d_{0},\alpha)}{P(y_{s+1}=d_{1}|w_{s},y_{s-1}=d_{0},\alpha)}\\
=  &  P(y_{s}=1|w_{s},y_{s-1}=d_{0},y_{s+1}=d_{1},\alpha),
\end{align*}
where the first, third, fourth, and last equalities use Bayes' theorem, and
the second and fifth equalities follow by Assumptions \href{assumptionSI}{SI}%
(a) and \hyperref[Assumption:HK]{A}(a).\footnote{$(y_{t-1},y_{t+1})\perp
(w_{s},y_{s-1},y_{s})|(w_{t},y_{s+1},\alpha)$ and $(y_{s},y_{s+1})\perp
w_{t}|(w_{s},y_{s-1},\alpha)$.} Using similar arguments yields the same
simplification for $\tau=t$.

For the case with $d_{1}=1$, we uses the same arguments for deriving
(\ref{eq:A1}) to write
\begin{align}
&  P(y_{\tau}=1|w_{\tau},y_{\tau-1},y_{\tau+1}=1,\alpha)\nonumber\\
=  &  P(y_{\tau}=1|w_{\tau},y_{\tau-1},y_{\tau+1}=1,\alpha,E_{\tau
+1,1})P(E_{\tau+1,1}|w_{\tau},y_{\tau-1},y_{\tau+1}=1,\alpha)\nonumber\\
=  &  F_{\epsilon|\alpha}(w_{\tau}+\gamma y_{\tau-1}+\alpha)P(E_{\tau
+1,1}|w_{\tau},y_{\tau-1},y_{\tau+1}=1,\alpha), \label{eq:A18}%
\end{align}
where the last equality follows from $E_{\tau+1,1}\subset\{y_{\tau+1}=1\}$,
Assumption \href{assumptionSI}{SI}(a), and Assumption \hyperref[Assumption:HK]{A}%
(a). Then, we use analogous arguments for proving Lemma \ref{Lemma:A2} to
deduce
\begin{align}
&  P(E_{\tau+1,1}|w_{\tau},y_{\tau-1},y_{\tau+1}=1,\alpha)\nonumber\\
=  &  P(E_{\tau+1,1}|w_{\tau},y_{\tau-1},y_{\tau+1}=1,\alpha,y_{\tau
}=1)P(y_{\tau}=1|w_{\tau},y_{\tau-1},y_{\tau+1}=1,\alpha)\nonumber\\
&  +P(E_{\tau+1,1}|w_{\tau},y_{\tau-1},y_{\tau+1}=1,\alpha,y_{\tau
}=0)[1-P(y_{\tau}=1|w_{\tau},y_{\tau-1},y_{\tau+1}=1,\alpha)]\nonumber\\
=  &  P(y_{\tau}=1|w_{\tau},y_{\tau-1},y_{\tau+1}=1,\alpha)\nonumber\\
&  +\frac{P(E_{\tau+1,1}|w_{\tau},y_{\tau-1},\alpha,y_{\tau}=0)}%
{P(E_{\tau+1,1}\cup E_{\tau+1,2}|w_{\tau},y_{\tau-1},\alpha,y_{\tau}%
=0)}[1-P(y_{\tau}=1|w_{\tau},y_{\tau-1},y_{\tau+1}=1,\alpha)]\nonumber\\
=  &  P(y_{\tau}=1|w_{\tau},y_{\tau-1},y_{\tau+1}=1,\alpha)\nonumber\\
&  +\frac{P(E_{\tau+1,1}|\alpha)}{P(E_{\tau+1,1}\cup E_{\tau+1,2}|\alpha
)}[1-P(y_{\tau}=1|w_{\tau},y_{\tau-1},y_{\tau+1}=1,\alpha)], \label{eq:A19}%
\end{align}
where the last equality follows from Assumptions \href{assumptionSI}{SI}(a)
and \hyperref[Assumption:HK]{A}(a).

Combine (\ref{eq:A17}), (\ref{eq:A18}) and (\ref{eq:A19}) to solve
\begin{align*}
&  P(y_{\tau}=1|w_{s},w_{t},y_{s-1}=y_{t-1},y_{s+1}=y_{t+1}=1,\alpha)\\
=  &  P(y_{\tau}=1|w_{\tau},y_{\tau-1},y_{\tau+1}=1,\alpha)=\frac
{\phi_{1\alpha}\psi(w_{\tau})}{1-\psi(w_{\tau})+\phi_{1\alpha}\psi(w_{\tau}%
)}\equiv\mathcal{G}_{1}(w_{\tau}),
\end{align*}
where $\phi_{1\alpha}\equiv P(E_{\tau+1,1}|\alpha)/P(E_{\tau+1,1}\cup
E_{\tau+1,2}|\alpha)$ is a positive constant for any given $\alpha$. The
monotonic relation stated in the lemma is then established by verifying the
monotonicity of $\mathcal{G}_{1}(\cdot)$.

For the case with $d_{1}=0$, using the same arguments for deriving
(\ref{eq:A2}) yields
\begin{align}
&  P(y_{\tau}=1|w_{\tau},y_{\tau-1},y_{\tau+1}=0,\alpha)\nonumber\\
=  &  P(E_{\tau+1,2}|w_{\tau},y_{\tau-1},y_{\tau+1}=0,\alpha)\nonumber\\
&  +P(y_{\tau}=1|w_{\tau},y_{\tau-1},y_{\tau+1}=0,\alpha,E_{\tau
+1,3})P(E_{\tau+1,3}|w_{\tau},y_{\tau-1},y_{\tau+1}=0,\alpha)\nonumber\\
=  &  P(E_{\tau+1,2}|w_{\tau},y_{\tau-1},y_{\tau+1}=0,\alpha)+F_{\epsilon
|\alpha}(w_{\tau}+\gamma y_{\tau-1}+\alpha)P(E_{\tau+1,3}|w_{\tau},y_{\tau
-1},y_{\tau+1}=0,\alpha), \label{eq:A20}%
\end{align}
where the last equality follows by $E_{\tau+1,3}\subset\{y_{\tau+1}=0\}$,
Assumption \href{assumptionSI}{SI}(a), and Assumption \hyperref[Assumption:HK]{A}(a).

Use analogous arguments for proving Lemma \ref{Lemma:A2} to obtain
\begin{align}
&  P(E_{\tau+1,2}|w_{\tau},y_{\tau-1},y_{\tau+1}=0,\alpha)\nonumber\\
=  &  \frac{P(E_{\tau+1,2}|\alpha)}{P(E_{\tau+1,2}\cup E_{\tau+1,3}|\alpha
)}P(y_{\tau}=1|w_{\tau},y_{\tau-1},y_{\tau+1}=0,\alpha), \label{eq:A21}%
\end{align}
and
\begin{align}
&  P(E_{\tau+1,3}|w_{\tau},y_{\tau-1},y_{\tau+1}=0,\alpha)\nonumber\\
=  &  1-P(y_{\tau}=1|w_{\tau},y_{\tau-1},y_{\tau+1}=0,\alpha)\nonumber\\
&  +\frac{P(E_{\tau+1,3}|\alpha)}{P(E_{\tau+1,2}\cup E_{\tau+1,3}|\alpha
)}P(y_{\tau}=1|w_{\tau},y_{\tau-1},y_{\tau+1}=0,\alpha). \label{eq:A22}%
\end{align}

Combine (\ref{eq:A17}), (\ref{eq:A20}), (\ref{eq:A21}), and (\ref{eq:A22}) to
obtain
\begin{align*}
&  P(y_{\tau}=1|w_{s},w_{t},y_{s-1}=y_{t-1},y_{s+1}=y_{t+1}=1,\alpha)\\
=  &  P(y_{\tau}=1|w_{\tau},y_{\tau-1},y_{\tau+1}=1,\alpha)=\frac{\psi
(w_{\tau})}{\psi(w_{\tau})+\phi_{0\alpha}-\phi_{0\alpha}\psi(w_{\tau})}%
\equiv\mathcal{G}_{0}(w_{\tau}),
\end{align*}
where $\phi_{0\alpha}\equiv P(E_{\tau+1,3}|\alpha)/P(E_{\tau+1,2}\cup
E_{\tau+1,3}|\alpha)$ is a positive constant for any given $\alpha$. Note that
$\mathcal{G}_{0}(w_{\tau})$ is an increasing function, from which the
monotonic relation stated in the lemma is established. Putting all these
results together completes the proof.
\end{proof}

\begin{proof}
[Proof of Lemma \ref{LE:xi_ib}]\noindent\textbf{Preparation.} Relating to the
notations in \cite*{SeoOtsu2018}, $h_{n}=1$ (in their notations) for our
estimator $\hat{\beta}.$ $\xi_{i}\left(  b\right)  $ only takes value $-1,0,$
and $1,$ so it is bounded.\ Proposition \ref{Lemma:beta}\ shows that $\beta$
it the unique solution to $\max_{b\in\mathcal{B}}\mathbb{E}\left(  \xi
_{i}\left(  b\right)  \right)  .$ The following calculation can help
understand this result.%
\begin{align*}
\mathbb{E}\left(  \xi_{i}\left(  b\right)  \right)   &  =\mathbb{E}\left\{
\mathbb{E}\left[  1\left[  y_{i0}=y_{i2}=y_{i4}\right]  \left(  y_{i3}%
-y_{i1}\right)  |x_{i1},x_{i3}\right]  \left(  1\left[  x_{i31}^{\prime
}b>0\right]  -1\left[  x_{i31}^{\prime}\beta>0\right]  \right)  \right\} \\
&  =\mathbb{E}\left\{  \left(  \mathbb{E}\left[  1\left[  y_{i0}=y_{i2}%
=y_{i4}\right]  \left(  y_{i3}-y_{i1}\right)  |y_{i0}=y_{i2}=y_{i4}%
,x_{i1},x_{i3}\right]  P\left(  y_{i0}=y_{i2}=y_{i4}|x_{i1},x_{i3}\right)
\right.  \right. \\
&  \left.  +\mathbb{E}\left[  1\left[  y_{i0}=y_{i2}=y_{i4}\right]  \left(
y_{i3}-y_{i1}\right)  |y_{i0}=y_{i2}=y_{i4},x_{i1},x_{i3}\right]  P\left(
\overline{y_{i0}=y_{i2}=y_{i4}}|x_{i1},x_{i3}\right)  \right) \\
&  \left(  1\left[  x_{i31}^{\prime}b>0\right]  -1\left[  x_{i31}^{\prime
}\beta>0\right]  \right)  \}\\
&  =\mathbb{E}\left\{  \mathbb{E}\left[  \left(  y_{i3}-y_{i1}\right)
|y_{i0}=y_{i2}=y_{i4},x_{i1},x_{i3}\right]  P\left(  y_{i0}=y_{i2}%
=y_{i4}|x_{i1},x_{i3}\right)  \right. \\
&  \left.  \left(  1\left[  x_{i31}^{\prime}b>0\right]  -1\left[
x_{i31}^{\prime}\beta>0\right]  \right)  \right\} \\
&  \equiv\mathbb{E}\left\{  \mathbb{E}\left[  \left(  y_{i3}-y_{i1}\right)
|y_{i0}=y_{i2}=y_{i4},x_{i1},x_{i3}\right]  \varphi\left(  x_{i1}%
,x_{i3}\right)  \left(  1\left[  x_{i31}^{\prime}b>0\right]  -1\left[
x_{i31}^{\prime}\beta>0\right]  \right)  \right\} \\
&  =\mathbb{E}\left\{  \left(  \mathbb{E}\left[  y_{i3}|y_{i0}=y_{i2}%
=y_{i4},x_{i1},x_{i3}\right]  -\mathbb{E}\left[  y_{i1}|y_{i0}=y_{i2}%
=y_{i4},x_{i1},x_{i3}\right]  \right)  \right. \\
&  \left.  \varphi\left(  x_{i1},x_{i3}\right)  \left(  1\left[
x_{i31}^{\prime}b>0\right]  -1\left[  x_{i31}^{\prime}\beta>0\right]  \right)
\right\} \\
&  =\mathbb{E}\left\{  \left(  \mathbb{E}\left[  y_{i3}|y_{i2}=y_{i4}%
,x_{i3}\right]  -\mathbb{E}\left[  y_{i1}|y_{i0}=y_{i2},x_{i1}\right]
\right)  \right. \\
&  \left.  \varphi\left(  x_{i1},x_{i3}\right)  \left(  1\left[
x_{i31}^{\prime}b>0\right]  -1\left[  x_{i31}^{\prime}\beta>0\right]  \right)
\right\}  ,
\end{align*}
where in the second equality $\overline{A}$ denotes the complement of the set
$A,$
\[
\varphi\left(  x_{i1},x_{i3}\right)  \equiv P\left(  y_{i0}=y_{i2}%
=y_{i4}|x_{i1},x_{i3}\right)
\]
in the fourth equality,\ and the sixth equality follows the same argument as
in the proof of Proposition \ref{Lemma:beta}.

By the stationary condition, the following is true
\[
\mathbb{E}\left[  y_{i3}|y_{i2}=y_{i4},x_{i3}=x\right]  =\mathbb{E}\left[
y_{i1}|y_{i0}=y_{i2},x_{i1}=x\right]  .
\]
Let%
\[
\phi\left(  x\right)  \equiv\mathbb{E}\left[  y_{i3}|y_{i2}=y_{i4}%
,x_{i3}=x\right]  =\mathbb{E}\left[  y_{i1}|y_{i0}=y_{i2},x_{i1}=x\right]  .
\]
With the introduction of the above notation,
\begin{equation}
\mathbb{E}\left(  \xi_{i}\left(  b\right)  \right)  =\mathbb{E}\left\{
\varphi\left(  x_{i1},x_{i3}\right)  \left(  \phi\left(  x_{i3}\right)
-\phi\left(  x_{i1}\right)  \right)  \left(  1\left[  x_{i31}^{\prime
}b>0\right]  -1\left[  x_{i31}^{\prime}\beta>0\right]  \right)  \right\}  .
\label{EQ:Exi_ib}%
\end{equation}
From the results in the proof of Proposition \ref{Lemma:beta}, %
$\phi\left(  x_{i3}\right)  -\phi\left(  x_{i1}\right)  >0$ if $x_{i31}%
^{\prime}\beta>0,$ $\phi\left(  x_{i3}\right)  -\phi\left(  x_{i1}\right)  =0$
if $x_{i31}^{\prime}\beta=0$, and $\phi\left(  x_{i3}\right)  -\phi\left(
x_{i1}\right)  <0$ if $x_{i31}^{\prime}\beta<0$\textbf{. }$\varphi\left(
x_{i1},x_{i3}\right)  $ is a conditional probability, so $\varphi\left(
x_{i1},x_{i3}\right)  \geq0$. The above observations imply that $\mathbb{E}%
\left(  \xi_{i}\left(  b\right)  \right)  $ is nonpositive and is equal to $0$
if $b=\beta$. Assumption \hyperref[Assumption:HK]{A} ensures that the
solution is unique. To {simplify} notations, let
\begin{equation}
\kappa\left(  x_{i31}\right)  \equiv\mathbb{E}\left[  \varphi\left(
x_{i1},x_{i3}\right)  \left(  \phi\left(  x_{i3}\right)  -\phi\left(
x_{i1}\right)  \right)  |x_{i31}\right]  . \label{EQ:kappa}%
\end{equation}
Easy to see that $\kappa$ defined here is equal to the $\kappa$ in the body of
Lemma \ref{LE:asym_xi}. The above discussion implies $\kappa\left(
x_{i31}\right)  $ has the same sign as $x_{i31}^{\prime}\beta.$%
\textbf{\medskip}

\noindent\textbf{On Assumption M.i in \cite*{SeoOtsu2018}}. We now try to get
the derivatives of $\mathbb{E}\left(  \xi_{i}\left(  b\right)  \right)  $ with
respect to $b$. $\mathbb{E}\left(  \xi_{i}\left(  b\right)  \right)  $ can be
rewritten as%
\[
\mathbb{E}\left(  \xi_{i}\left(  b\right)  \right)  =\mathbb{E}\left\{
\kappa\left(  x_{i31}\right)  \left(  1\left[  x_{i31}^{\prime}b>0\right]
-1\left[  x_{i31}^{\prime}\beta>0\right]  \right)  \right\}  .
\]
Following the same idea in Section 5 and Section 6.4 of \cite*{KimPollard1990}%
\textbf{ }and Section B.1 of\textbf{ }\cite*{SeoOtsu2018}\textbf{, }the above
expectation can be calculated {using} the classical differential geometry.
Since the results here are obtained using essentially the same argument, we
omit similar details. Define the following {mapping:}
\[
T_{b}=\left(  I-\left\Vert b\right\Vert _{2}^{-2}bb^{\prime}\right)  \left(
I-\beta\beta^{\prime}\right)  +\left\Vert b\right\Vert _{2}^{-2}b\beta
^{\prime},
\]
where $T_{b}$ maps the region $\left\{  x_{31}:x_{31}^{\prime}b>0\right\}  $
onto $\left\{  x_{31}:x_{31}^{\prime}\beta>0\right\}  ,$ taking the boundary
of $\left\{  x_{31}:x_{31}^{\prime}b>0\right\}  $ onto the boundary of
$\left\{  x_{31}:x_{31}^{\prime}\beta>0\right\}  .$ Equations (5.2) and (5.3)
in\textbf{ }\cite*{KimPollard1990}\textbf{ }imply\textbf{ }%
\[
\frac{\partial}{\partial b}\mathbb{E}\left(  \xi_{i}\left(  b\right)  \right)
=\left\Vert b\right\Vert _{2}^{-2}b^{\prime}\beta\left(  I-\left\Vert
b\right\Vert _{2}^{-2}bb^{\prime}\right)  \int1\left[  x_{31}^{\prime}%
\beta=0\right]  \kappa\left(  T_{b}x_{31}\right)  x_{31}f_{x_{31}}\left(
T_{b}x_{31}\right)  d\sigma_{0},
\]
where $f_{x_{31}}\left(  x_{31}\right)  $ is the density function of $x_{i31}$
and $\sigma_{0}$ is the surface measure of the boundary of $\left\{
x_{31}:x_{31}^{\prime}\beta>0\right\}  .$

$\left.  \frac{\partial}{\partial b}\mathbb{E}\left(  \xi_{i}\left(  b\right)
\right)  \right\vert _{b=\beta}=0,$ by $T_{\beta}x_{31}=x_{31}$ and $1\left[
x_{31}^{\prime}\beta=0\right]  \kappa\left(  x_{31}\right)  =0.$ Consequently,
the nonzero component of the second derivative of $\mathbb{E}\left(  \xi
_{i}\left(  b\right)  \right)  $ only comes from the derivative of
$\kappa\left(  T_{b}x_{31}\right)  .$ Notice that $\left.  \frac{\partial
}{\partial b}\kappa\left(  T_{b}x_{31}\right)  \right\vert _{b=\beta}=-\left(
\frac{\partial\kappa\left(  x_{31}\right)  }{\partial x_{31}}^{\prime}%
\beta\right)  x_{31},$ we have%
\[
\left.  \frac{\partial^{2}\mathbb{E}\left(  \xi_{i}\left(  b\right)  \right)
}{\partial b\partial b^{\prime}}\right\vert _{b=\beta}=-\int1\left[
x_{31}^{\prime}\beta=0\right]  \left(  \frac{\partial\kappa\left(
x_{31}\right)  }{\partial x_{31}}^{\prime}\beta\right)  f_{x_{31}}\left(
x_{31}\right)  x_{31}x_{31}^{\prime}d\sigma_{0}.
\]
Combining these results on the derivatives of $\mathbb{E}\left(  \xi
_{i}\left(  b\right)  \right)  $ implies that Assumption M.i in
\cite*{SeoOtsu2018} is satisfied with the matrix%
\begin{equation}
V_{1}\equiv-\int1\left[  x_{31}^{\prime}\beta=0\right]  \left(  \frac
{\partial\kappa\left(  x_{31}\right)  }{\partial x_{31}}^{\prime}\beta\right)
f_{x_{31}}\left(  x_{31}\right)  x_{31}x_{31}^{\prime}d\sigma_{0}.
\label{EQ:V1_def}%
\end{equation}
By definition,
\[
\left.  \frac{\partial\kappa\left(  x_{31}\right)  }{\partial x_{31}}^{\prime
}\beta\right\vert _{x_{31}^{\prime}\beta=0}=\left.  \lim_{h\rightarrow0}%
\frac{\kappa\left(  x_{31}+h\beta\right)  -\kappa\left(  x_{31}\right)  }%
{h}\right\vert _{x_{31}^{\prime}\beta=0}.
\]
Notice that $\left(  x_{31}+h\beta\right)  ^{\prime}\beta=h\left\Vert
\beta\right\Vert $ if $x_{31}^{\prime}\beta=0.$ Similar to the discussion
under equation (\ref{EQ:Exi_ib}), for $x_{31}$ satisfied with $x_{31}^{\prime
}\beta=0$, $\kappa\left(  x_{31}+h\beta\right)  \geq0=\kappa\left(
x_{31}\right)  $ if $h>0$ and $\kappa\left(  x_{31}+h\beta\right)
\leq0=\kappa\left(  x_{31}\right)  $ if $h<0$. Thus, $\left.  \frac
{\partial\kappa\left(  x_{31}\right)  }{\partial x_{31}}^{\prime}%
\beta\right\vert _{x_{31}^{\prime}\beta=0}\geq0,$ and $V_{1}$ is negative
semidefinite.\textbf{\medskip}

\noindent\textbf{On Assumption M.ii in \cite*{SeoOtsu2018}. }Note%
\[
\xi_{i}\left(  b_{1}\right)  -\xi_{i}\left(  b_{2}\right)  =1\left[
y_{i0}=y_{i2}=y_{i4}\right]  \left(  y_{i3}-y_{i1}\right)  \left(  1\left[
x_{i31}^{\prime}b_{1}>0\right]  -1\left[  x_{i31}^{\prime}b_{2}>0\right]
\right)
\]
and
\begin{equation}
\left(  \xi_{i}\left(  b_{1}\right)  -\xi_{i}\left(  b_{2}\right)  \right)
^{2}=1\left[  y_{i0}=y_{i2}=y_{i4}\right]  \left\vert y_{i3}-y_{i1}\right\vert
\left\vert 1\left[  x_{i31}^{\prime}b_{1}>0\right]  -1\left[  x_{i31}^{\prime
}b_{2}>0\right]  \right\vert , \label{EQ:xi^2}%
\end{equation}
this condition can be verified by the following calculation,
\begin{align*}
&  \left[  \mathbb{E}\left(  \xi_{i}\left(  b_{1}\right)  -\xi_{i}\left(
b_{2}\right)  \right)  ^{2}\right]  ^{1/2}\\
&  =\left[  \mathbb{E}\left\{  \mathbb{E}\left[  \left\vert \varphi\left(
x_{i1},x_{i3}\right)  \left(  \phi\left(  x_{i3}\right)  -\phi\left(
x_{i1}\right)  \right)  \right\vert |x_{i31}\right]  \left\vert 1\left[
x_{i31}^{\prime}b_{1}>0\right]  -1\left[  x_{i31}^{\prime}b_{2}>0\right]
\right\vert \right\}  \right]  ^{1/2}\\
&  \geq\mathbb{E}\left\{  \mathbb{E}\left[  \left\vert \varphi\left(
x_{i1},x_{i3}\right)  \left(  \phi\left(  x_{i3}\right)  -\phi\left(
x_{i1}\right)  \right)  \right\vert |x_{i31}\right]  \left\vert 1\left[
x_{i31}^{\prime}b_{1}>0\right]  -1\left[  x_{i31}^{\prime}b_{2}>0\right]
\right\vert \right\} \\
&  \geq c_{1}\mathbb{E}\left\vert 1\left[  x_{i31}^{\prime}b_{1}>0\right]
-1\left[  x_{i31}^{\prime}b_{2}>0\right]  \right\vert \\
&  \geq c_{2}\left\Vert b_{1}-b_{2}\right\Vert _{2},
\end{align*}
where the second line holds because {the} value of the term in that line is
smaller than 1, and a positive $c_{1}$ and $c_{2}$ can be guaranteed by
Assumption \hyperref[Assumption:HK]{A}.\textbf{\medskip}

\noindent\textbf{On Assumption M.iii in \cite*{SeoOtsu2018}. }This condition
can be similarly verified by%
\begin{align*}
&  \mathbb{E}\left[  \sup_{b_{1},b_{2}\in\mathcal{B}:\left\Vert b_{1}%
-b_{2}\right\Vert <\varepsilon}\left\vert \xi_{i}\left(  b_{1}\right)
-\xi_{i}\left(  b_{2}\right)  \right\vert ^{2}\right] \\
&  =\mathbb{E}\left\{  \sup_{b_{1},b_{2}\in\mathcal{B}:\left\Vert b_{1}%
-b_{2}\right\Vert <\varepsilon}\mathbb{E}\left[  \left\vert \varphi\left(
x_{i1},x_{i3}\right)  \left(  \phi\left(  x_{i3}\right)  -\phi\left(
x_{i1}\right)  \right)  \right\vert |x_{i31}\right]  \left\vert 1\left[
x_{i31}^{\prime}b_{1}>0\right]  -1\left[  x_{i31}^{\prime}b_{2}>0\right]
\right\vert \right\} \\
&  \leq c_{3}\mathbb{E}\left\{  \sup_{b_{1}\in\mathcal{B}:\left\Vert
b_{1}-b_{2}\right\Vert <\varepsilon}\left\vert 1\left[  x_{i31}^{\prime}%
b_{1}>0\right]  -1\left[  x_{i31}^{\prime}b_{2}>0\right]  \right\vert \right\}
\\
&  \leq c_{4}\varepsilon,
\end{align*}
where third line holds because $\varphi$ and $\phi$ are conditional
probability and are bounded, and the last line holds since the density of
$x_{31}$ is assumed to be bounded in Assumption \ref{A:boundedDensity}%
\textbf{. }
\end{proof}

\begin{proof}
[Proof of Lemma \ref{LE:ita_ib}]The objective function in this lemma is very
similar to the one in HK\textbf{. }The only difference is that \textbf{ }HK
put $x_{32}$ in the kernel $\mathcal{K}_{h_{n}}\left(  \cdot\right)  $ while
we put $x_{32}^{\prime}b$ and $x_{43}^{\prime}b$\ instead.

\cite*{SeoOtsu2018} verified all the technical conditions needed for the
estimator in HK and derived its asymptotics in Section B.1. Assumptions
\hyperref[Assumption:HK]{A} and \ref{A:boundedDensity}--\ref{A:hn} can
imply the technical conditions assumed in Section B.1 of \cite*{SeoOtsu2018},
and the conclusion follows.
\end{proof}

\begin{proof}
[Proof of Lemma \ref{LE:asym_xi}]Note that%
\[
Z_{n,1}\left(  \boldsymbol{s}\right)  =n^{2/3}\cdot n^{-1}\sum_{i=1}^{n}%
\xi_{i}\left(  \beta+\boldsymbol{s}n^{-1/3}\right)  =n^{1/6}\mathbb{G}%
_{n}\left(  \xi_{i}\left(  \beta+\boldsymbol{s}n^{-1/3}\right)  \right)
+n^{2/3}\mathbb{E}\left(  \xi_{i}\left(  \beta+\boldsymbol{s}n^{-1/3}\right)
\right)  ,
\]
where $\mathbb{G}_{n}\left(  \xi_{i}\left(  \beta+\boldsymbol{s}%
n^{-1/3}\right)  \right)  =n^{-1/2}\sum_{i=1}^{n}\left[  \xi_{i}\left(
\beta+\boldsymbol{s}n^{-1/3}\right)  -\mathbb{E}\left(  \xi_{i}\left(
\beta+\boldsymbol{s}n^{-1/3}\right)  \right)  \right]  $.

The mean of $Z_{n,1}\left(  \boldsymbol{s}\right)  $ is $n^{2/3}%
\mathbb{E}\left(  \xi_{i}\left(  \beta+\boldsymbol{s}n^{-1/3}\right)  \right)
.$ With Assumptions \hyperref[Assumption:HK]{A} and \ref{A:boundedDensity},
some calculation in the proof of Lemma \ref{LE:xi_ib} yields%
\begin{align*}
&  n^{2/3}\mathbb{E}\left(  \xi_{i}\left(  \beta+\boldsymbol{s}n^{-1/3}%
\right)  \right) \\
&  =n^{2/3}\left\{  \mathbb{E}\left(  \xi_{i}\left(  \beta\right)  \right)
+n^{-1/3}\left.  \frac{\partial\mathbb{E}\left(  \xi_{i}\left(  b\right)
\right)  }{\partial b}\right\vert _{b=\beta}^{\prime}\boldsymbol{s}+\frac
{1}{2}n^{-2/3}\boldsymbol{s}^{\prime}\left.  \frac{\partial^{2}\mathbb{E}%
\left(  \xi_{i}\left(  b\right)  \right)  }{\partial b\partial b^{\prime}%
}\right\vert _{b=\beta}\boldsymbol{s}+o\left(  n^{-2/3}\right)  \right\} \\
&  =\frac{1}{2}\boldsymbol{s}^{\prime}V_{1}\boldsymbol{s}+o\left(  1\right)  ,
\end{align*}
where $V_{1}$ is defined in equation (\ref{EQ:V1}).

By definition, $H_{1}\left(  \boldsymbol{s},\boldsymbol{t}\right)
=\lim_{\alpha\rightarrow\infty}\alpha\mathbb{E}\left[  \xi_{i}\left(
\beta+\boldsymbol{s}/\alpha\right)  \xi_{i}\left(  \beta+\boldsymbol{t}%
/\alpha\right)  \right]  $ is the covariance kernel for the limiting
distribution of $Z_{n,1}\left(  \boldsymbol{s}\right)  $. To obtain $H_{1}%
,$\ define%
\begin{align*}
L_{1}\left(  \boldsymbol{s}-\boldsymbol{t}\right)   &  \equiv\lim
_{\alpha\rightarrow\infty}\alpha\mathbb{E}\left[  \left(  \xi_{i}\left(
\beta+\boldsymbol{s}/\alpha\right)  -\xi_{i}\left(  \beta+\boldsymbol{t}%
/\alpha\right)  \right)  ^{2}\right]  ,\\
L_{1}\left(  \boldsymbol{s}\right)   &  \equiv\lim_{\alpha\rightarrow\infty
}\alpha\mathbb{E}\left[  \left(  \xi_{i}\left(  \beta+\boldsymbol{s}%
/\alpha\right)  -\xi_{i}\left(  \beta\right)  \right)  ^{2}\right]  ,
\end{align*}
and
\[
L_{1}\left(  \boldsymbol{t}\right)  \equiv\lim_{\alpha\rightarrow\infty}%
\alpha\mathbb{E}\left[  \left(  \xi_{i}\left(  \beta+\boldsymbol{t}%
/\alpha\right)  -\xi_{i}\left(  \beta\right)  \right)  ^{2}\right]  .
\]
Notice that $\xi_{i}\left(  \beta\right)  =0$, the relationship between
$H_{1}$ and $L_{1}$ is%
\begin{equation}
H_{1}\left(  \boldsymbol{s},\boldsymbol{t}\right)  =\frac{1}{2}\left[
L_{1}\left(  \boldsymbol{s}\right)  +L_{1}\left(  \boldsymbol{t}\right)
-L_{1}\left(  \boldsymbol{s}-\boldsymbol{t}\right)  \right]  . \label{EQ:H1}%
\end{equation}
From equations (\ref{EQ:Exi_ib}) and (\ref{EQ:xi^2}),
\begin{align*}
&  \alpha\mathbb{E}\left[  \left(  \xi_{i}\left(  \beta+\boldsymbol{s}%
/\alpha\right)  -\xi_{i}\left(  \beta+\boldsymbol{t}/\alpha\right)  \right)
^{2}\right] \\
&  =\alpha\mathbb{E}\left\{  \mathbb{E}\left[  \left\vert \varphi\left(
x_{i1},x_{i3}\right)  \left(  \phi\left(  x_{i3}\right)  -\phi\left(
x_{i1}\right)  \right)  \right\vert |x_{i31}\right]  \left\vert 1\left[
x_{i31}^{\prime}\left(  \beta+\boldsymbol{s}/\alpha\right)  >0\right]
-1\left[  x_{i31}^{\prime}\left(  \beta+\boldsymbol{t}/\alpha\right)
>0\right]  \right\vert \right\} \\
&  \equiv\alpha\mathbb{E}\left\{  \psi\left(  x_{i31}\right)  \left\vert
1\left[  x_{i31}^{\prime}\left(  \beta+\boldsymbol{s}/\alpha\right)
>0\right]  -1\left[  x_{i31}^{\prime}\left(  \beta+\boldsymbol{t}%
/\alpha\right)  >0\right]  \right\vert \right\}  .
\end{align*}
where {in the third line, we simplify notations by letting}
\[
\psi\left(  x_{i31}\right)  \equiv\mathbb{E}\left[  \left\vert \varphi\left(
x_{i1},x_{i3}\right)  \left(  \phi\left(  x_{i3}\right)  -\phi\left(
x_{i1}\right)  \right)  \right\vert |x_{i31}\right]  .
\]
{It is not} hard to see that $\psi$ defined here is equal to the $\psi$ in the body of
this lemma. Following \cite*{KimPollard1990}\textbf{, }we\textbf{ }decompose
$x_{31}$ into $\varpi\beta+x_{\beta},$ with $x_{\beta}$ orthogonal to $\beta.$
The decomposition leads to
\begin{align*}
&  \alpha\mathbb{E}\left[  \left(  \xi_{i}\left(  \beta+\boldsymbol{s}%
/\alpha\right)  -\xi_{i}\left(  \beta+\boldsymbol{t}/\alpha\right)  \right)
^{2}\right] \\
&  =\alpha\mathbb{E}\left\{  \psi\left(  x_{i31}\right)  \left\vert 1\left[
x_{i31}^{\prime}\left(  \beta+\boldsymbol{s}/\alpha\right)  >0\right]
-1\left[  x_{i31}^{\prime}\left(  \beta+\boldsymbol{t}/\alpha\right)
>0\right]  \right\vert \right\} \\
&  =\alpha\int_{\mathbb{R}^{K-1}}\int_{\mathbb{R}}\psi\left(  \varpi
\beta+x_{\beta}\right)  \left\vert 1\left[  x_{\beta}^{\prime}\boldsymbol{s}%
/\alpha+\varpi+\varpi\beta^{\prime}\boldsymbol{s}/\alpha>0\right]  -1\left[
x_{\beta}^{\prime}\boldsymbol{t}/\alpha+\varpi+\varpi\beta^{\prime
}\boldsymbol{t}/\alpha>0\right]  \right\vert \\
&  f_{x_{31}}\left(  \varpi\beta+x_{\beta}\right)  d\varpi dx_{\beta}\\
&  =\alpha\int_{\mathbb{R}^{K-1}}\int_{\mathbb{R}}\psi\left(  \varpi
\beta+x_{\beta}\right)  1\left[  \frac{-x_{\beta}^{\prime}\boldsymbol{s}%
/\alpha}{1+\beta^{\prime}\boldsymbol{s}/\alpha}<\varpi\leq\frac{-x_{\beta
}^{\prime}\boldsymbol{t}/\alpha}{1+\beta^{\prime}\boldsymbol{t}/\alpha
}\right]  f_{x_{31}}\left(  \varpi\beta+x_{\beta}\right)  d\varpi dx_{\beta}\\
&  +\alpha\int_{\mathbb{R}^{K-1}}\int_{\mathbb{R}}\psi\left(  \varpi
\beta+x_{\beta}\right)  1\left[  \frac{-x_{\beta}^{\prime}\boldsymbol{t}%
/\alpha}{1+\beta^{\prime}\boldsymbol{t}/\alpha}<\varpi\leq\frac{-x_{\beta
}^{\prime}\boldsymbol{s}/\alpha}{1+\beta^{\prime}\boldsymbol{s}/\alpha
}\right]  f_{x_{31}}\left(  \varpi\beta+x_{\beta}\right)  d\varpi dx_{\beta}\\
&  =\int_{\mathbb{R}^{K-1}}\int_{\mathbb{R}}\psi\left(  u/\alpha\beta
+x_{\beta}\right)  1\left[  \frac{-x_{\beta}^{\prime}\boldsymbol{s}}%
{1+\beta^{\prime}\boldsymbol{s}/\alpha}<u\leq\frac{-x_{\beta}^{\prime
}\boldsymbol{t}}{1+\beta^{\prime}\boldsymbol{t}/\alpha}\right]  f_{x_{31}%
}\left(  \left(  u/\alpha\right)  \beta+x_{\beta}\right)  dudx_{\beta}\\
&  +\int_{\mathbb{R}^{K-1}}\int_{\mathbb{R}}\psi\left(  u/\alpha\beta
+x_{\beta}\right)  1\left[  \frac{-x_{\beta}^{\prime}\boldsymbol{t}}%
{1+\beta^{\prime}\boldsymbol{t}/\alpha}<u\leq\frac{-x_{\beta}^{\prime
}\boldsymbol{s}}{1+\beta^{\prime}\boldsymbol{s}/\alpha}\right]  f_{x_{31}%
}\left(  \left(  u/\alpha\right)  \beta+x_{\beta}\right)  dudx_{\beta},
\end{align*}
where the fourth equality {follows} by the change of variables {$u=\alpha\varpi$.}
As $\alpha\rightarrow\infty,$
\[
L_{1}\left(  \boldsymbol{s}-\boldsymbol{t}\right)  =\int_{\mathbb{R}^{K-1}%
}\psi\left(  x_{\beta}\right)  \left\vert x_{\beta}^{\prime}\left(
\boldsymbol{s}-\boldsymbol{t}\right)  \right\vert f_{x_{31}}\left(  x_{\beta
}\right)  dx_{\beta},
\]
Similarly,
\[
L_{1}\left(  \boldsymbol{s}\right)  =\int_{\mathbb{R}^{K-1}}\psi\left(
x_{\beta}\right)  \left\vert x_{\beta}^{\prime}\boldsymbol{s}\right\vert
f_{x_{31}}\left(  x_{\beta}\right)  dx_{\beta}%
\]
and%
\[
L_{1}\left(  \boldsymbol{t}\right)  =\int_{\mathbb{R}^{K-1}}\psi\left(
x_{\beta}\right)  \left\vert x_{\beta}^{\prime}\boldsymbol{t}\right\vert
f_{x_{31}}\left(  x_{\beta}\right)  dx_{\beta}.
\]
Substituting those $L_{1}$ into equation (\ref{EQ:H1}) yields%
\[
H_{1}\left(  \boldsymbol{s},\boldsymbol{t}\right)  =\frac{1}{2}\int%
_{\mathbb{R}^{K-1}}\psi\left(  x_{\beta}\right)  \left[  \left\vert x_{\beta
}^{\prime}\boldsymbol{s}\right\vert +\left\vert x_{\beta}^{\prime
}\boldsymbol{t}\right\vert -\left\vert x_{\beta}^{\prime}\left(
\boldsymbol{s}-\boldsymbol{t}\right)  \right\vert \right]  f_{x_{31}}\left(
x_{\beta}\right)  dx_{\beta}.
\]

\end{proof}

\begin{proof}
[Proof of Lemma \ref{LE:equi}]Note%
\begin{align}
\hat{Z}_{n,2}\left(  s\right)   &  =\left(  nh_{n}\right)  ^{2/3}\cdot
n^{-1}\sum_{i=1}^{n}\varsigma_{ni}\left(  \gamma+s\left(  nh_{n}\right)
^{-1/3},\hat{\beta}\right) \nonumber\\
&  =n^{1/6}h_{n}^{2/3}\mathbb{G}_{n}\left(  \varsigma_{ni}\left(
\gamma+s\left(  nh_{n}\right)  ^{-1/3},\hat{\beta}\right)  \right)  +\left(
nh_{n}\right)  ^{2/3}\mathbb{E}_{n}\left(  \varsigma_{ni}\left(
\gamma+s\left(  nh_{n}\right)  ^{-1/3},\hat{\beta}\right)  \right) \nonumber\\
&  =n^{1/6}h_{n}^{2/3}\mathbb{G}_{n}\left(  \varsigma_{ni}\left(
\gamma+s\left(  nh_{n}\right)  ^{-1/3},\beta\right)  \right)  +\left(
nh_{n}\right)  ^{2/3}\mathbb{E}\left(  \varsigma_{ni}\left(  \gamma+s\left(
nh_{n}\right)  ^{-1/3},\beta\right)  \right) \nonumber\\
&  +n^{1/6}h_{n}^{2/3}\mathbb{G}_{n}\left(  \varsigma_{ni}\left(
\gamma+s\left(  nh_{n}\right)  ^{-1/3},\hat{\beta}\right)  -\varsigma
_{ni}\left(  \gamma+s\left(  nh_{n}\right)  ^{-1/3},\beta\right)  \right)
\nonumber\\
&  +\left(  nh_{n}\right)  ^{2/3}\mathbb{E}_{n}\left[  \varsigma_{ni}\left(
\gamma+s\left(  nh_{n}\right)  ^{-1/3},\hat{\beta}\right)  -\varsigma
_{ni}\left(  \gamma+s\left(  nh_{n}\right)  ^{-1/3},\beta\right)  \right]  ,
\label{EQ:Zns2}%
\end{align}
where $\mathbb{G}_{n}\left(  \varsigma_{ni}\left(  r,b\right)  \right)
=n^{-1/2}\sum_{i=1}^{n}\left(  \varsigma_{ni}\left(  r,b\right)
-\mathbb{E}_{n}\left(  \varsigma_{ni}\left(  r,b\right)  \right)  \right)  $.

We {first} deal with the term in the fourth line of equation (\ref{EQ:Zns2}).
Lemma \ref{LE:ita_ib} verifies the technical conditions in \cite*{SeoOtsu2018}.
Thus we can applying the result of Lemma M in\textbf{ }\cite*{SeoOtsu2018} on
$\varsigma$ and it yields\footnote{It holds by setting the $\delta$ in Lemma M
of \cite*{SeoOtsu2018} as $n^{-1/3}.$}%
\begin{align*}
&  \mathbb{E}\left\{  \sup_{\left\vert s\right\vert \leq C,\left\Vert
b-\beta\right\Vert _{2}\leq Mn^{-1/3}}n^{1/6}h_{n}^{2/3}\left\vert
\mathbb{G}_{n}\left[  \left(  \varsigma_{ni}\left(  \gamma+s\left(
nh_{n}\right)  ^{-1/3},b\right)  -\varsigma_{ni}\left(  \gamma+s\left(
nh_{n}\right)  ^{-1/3},\beta\right)  \right)  \right]  \right\vert \right\} \\
&  =n^{1/6}h_{n}^{1/6}\mathbb{E}\left\{  \sup_{\left\vert s\right\vert \leq
C,\left\Vert b-\beta\right\Vert _{2}\leq Mn^{-1/3}}\left\vert \mathbb{G}%
_{n}\left[  h_{n}^{1/2}\left(  \varsigma_{ni}\left(  \gamma+s\left(
nh_{n}\right)  ^{-1/3},b\right)  -\varsigma_{ni}\left(  \gamma+s\left(
nh_{n}\right)  ^{-1/3},\beta\right)  \right)  \right]  \right\vert \right\} \\
&  \leq cn^{1/6}h_{n}^{1/6}n^{-1/6}=o\left(  1\right)  ,
\end{align*}
for some positive $c,$\ any positive constants $M$ and $C$. By Markov's
inequality, the above yields%
\[
\sup_{\left\vert s\right\vert \leq C,\left\Vert b-\beta\right\Vert _{2}\leq
Mn^{-1/3}}n^{1/6}h_{n}^{2/3}\left\vert \mathbb{G}_{n}\left[  \left(
\varsigma_{ni}\left(  \gamma+s\left(  nh_{n}\right)  ^{-1/3},b\right)
-\varsigma_{ni}\left(  \gamma+s\left(  nh_{n}\right)  ^{-1/3},\beta\right)
\right)  \right]  \right\vert =o_{P}\left(  1\right)  .
\]
Since $\hat{\beta}-\beta=O_{P}\left(  n^{-1/3}\right)  ,$ we can take $M$
large enough so that $P\left(  \left\Vert \hat{\beta}-\beta\right\Vert
_{2}>Mn^{-1/3}\right)  <\varepsilon$ for any small $\varepsilon>0.$ For any
small $\delta>0$,
\begin{align*}
&  P\left(  \sup_{\left\vert s\right\vert \leq C}n^{1/6}h_{n}^{2/3}\left\vert
\mathbb{G}_{n}\left[  \left(  \varsigma_{ni}\left(  \gamma+s\left(
nh_{n}\right)  ^{-1/3},\hat{\beta}\right)  -\varsigma_{ni}\left(
\gamma+s\left(  nh_{n}\right)  ^{-1/3},\beta\right)  \right)  \right]
\right\vert \geq\delta\right) \\
&  =P\left(  \left\{  \sup_{\left\vert s\right\vert \leq C}n^{1/6}h_{n}%
^{2/3}\left\vert \mathbb{G}_{n}\left[  \left(  \varsigma_{ni}\left(
\gamma+s\left(  nh_{n}\right)  ^{-1/3},\hat{\beta}\right)  -\varsigma
_{ni}\left(  \gamma+s\left(  nh_{n}\right)  ^{-1/3},\beta\right)  \right)
\right]  \right\vert \geq\delta\right\}  \right. \\
&  \left.  \cap\left\{  \left\Vert \hat{\beta}-\beta\right\Vert _{2}\leq
Mn^{-1/3}\right\}  \right) \\
&  +P\left(  \left\{  \sup_{\left\vert s\right\vert \leq C}n^{1/6}h_{n}%
^{2/3}\left\vert \mathbb{G}_{n}\left[  \left(  \varsigma_{ni}\left(
\gamma+s\left(  nh_{n}\right)  ^{-1/3},\hat{\beta}\right)  -\varsigma
_{ni}\left(  \gamma+s\left(  nh_{n}\right)  ^{-1/3},\beta\right)  \right)
\right]  \right\vert \geq\delta\right\}  \right. \\
&  \left.  \cap\left\{  \left\Vert \hat{\beta}-\beta\right\Vert _{2}%
>Mn^{-1/3}\right\}  \right) \\
&  \leq P\left(  \sup_{\left\vert s\right\vert \leq C,\left\Vert
b-\beta\right\Vert _{2}\leq Mn^{-1/3}}n^{1/6}h_{n}^{2/3}\left\vert
\mathbb{G}_{n}\left[  \left(  \varsigma_{ni}\left(  \gamma+s\left(
nh_{n}\right)  ^{-1/3},b\right)  -\varsigma_{ni}\left(  \gamma+s\left(
nh_{n}\right)  ^{-1/3},\beta\right)  \right)  \right]  \right\vert \geq
\delta\right)  +\varepsilon.
\end{align*}
Because the first term in the last line can be
arbitrary small as $n\rightarrow\infty$, for $n$ large enough,
\[
P\left(  \sup_{\left\vert s\right\vert \leq C}n^{1/6}h_{n}^{2/3}\left\vert
\mathbb{G}_{n}\left[  \left(  \varsigma_{ni}\left(  \gamma+s\left(
nh_{n}\right)  ^{-1/3},\hat{\beta}\right)  -\varsigma_{ni}\left(
\gamma+s\left(  nh_{n}\right)  ^{-1/3},\beta\right)  \right)  \right]
\right\vert \geq\delta\right)  \leq2\varepsilon,
\]
{holds for any small $\delta>0$.} This implies%
\begin{equation}
\sup_{\left\vert s\right\vert \leq C}n^{1/6}h_{n}^{2/3}\left\vert
\mathbb{G}_{n}\left[  \left(  \varsigma_{ni}\left(  \gamma+s\left(
nh_{n}\right)  ^{-1/3},\hat{\beta}\right)  -\varsigma_{ni}\left(
\gamma+s\left(  nh_{n}\right)  ^{-1/3},\beta\right)  \right)  \right]
\right\vert =o_{P}\left(  1\right)  . \label{EQ:max}%
\end{equation}

For the fourth term in equation (\ref{EQ:Zns2}), with $\hat{\beta}-\beta
=O_{P}\left(  n^{-1/3}\right)  $ and $h_{n}\rightarrow0,$ the expansion in
equation (\ref{EQ:expansion2}) implies%
\begin{equation}
\left(  nh_{n}\right)  ^{2/3}\mathbb{E}_{n}\left(  \varsigma_{ni}\left(
\gamma+s\left(  nh_{n}\right)  ^{-1/3},\hat{\beta}\right)  \right)  =\left(
nh_{n}\right)  ^{2/3}\mathbb{E}\left(  \varsigma_{ni}\left(  \gamma+s\left(
nh_{n}\right)  ^{-1/3},\beta\right)  \right)  +o_{P}\left(  1\right)  ,
\label{EQ:max2}%
\end{equation}
uniformly over $\left\vert s\right\vert \leq C$. Substituting the results of
equations (\ref{EQ:max}) and (\ref{EQ:max2}) into equation (\ref{EQ:Zns2})
yields,
\begin{align*}
\hat{Z}_{n,2}\left(  s\right)   &  =n^{1/6}h_{n}^{2/3}\mathbb{G}_{n}\left(
\varsigma_{ni}\left(  \gamma+s\left(  nh_{n}\right)  ^{-1/3},\beta\right)
\right)  +\left(  nh_{n}\right)  ^{2/3}\mathbb{E}\left(  \varsigma_{ni}\left(
\gamma+s\left(  nh_{n}\right)  ^{-1/3},\beta\right)  \right)  +o_{P}\left(
1\right) \\
&  =Z_{n,2}\left(  s\right)  +o_{P}\left(  1\right)  ,
\end{align*}
where the small order term holds uniformly over $\left\vert s\right\vert
\leq C$ for any positive $C$. The claim is proved.
\end{proof}

\begin{proof}
[Proof of Lemma \ref{LE:asym_ita}]We could prove the first claim in this lemma
by the Taylor expansion of $\mathbb{E}\left(  \varsigma_{ni}\left(
r,\beta\right)  \right)  $ with respect to $r$ around $\gamma.$ We show a more
general result instead; we derive the Taylor expansion of $\mathbb{E}\left(
\varsigma_{ni}\left(  r,b\right)  \right)  $ with respect to $\left(
r,b\right)  $ around $\left(  \gamma,\beta\right)  .$ This more general result
is useful for understanding Lemma \ref{LE:asym_ita}\ and part of the
derivation in Lemma \ref{LE:equi}.

Recall that%
\begin{align*}
\varsigma_{ni}\left(  r,b\right)   &  \equiv\mathcal{K}_{h_{n}}\left(
x_{i32}^{\prime}b\right)  \left(  y_{i2}-y_{i1}\right)  \left(  1\left[
x_{i21}^{\prime}b+r\left(  y_{i3}-y_{i0}\right)  >0\right]  -1\left[
x_{i21}^{\prime}\beta+\gamma\left(  y_{i3}-y_{i0}\right)  >0\right]  \right)
\\
&  +\mathcal{K}_{h_{n}}\left(  x_{i43}^{\prime}b\right)  \left(  y_{i3}%
-y_{i2}\right)  \left(  1\left[  x_{i32}^{\prime}b+r\left(  y_{i4}%
-y_{i1}\right)  >0\right]  -1\left[  x_{i32}^{\prime}\beta+\gamma\left(
y_{i4}-y_{i1}\right)  >0\right]  \right)  .
\end{align*}
To ease of notations, let
\begin{align*}
\vartheta_{1}\left(  r,b\right)   &  \equiv\left(  y_{2}-y_{1}\right)  \left(
1\left[  x_{21}^{\prime}b+r\left(  y_{3}-y_{0}\right)  >0\right]  -1\left[
x_{21}^{\prime}\beta+\gamma\left(  y_{3}-y_{0}\right)  >0\right]  \right)  ,\\
\vartheta_{2}\left(  r,b\right)   &  \equiv\left(  y_{3}-y_{2}\right)  \left(
1\left[  x_{32}^{\prime}b+r\left(  y_{4}-y_{1}\right)  >0\right]  -1\left[
x_{32}^{\prime}\beta+\gamma\left(  y_{4}-y_{1}\right)  >0\right]  \right)  .
\end{align*}
We deal with the first component in $\varsigma_{ni}\left(  r,b\right)  $ first
and the second term can be done analogously. First,%
\begin{align*}
&  \mathbb{E}\left[  \mathcal{K}_{h_{n}}\left(  x_{32}^{\prime}b\right)
\vartheta_{1}\left(  r,b\right)  \right] \\
&  =\int_{\mathbb{R}^{K}}\mathbb{E}\left[  \vartheta_{1}\left(  r,b\right)
|x_{32}=x\right]  \mathcal{K}_{h_{n}}\left(  x^{\prime}b\right)  f_{x_{32}%
}\left(  x\right)  dx\\
&  =\int_{\mathbb{R}^{K}}\mathbb{E}\left[  \vartheta_{1}\left(  r,b\right)
|x_{32}=x\right]  \frac{1}{h_{n}}\mathcal{K}\left(  \frac{x^{\prime}b}{h_{n}%
}\right)  f_{x_{32}}\left(  x\right)  dx.
\end{align*}
Decompose $x_{32}$ into $x_{32}=\varpi b+x_{b},$ where $x_{b}$ is orthogonal
to $b.$ That yields%
\begin{align}
&  \mathbb{E}\left[  \mathcal{K}_{h_{n}}\left(  x_{32}^{\prime}b\right)
\vartheta_{1}\left(  r,b\right)  \right]  =\int_{\mathbb{R}^{K-1}}%
\int_{\mathbb{R}}\mathbb{E}\left[  \vartheta_{1}\left(  r,b\right)
|x_{32}=\varpi b+x_{b}\right]  \frac{1}{h_{n}}\mathcal{K}\left(  \frac{\varpi
}{h_{n}}\right)  f_{x_{32}}\left(  \varpi b+x_{b}\right)  d\varpi
dx_{b}\label{EQ:expansion}\\
&  =\int_{\mathbb{R}^{K-1}}\int_{\mathbb{R}}\mathbb{E}\left[  \vartheta
_{1}\left(  r,b\right)  \left\vert x_{32}=uh_{n}b+x_{b}\right.  \right]
\mathcal{K}\left(  u\right)  f_{x_{32}}\left(  uh_{n}b+x_{b}\right)
dudx_{b}\nonumber\\
&  =\int_{\mathbb{R}^{K-1}}\mathbb{E}\left[  \vartheta_{1}\left(  r,b\right)
\left\vert x_{32}=x_{b}\right.  \right]  f_{x_{32}}\left(  x_{b}\right)
dx_{b}\nonumber\\
&  +\frac{h_{n}^{2}}{2}\int_{\mathbb{R}^{K-1}}\int_{\mathbb{R}}u^{2}%
\mathcal{K}\left(  u\right)  \left.  \frac{\partial^{2}\left(  \mathbb{E}%
\left[  \vartheta_{1}\left(  r,b\right)  \left\vert x_{32}=tb+x_{b}\right.
\right]  f_{x_{32}}\left(  tb+x_{b}\right)  \right)  }{\partial t^{2}%
}\right\vert _{t=t_{u}}du\nonumber
\end{align}
where in the first line we use the fact $\left\Vert b\right\Vert _{2}=1$, the
second line holds by the change of variables $u=\frac{\varpi}{h_{n}}$, and
last two lines hold by the Taylor expansion and $t_{u}$ is some value between
$0$ and $uh_{n}$. The bias term is of order $h_{n}^{2}$\ by Assumption
\ref{A:boundedDensity} and the symmetry and boundedness conditions of
$\mathcal{K}$ in Assumption \ref{A:kernel}. By $nh_{n}^{4}\rightarrow0$ in
Assumption \ref{A:hn}$,$ the bias term is $o\left(  \left(  nh_{n}\right)
^{-2/3}\right)  $ and asymptotically negligible.

Similar results can be obtained for $\mathbb{E}\left[  \mathcal{K}_{h_{n}%
}\left(  x_{43}^{\prime}b\right)  \vartheta_{2}\left(  r,b\right)  \right]  $.

To {summarize},
\begin{align}
\mathbb{E}\left(  \varsigma_{ni}\left(  r,b\right)  \right)   &
=\int_{\mathbb{R}^{K-1}}\mathbb{E}\left[  \vartheta_{1}\left(  r,b\right)
\left\vert x_{32}=x_{b}\right.  \right]  f_{x_{32}}\left(  x_{b}\right)
dx_{b}\label{EQ:xi_bias}\\
&  +\int_{\mathbb{R}^{K-1}}\mathbb{E}\left[  \vartheta_{2}\left(  r,b\right)
\left\vert x_{43}=x_{b}\right.  \right]  f_{x_{43}}\left(  x_{b}\right)
dx_{b}+o\left(  \left(  nh_{n}\right)  ^{-2/3}\right)  .\nonumber
\end{align}
As a result, to prove the assertion in the lemma, it is enough to derive the
first and second derivatives of {the} leading term in the {above.}

Notice that%
\[
\left.  \vartheta_{1}\right\vert _{\left(  r,b\right)  =\left(  \gamma
,\beta\right)  }=0.
\]
Consequently, only the derivative of $E\left[  \vartheta_{1}\left(
r,b\right)  \left\vert x_{32}=x_{b}\right.  \right]  $ with respect to $b$ in
$\vartheta_{1}$ will appear in%
\[
\left.  \frac{\partial}{\partial b}\int_{\mathbb{R}^{K-1}}\mathbb{E}\left[
\vartheta_{1}\left(  r,b\right)  \left\vert x_{32}=x_{b}\right.  \right]
f_{x_{32}}\left(  x_{b}\right)  dx_{b}\right\vert _{r=\gamma,b=\beta}.
\]
That leads to%
\begin{align*}
&  \left.  \frac{\partial}{\partial b}\int_{\mathbb{R}^{K-1}}\mathbb{E}\left[
\vartheta_{1}\left(  r,b\right)  \left\vert x_{32}=x_{b}\right.  \right]
f_{x_{32}}\left(  x_{b}\right)  dx_{b}\right\vert _{r=\gamma,b=\beta}\\
&  =\int_{\mathbb{R}^{K-1}}\left.  \frac{\partial}{\partial b}\mathbb{E}%
\left[  \vartheta_{1}\left(  r,b\right)  \left\vert x_{32}=x_{\beta}\right.
\right]  \right\vert _{\left(  r,b\right)  =\left(  \gamma,\beta\right)
}f_{x_{32}}\left(  x_{\beta}\right)  dx_{\beta}.
\end{align*}
By similar derivation as for the derivatives of $\mathbb{E}\left(  \xi
_{i}\left(  b\right)  \right)  $, we have%
\begin{align*}
&  \left.  \frac{\partial\mathbb{E}\left[  \vartheta_{1}\left(  r,b\right)
\left\vert x_{32}=x_{\beta}\right.  \right]  }{\partial\left(  r,b^{\prime
}\right)  ^{\prime}}\right\vert _{r=\gamma,b=\beta}\\
&  =\int1\left[  x_{21}^{\prime}\beta+\gamma y_{30}=0\right]  \mathbb{E}%
\left(  y_{21}|x_{21},y_{30},x_{32}=x_{\beta}\right)  \left(
\begin{array}
[c]{c}%
y_{30}\\
x_{21}%
\end{array}
\right)  f\left(  x_{21},y_{30}|x_{32}=x_{\beta}\right)  d\sigma_{0},
\end{align*}
where $\sigma_{0}$ is the surface measure of $\left\{  \left(  x_{21}%
,y_{30}\right)  :x_{21}^{\prime}\beta+\gamma y_{30}=0\right\}  $.

$\mathbb{E}\left(  y_{21}|x_{21},y_{30},x_{32}^{\prime}\beta=0\right)  =0$
along $x_{21}^{\prime}\beta+\gamma y_{30}=0$ by Proposition \ref{Lemma:gamma}.
Thus, the derivative above is equal to $0$ and
\[
\left.  \frac{\partial}{\partial\left(  r,b^{\prime}\right)  ^{\prime}}%
\int_{\mathbb{R}^{K-1}}\mathbb{E}\left[  \vartheta_{1}\left(  r,b\right)
\left\vert x_{32}=x_{b}\right.  \right]  f_{x_{32}}\left(  x_{b}\right)
dx_{b}\right\vert _{r=\gamma,b=\beta}=0.
\]
The fact $\mathbb{E}\left(  y_{21}|x_{21},y_{30},x_{32}^{\prime}%
\beta=0\right)  =0$ along $x_{21}^{\prime}\beta+\gamma y_{30}=0$ implies that
only the second derivatives of $\mathbb{E}\left[  \vartheta_{1}\left(
r,b\right)  \left\vert x_{32}=x_{\beta}\right.  \right]  $ contribute to the
second derivative. By similar derivation as for the second derivative of
$\mathbb{E}\left(  \xi_{i}\left(  b\right)  \right)  ,$%
\begin{align*}
&  \left.  \frac{\partial^{2}\mathbb{E}\left[  \vartheta_{1}\left(
r,b\right)  \left\vert x_{32}=x_{\beta}\right.  \right]  }{\partial\left(
r,b^{\prime}\right)  ^{\prime}\partial\left(  r,b^{\prime}\right)
}\right\vert _{r=\gamma,b=\beta}\\
&  =-\int1\left[  x_{21}^{\prime}\beta+\gamma y_{30}=0\right]  \left(
\frac{\partial\mathbb{E}\left(  y_{21}|x_{21},y_{30},x_{32}=x_{\beta}\right)
}{\partial\left(  y_{30},x_{21}^{\prime}\right)  ^{\prime}}^{\prime}\left(
\begin{array}
[c]{c}%
\gamma\\
\beta
\end{array}
\right)  \right) \\
&  f\left(  x_{21},y_{30}|x_{32}=x_{\beta}\right)  \left(
\begin{array}
[c]{c}%
y_{30}\\
x_{21}%
\end{array}
\right)  \left(
\begin{array}
[c]{cc}%
y_{30} & x_{21}^{\prime}%
\end{array}
\right)  d\sigma_{0}.
\end{align*}
Therefore%
\begin{align*}
&  \left.  \frac{\partial^{2}}{\partial\left(  r,b^{\prime}\right)  ^{\prime
}\partial\left(  r,b^{\prime}\right)  }\int_{\mathbb{R}^{K-1}}\mathbb{E}%
\left[  \vartheta_{1}\left(  r,b\right)  \left\vert x_{32}=x_{b}\right.
\right]  f_{x_{32}}\left(  x_{b}\right)  dx_{b}\right\vert _{r=\gamma,b=\beta
}\\
&  =-\int_{\mathbb{R}^{K-1}}\int1\left[  x_{21}^{\prime}\beta+\gamma
y_{30}=0\right]  \left(  \frac{\partial\mathbb{E}\left(  y_{21}|x_{21}%
,y_{30},x_{32}=x_{\beta}\right)  }{\partial\left(  y_{30},x_{21}^{\prime
}\right)  ^{\prime}}^{\prime}\left(
\begin{array}
[c]{c}%
\gamma\\
\beta
\end{array}
\right)  \right) \\
&  f\left(  x_{21},y_{30}|x_{32}=x_{\beta}\right)  \left(
\begin{array}
[c]{c}%
y_{30}\\
x_{21}%
\end{array}
\right)  \left(
\begin{array}
[c]{cc}%
y_{30} & x_{21}^{\prime}%
\end{array}
\right)  d\sigma_{0}f_{x_{32}}\left(  x_{\beta}\right)  dx_{\beta}\\
&  \equiv-\tilde{V}_{21}.
\end{align*}
Similarly,%
\begin{align*}
&  \left.  \frac{\partial^{2}}{\partial\left(  r,b^{\prime}\right)  ^{\prime
}\partial\left(  r,b^{\prime}\right)  }\int_{\mathbb{R}^{K-1}}\mathbb{E}%
\left[  \vartheta_{2}\left(  r,b\right)  \left\vert x_{43}=x_{b}\right.
\right]  f_{x_{43}}\left(  x_{b}\right)  dx_{b}\right\vert _{r=\gamma,b=\beta
}\\
&  =-\int_{\mathbb{R}^{K-1}}\int1\left[  x_{32}^{\prime}\beta+\gamma
y_{41}=0\right]  \left(  \frac{\partial\mathbb{E}\left(  y_{32}|x_{32}%
,y_{41},x_{43}=x_{\beta}\right)  }{\partial\left(  y_{41},x_{32}^{\prime
}\right)  ^{\prime}}^{\prime}\left(
\begin{array}
[c]{c}%
\gamma\\
\beta
\end{array}
\right)  \right) \\
&  f\left(  x_{32},y_{41}|x_{43}=x_{\beta}\right)  \left(
\begin{array}
[c]{c}%
y_{41}\\
x_{32}%
\end{array}
\right)  \left(
\begin{array}
[c]{cc}%
y_{41} & x_{32}^{\prime}%
\end{array}
\right)  d\sigma_{0}f_{x_{43}}\left(  x_{\beta}\right)  dx_{\beta}\\
&  \equiv-\tilde{V}_{22}.
\end{align*}
Let
\begin{equation}
\tilde{V}_{2}\equiv\tilde{V}_{21}+\tilde{V}_{22}. \label{EQ:V2}%
\end{equation}

By the Taylor expansion, Assumption \ref{A:boundedDensity}, and equation
(\ref{EQ:xi_bias}),
\begin{equation}
\mathbb{E}\left(  \varsigma_{ni}\left(  r,b\right)  \right)  =-\frac{1}%
{2}\left(  r-\gamma,\left(  b-\beta\right)  ^{\prime}\right)  \tilde{V}%
_{2}\left(
\begin{array}
[c]{c}%
r-\gamma\\
b-\beta
\end{array}
\right)  +o\left(  \left\Vert \left(
\begin{array}
[c]{c}%
r-\gamma\\
b-\beta
\end{array}
\right)  \right\Vert _{2}^{2}\right)  +o\left(  \left(  nh_{n}\right)
^{-2/3}\right)  . \label{EQ:expansion2}%
\end{equation}

We define $V_{2}$ as the first diagonal of $\tilde{V}_{2}$, that is%
\[
V_{2}\equiv e_{1}^{\prime}\tilde{V}_{2}e_{1},
\]
where $e_{1}$ is a $\left(  K+1\right)  \times1$ vector with the first element
as $1$ and the rest as $0.$ Not hard to see that%
\begin{align}
V_{2}  &  =-\int_{\mathbb{R}^{K-1}}\int1\left[  x_{21}^{\prime}\beta+\gamma
y_{30}=0\right]  \left(  \frac{\partial\mathbb{E}\left(  y_{21}|x_{21}%
,y_{30},x_{32}=x_{\beta}\right)  }{\partial\left(  y_{30},x_{21}^{\prime
}\right)  ^{\prime}}^{\prime}\left(
\begin{array}
[c]{c}%
\gamma\\
\beta
\end{array}
\right)  \right) \\
&  f\left(  x_{21},y_{30}|x_{32}=x_{\beta}\right)  \left\vert y_{30}%
\right\vert d\sigma_{0}f_{x_{32}}\left(  x_{\beta}\right)  dx_{\beta
}\nonumber\\
&  -\int_{\mathbb{R}^{K-1}}\int1\left[  x_{32}^{\prime}\beta+\gamma
y_{41}=0\right]  \left(  \frac{\partial\mathbb{E}\left(  y_{32}|x_{32}%
,y_{41},x_{43}=x_{\beta}\right)  }{\partial\left(  y_{41},x_{32}^{\prime
}\right)  ^{\prime}}^{\prime}\left(
\begin{array}
[c]{c}%
\gamma\\
\beta
\end{array}
\right)  \right) \nonumber\\
&  f\left(  x_{32},y_{41}|x_{43}=x_{\beta}\right)  \left\vert y_{41}%
\right\vert d\sigma_{0}f_{x_{43}}\left(  x_{\beta}\right)  dx_{\beta
}.\nonumber
\end{align}
Note that $V_{2}$ is a scalar.
\[
\frac{\partial\mathbb{E}\left(  y_{21}|x_{21},y_{30},x_{32}=x_{\beta}\right)
}{\partial\left(  y_{30},x_{21}^{\prime}\right)  ^{\prime}}^{\prime}\left(
\begin{array}
[c]{c}%
\gamma\\
\beta
\end{array}
\right)  \geq0\text{ and }\frac{\partial\mathbb{E}\left(  y_{32}|x_{32}%
,y_{41},x_{43}=\beta\right)  }{\partial\left(  y_{41},x_{32}^{\prime}\right)
^{\prime}}^{\prime}\left(
\begin{array}
[c]{c}%
\gamma\\
\beta
\end{array}
\right)  \geq0
\]
hold for the same reason as in the discussion under equation (\ref{EQ:V1_def}). Thus, $V_{2}\leq 0$

Using equation (\ref{EQ:expansion2}),%
\[
\lim_{n\rightarrow\infty}\left(  nh_{n}\right)  ^{2/3}\mathbb{E}_{n}\left(
\varsigma_{ni}\left(  \gamma+s\left(  nh_{n}\right)  ^{-1/3},\beta\right)
\right)  =\frac{1}{2}V_{2}s^{2}.
\]

Now, we turn to the covariance kernel. Note%
\[
H_{2}\left(  s,t\right)  =\lim_{n\rightarrow\infty}\left(  nh_{n}\right)
^{1/3}\mathbb{E}\left(  h_{n}\varsigma_{ni}\left(  \gamma+s\left(
nh_{n}\right)  ^{-1/3},\beta\right)  \varsigma_{ni}\left(  \gamma+t\left(
nh_{n}\right)  ^{-1/3},\beta\right)  \right)  .
\]
Similar for the calculation of $H_{1}$ in Lemma \ref{LE:xi_ib}, define%
\begin{align*}
L_{2}\left(  s-t\right)   &  \equiv\lim_{n\rightarrow\infty}\left(
nh_{n}\right)  ^{1/3}\mathbb{E}\left[  h_{n}\left(  \varsigma_{ni}\left(
\gamma+s\left(  nh_{n}\right)  ^{-1/3},\beta\right)  -\varsigma_{ni}\left(
\gamma+t\left(  nh_{n}\right)  ^{-1/3},\beta\right)  \right)  ^{2}\right]  ,\\
L_{2}\left(  s\right)   &  \equiv\lim_{n\rightarrow\infty}\left(
nh_{n}\right)  ^{1/3}\mathbb{E}\left[  h_{n}\left(  \varsigma_{ni}\left(
\gamma+s\left(  nh_{n}\right)  ^{-1/3},\beta\right)  -\varsigma_{ni}\left(
\gamma,\beta\right)  \right)  ^{2}\right]  ,\\
L_{2}\left(  t\right)   &  \equiv\lim_{n\rightarrow\infty}\left(
nh_{n}\right)  ^{1/3}\mathbb{E}\left[  h_{n}\left(  \varsigma_{ni}\left(
\gamma+t\left(  nh_{n}\right)  ^{-1/3},\beta\right)  -\varsigma_{ni}\left(
\gamma,\beta\right)  \right)  ^{2}\right]  .
\end{align*}
Since $\varsigma_{ni}\left(  \gamma,\beta\right)  =0,$ $H_{2}\left(
s,t\right)  =\frac{1}{2}\left[  L_{2}\left(  s\right)  +L_{2}\left(  t\right)
-L_{2}\left(  s-t\right)  \right]  .$

The following calculation is useful for $L_{2}\left(  s-t\right)  .$%
\begin{align*}
&  \mathbb{E}\left[  h_{n}\left(  \varsigma_{ni}\left(  r_{1},\beta\right)
-\varsigma_{ni}\left(  r_{2},\beta\right)  \right)  ^{2}\right] \\
&  =\mathbb{E}\left\{  h_{n}\left[  \mathcal{K}_{h_{n}}\left(  x_{i32}%
^{\prime}\beta\right)  \left(  \vartheta_{1}\left(  r_{1},\beta\right)
-\vartheta_{1}\left(  r_{2},\beta\right)  \right)  +\mathcal{K}_{h_{n}}\left(
x_{i43}^{\prime}\beta\right)  \left(  \vartheta_{2}\left(  r_{1},\beta\right)
-\vartheta_{2}\left(  r_{2},\beta\right)  \right)  \right]  ^{2}\right\} \\
&  =\mathbb{E}\left\{  h_{n}\mathcal{K}_{h_{n}}\left(  x_{i32}^{\prime}%
\beta\right)  ^{2}\left\vert \vartheta_{1}\left(  r_{1},\beta\right)
-\vartheta_{1}\left(  r_{2},\beta\right)  \right\vert +h_{n}\mathcal{K}%
_{h_{n}}\left(  x_{i43}^{\prime}\beta\right)  ^{2}\left\vert \vartheta
_{2}\left(  r_{1},\beta\right)  -\vartheta_{2}\left(  r_{2},\beta\right)
\right\vert \right. \\
&  \left.  +2h_{n}\mathcal{K}_{h_{n}}\left(  x_{i32}^{\prime}\beta\right)
\mathcal{K}_{h_{n}}\left(  x_{i43}^{\prime}\beta\right)  \left(  \vartheta
_{1}\left(  r_{1},\beta\right)  -\vartheta_{1}\left(  r_{2},\beta\right)
\right)  \left(  \vartheta_{2}\left(  r_{1},\beta\right)  -\vartheta
_{2}\left(  r_{2},\beta\right)  \right)  \right\} \\
&  \equiv\mathbb{E}\left\{  h_{n}\mathcal{K}_{h_{n}}\left(  x_{i32}^{\prime
}\beta\right)  ^{2}\left\vert \vartheta_{1}\left(  r_{1},\beta\right)
-\vartheta_{1}\left(  r_{2},\beta\right)  \right\vert +h_{n}\mathcal{K}%
_{h_{n}}\left(  x_{i43}^{\prime}\beta\right)  ^{2}\left\vert \vartheta
_{2}\left(  r_{1},\beta\right)  -\vartheta_{2}\left(  r_{2},\beta\right)
\right\vert \right\}  +R_{n}.
\end{align*}
where $R_{n}$ denotes the term in the fourth line and will be shown to be
asymptotic negligible.

The first term in the above can be calculated as follows,
\begin{align*}
&  \mathbb{E}\left\{  h_{n}\mathcal{K}_{h_{n}}\left(  x_{i32}^{\prime}%
\beta\right)  ^{2}\left\vert \vartheta_{1}\left(  r_{1},\beta\right)
-\vartheta_{1}\left(  r_{2},\beta\right)  \right\vert \right\} \\
&  =\int_{\mathbb{R}^{K}}\mathbb{E}\left[  \left\vert \vartheta_{1}\left(
r_{1},\beta\right)  -\vartheta_{1}\left(  r_{2},\beta\right)  \right\vert
|x_{32}=x\right]  \frac{1}{h_{n}}\mathcal{K}\left(  \frac{x^{\prime}\beta
}{h_{n}}\right)  ^{2}f_{x_{32}}\left(  x\right)  dx.
\end{align*}
Decompose $x_{32}$ into $x_{32}=\varpi\beta+x_{\beta},$ where $x_{\beta}$ is
orthogonal to $\beta.$ Continue the expression in the above with this
decomposition,%
\begin{align*}
&  \mathbb{E}\left\{  h_{n}\mathcal{K}_{h_{n}}\left(  x_{32}^{\prime}%
\beta\right)  ^{2}\left\vert \vartheta_{1}\left(  r_{1},\beta\right)
-\vartheta_{1}\left(  r_{2},\beta\right)  \right\vert \right\} \\
&  =\int_{\mathbb{R}^{K-1}}\int_{\mathbb{R}}\mathbb{E}\left[  \left\vert
\vartheta_{1}\left(  r_{1},\beta\right)  -\vartheta_{1}\left(  r_{2}%
,\beta\right)  \right\vert |x_{32}=\varpi\beta+x_{\beta}\right]  \frac
{1}{h_{n}}\mathcal{K}\left(  \frac{\varpi}{h_{n}}\right)  ^{2}f_{x_{32}%
}\left(  \varpi\beta+x_{\beta}\right)  d\varpi dx_{\beta}\\
&  =\int_{\mathbb{R}^{K-1}}\int_{\mathbb{R}}\mathbb{E}\left[  \left\vert
\vartheta_{1}\left(  r_{1},\beta\right)  -\vartheta_{1}\left(  r_{2}%
,\beta\right)  \right\vert |x_{32}=uh_{n}\beta+x_{\beta}\right]
\mathcal{K}\left(  u\right)  ^{2}f_{x_{32}}\left(  uh_{n}\beta+x_{\beta
}\right)  dudx_{\beta}\\
&  =\mathcal{\bar{K}}_{2}\int_{\mathbb{R}^{K-1}}\mathbb{E}\left[  \left\vert
\vartheta_{1}\left(  r_{1},\beta\right)  -\vartheta_{1}\left(  r_{2}%
,\beta\right)  \right\vert |x_{32}=x_{\beta}\right]  f_{x_{32}}\left(
x_{\beta}\right)  dx_{\beta}+O\left(  h_{n}^{2}\right)
\end{align*}
where in the third line we substitute $u=\varpi/h_{n},$ in the fourth line we
do Taylor expansion around $h_{n}=0,$ the bias term is of order $h_{n}^{2}$
for the same reason as in equation (\ref{EQ:expansion}), and $\mathcal{\bar
{K}}_{2}=\int_{\mathbb{R}}\mathcal{K}\left(  u\right)  ^{2}du.$ Using
Assumption \ref{A:hn}, $\left(  nh_{n}\right)  ^{2/3}h_{n}^{2}\rightarrow0,$
so the bias term is negligible. The rate of the above term can be seen from
\begin{align*}
&  \mathbb{E}\left[  \left\vert \vartheta_{1}\left(  r_{1},\beta\right)
-\vartheta_{1}\left(  r_{2},\beta\right)  \right\vert |x_{32}=x_{\beta}\right]
\\
&  =\int_{\mathbb{R}}\mathbb{E}\left[  \left\vert y_{21}\right\vert
|x_{21}^{\prime}\beta=\varpi,y_{30}\neq0,x_{32}=x_{\beta}\right]  \left\vert
1\left[  \varpi+r_{1}\left(  y_{3}-y_{0}\right)  >0\right]  -1\left[
\varpi+r_{2}\left(  y_{3}-y_{0}\right)  >0\right]  \right\vert \\
&  P\left(  y_{30}\neq0|x_{32}=x_{\beta},x_{21}^{\prime}\beta=\varpi\right)
f\left(  x_{21}^{\prime}\beta=\varpi|x_{32}=x_{\beta}\right)  d\varpi\\
&  =\left\vert \int_{-r_{1}}^{-r_{2}}\mathbb{E}\left[  \left\vert
y_{21}\right\vert |x_{21}^{\prime}\beta=\varpi,y_{30}=1,x_{32}=x_{\beta
}\right]  P\left(  y_{30}=1|x_{32}=x_{\beta},x_{21}^{\prime}\beta
=\varpi\right)  f\left(  x_{21}^{\prime}\beta=\varpi|x_{32}=x_{\beta}\right)
d\varpi\right\vert \\
&  +\left\vert \int_{r_{1}}^{r_{2}}\mathbb{E}\left[  \left\vert y_{21}%
\right\vert |x_{21}^{\prime}\beta=\varpi,y_{30}=-1,x_{32}=x_{\beta}\right]
P\left(  y_{30}=-1|x_{32}=x_{\beta},x_{21}^{\prime}\beta=\varpi\right)
f\left(  x_{21}^{\prime}\beta=\varpi|x_{32}=x_{\beta}\right)  d\varpi
\right\vert \\
&  \propto\left\vert r_{2}-r_{1}\right\vert .
\end{align*}
If $r_{1}=\gamma+s\left(  nh_{n}\right)  ^{-1/3}$ and $r_{2}=\gamma+t\left(
nh_{n}\right)  ^{-1/3},$ $\mathbb{E}\left[  \left\vert \vartheta_{1}\left(
r_{1},\beta\right)  -\vartheta_{1}\left(  r_{2},\beta\right)  \right\vert
|x_{32}=x_{\beta}\right]  \propto\left(  nh_{n}\right)  ^{-1/3}$ and
\begin{align*}
&  \lim_{n\rightarrow\infty}\left(  nh_{n}\right)  ^{1/3}\mathbb{E}\left[
\left\vert \vartheta_{1}\left(  r_{1},\beta\right)  -\vartheta_{1}\left(
r_{2},\beta\right)  \right\vert |x_{32}=x_{\beta}\right] \\
&  =\left\{  \mathbb{E}\left[  \left\vert y_{21}\right\vert |x_{21}^{\prime
}\beta=-\gamma,y_{30}=1,x_{32}=x_{\beta}\right]  P\left(  y_{30}%
=1|x_{32}=x_{\beta},x_{21}^{\prime}\beta=-\gamma\right)  f\left(
x_{21}^{\prime}\beta=-\gamma|x_{32}=x_{\beta}\right)  \right. \\
&  +\left.  \mathbb{E}\left[  \left\vert y_{21}\right\vert |x_{21}^{\prime
}\beta=\gamma,y_{30}=-1,x_{32}=x_{\beta}\right]  P\left(  y_{30}%
=-1|x_{32}=x_{\beta},x_{21}^{\prime}\beta=\gamma\right)  f\left(
x_{21}^{\prime}\beta=\gamma|x_{32}=x_{\beta}\right)  \right\} \\
&  \cdot\left\vert s-t\right\vert \\
&  =\left\{  \mathbb{E}\left[  \left\vert y_{21}\right\vert |x_{21}^{\prime
}\beta=-\gamma,y_{30}=1,x_{32}=x_{\beta}\right]  f\left(  y_{30}%
=1,x_{21}^{\prime}\beta=-\gamma|x_{32}=x_{\beta}\right)  \right. \\
&  \left.  +\mathbb{E}\left[  \left\vert y_{21}\right\vert |x_{21}^{\prime
}\beta=\gamma,y_{30}=-1,x_{32}=x_{\beta}\right]  f\left(  y_{30}%
=-1,x_{21}^{\prime}\beta=\gamma|x_{32}=x_{\beta}\right)  \right\}  \left\vert
s-t\right\vert
\end{align*}
Therefore%
\begin{align*}
&  \lim_{n\rightarrow\infty}\left(  nh_{n}\right)  ^{1/3}\mathbb{E}\left\{
h_{n}\mathcal{K}_{h_{n}}\left(  x_{i32}^{\prime}\beta\right)  ^{2}\left\vert
\vartheta_{1}\left(  \gamma+s\left(  nh_{n}\right)  ^{-1/3},\beta\right)
-\vartheta_{1}\left(  \gamma+t\left(  nh_{n}\right)  ^{-1/3},\beta\right)
\right\vert \right\} \\
&  =\left\vert s-t\right\vert \mathcal{\bar{K}}_{2}\int_{\mathbb{R}^{K-1}%
}\left\{  \mathbb{E}\left[  \left\vert y_{21}\right\vert |x_{21}^{\prime}%
\beta=-\gamma,y_{30}=1,x_{32}=x_{\beta}\right]  f\left(  y_{30}=1,x_{21}%
^{\prime}\beta=-\gamma|x_{32}=x_{\beta}\right)  \right. \\
&  +\left.  \mathbb{E}\left[  \left\vert y_{21}\right\vert |x_{21}^{\prime
}\beta=\gamma,y_{30}=-1,x_{32}=x_{\beta}\right]  f\left(  y_{30}%
=-1,x_{21}^{\prime}\beta=\gamma|x_{32}=x_{\beta}\right)  \right\}  f_{x_{32}%
}\left(  x_{\beta}\right)  dx_{\beta}.
\end{align*}
For the same reason,
\begin{align*}
&  \lim_{n\rightarrow\infty}\left(  nh_{n}\right)  ^{1/3}\mathbb{E}\left\{
h_{n}\mathcal{K}_{h_{n}}\left(  x_{i43}^{\prime}\beta\right)  ^{2}\left\vert
\vartheta_{2}\left(  \gamma+s\left(  nh_{n}\right)  ^{-1/3},\beta\right)
-\vartheta_{2}\left(  \gamma+t\left(  nh_{n}\right)  ^{-1/3},\beta\right)
\right\vert \right\} \\
&  =\left\vert s-t\right\vert \mathcal{\bar{K}}_{2}\int_{\mathbb{R}^{K-1}%
}\left\{  \mathbb{E}\left[  \left\vert y_{32}\right\vert |x_{32}^{\prime}%
\beta=-\gamma,y_{41}=1,x_{43}=x_{\beta}\right]  f\left(  y_{41}=1,x_{32}%
^{\prime}\beta=-\gamma|x_{43}=x_{\beta}\right)  \right. \\
&  \left.  +\mathbb{E}\left[  \left\vert y_{32}\right\vert |x_{32}^{\prime
}\beta=\gamma,y_{41}=-1,x_{43}=x_{\beta}\right]  f\left(  y_{41}%
=-1,x_{32}^{\prime}\beta=\gamma|x_{43}=x_{\beta}\right)  \right\}  f_{x_{43}%
}\left(  x_{\beta}\right)  dx_{\beta}.
\end{align*}
Similar derivation on $R_{n}$ $=2h_{n}\mathbb{E}\left[  \mathcal{K}_{h_{n}%
}\left(  x_{i32}^{\prime}\beta\right)  \mathcal{K}_{h_{n}}\left(
x_{i43}^{\prime}\beta\right)  \left(  \vartheta_{1}\left(  r_{1},\beta\right)
-\vartheta_{1}\left(  r_{2},\beta\right)  \right)  \left(  \vartheta
_{2}\left(  r_{1},\beta\right)  -\vartheta_{2}\left(  r_{2},\beta\right)
\right)  \right]  $ can show that $R_{n}\propto\left(  nh_{n}\right)
^{-2/3}h_{n}$ when $r_{1}=\gamma+s\left(  nh_{n}\right)  ^{-1/3}$ and
$r_{2}=\gamma+t\left(  nh_{n}\right)  ^{-1/3}.$ So $\left(  nh_{n}\right)
^{1/3}R_{n}\rightarrow0,$ as $n\rightarrow\infty.$

The results on $L_{2}\left(  s-t\right)  $ lead to
\begin{align*}
&  L_{2}\left(  s-t\right) \\
&  =\left\vert s-t\right\vert \mathcal{\bar{K}}_{2}\int_{\mathbb{R}^{K-1}%
}\left\{  \mathbb{E}\left[  \left\vert y_{21}\right\vert |x_{21}^{\prime}%
\beta=-\gamma,y_{30}=1,x_{32}=x_{\beta}\right]  f\left(  y_{30}=1,x_{21}%
^{\prime}\beta=-\gamma|x_{32}=x_{\beta}\right)  \right. \\
&  +\left.  \mathbb{E}\left[  \left\vert y_{21}\right\vert |x_{21}^{\prime
}\beta=\gamma,y_{30}=-1,x_{32}=x_{\beta}\right]  f\left(  y_{30}%
=-1,x_{21}^{\prime}\beta=\gamma|x_{32}=x_{\beta}\right)  \right\}  f_{x_{32}%
}\left(  x_{\beta}\right)  dx_{\beta}\\
&  +\left\vert s-t\right\vert \mathcal{\bar{K}}_{2}\int_{\mathbb{R}^{K-1}%
}\left\{  \mathbb{E}\left[  \left\vert y_{32}\right\vert |x_{32}^{\prime}%
\beta=-\gamma,y_{41}=1,x_{43}=x_{\beta}\right]  f\left(  y_{41}=1,x_{32}%
^{\prime}\beta=-\gamma|x_{43}=x_{\beta}\right)  \right. \\
&  \left.  +\mathbb{E}\left[  \left\vert y_{32}\right\vert |x_{32}^{\prime
}\beta=\gamma,y_{41}=-1,x_{43}=x_{\beta}\right]  f\left(  y_{41}%
=-1,x_{32}^{\prime}\beta=\gamma|x_{43}=x_{\beta}\right)  \right\}  f_{x_{43}%
}\left(  x_{\beta}\right)  dx_{\beta}.
\end{align*}
$L_{2}\left(  s\right)  $ and $L_{2}\left(  t\right)  $ can be obtained by%
\begin{align*}
L_{2}\left(  s\right)   &  =L_{2}\left(  s-0\right)  ,\\
L_{2}\left(  t\right)   &  =L_{2}\left(  t-0\right)  .
\end{align*}
As a result%
\[
H_{2}\left(  s,t\right)  =\frac{1}{2}\left[  L_{2}\left(  s\right)
+L_{2}\left(  t\right)  -L_{2}\left(  s-t\right)  \right]  ,
\]
which can be written as in equation (\ref{EQ:H2}).
\end{proof}

\begin{proof}
[Proof of Lemma \ref{Lemma:SD}] For the sake of brevity, we only prove the case $y_{s-1}=y_{s+1}=y_{t-1}=y_{t+1}=1$. The proofs for the {other} cases are similar. Denote
\[
C=\{y_{0}=d_{0},y_{1}=d_{1},...,y_{s-1}=1,y_{s}=d_{s},y_{s+1}=1,...,y_{t-1}%
=1,y_{t}=d_{t},y_{t+1}=1,...,y_{T}=d_{T}\}
\]
and $\varpi=(w_{1},...w_{s-1},w_{s+2},...,w_{t-1},w_{t+2},...,w_{T})$. Then,
by model (\ref{model}) and Assumption \hyperref[Assumption:HK]{A}(a)
\begin{align*}
&  P(C|(w_{s},w_{s+1},w_{t},w_{t+1})=(\omega_{0},\omega_{1},\omega_{0}%
^{\prime},\omega_{1}^{\prime}),\alpha)\\
=  &  \int P(C|\varpi,(w_{s},w_{s+1},w_{t},w_{t+1})=(\omega_{0},\omega
_{1},\omega_{0}^{\prime},\omega_{1}^{\prime}),\alpha)dF_{\varpi|(w_{s}%
,w_{s+1},w_{t},w_{t+1})=(\omega_{0},\omega_{1},\omega_{0}^{\prime},\omega
_{1}^{\prime}),\alpha}\\
=  &  \int p_{0}(w^{T},\alpha)^{d_{0}}(1-p_{0}(w^{T},\alpha))^{1-d_{0}}\times
F_{\epsilon|\alpha}(w_{1}+\gamma d_{0}+\alpha)^{d_{1}}(1-F_{\epsilon|\alpha
}(w_{1}+\gamma d_{0}+\alpha))^{1-d_{1}}\times\cdots\\
&  \times F_{\epsilon|\alpha}(\omega_{s-1}+\gamma d_{s-2}+\alpha
)F_{\epsilon|\alpha}(\omega_{0}+\gamma+\alpha)^{d_{s}}(1-F_{\epsilon|\alpha
}(\omega_{0}+\gamma+\alpha))^{1-d_{s}}F_{\epsilon|\alpha}(\omega_{1}+\gamma
d_{s}+\alpha)\times\cdots\\
&  \times F_{\epsilon|\alpha}(\omega_{t-1}+\gamma d_{t-2}+\alpha
)F_{\epsilon|\alpha}(\omega_{0}'+\gamma+\alpha)^{d_{t}}(1-F_{\epsilon|\alpha}(\omega_{0}'+\gamma+\alpha))^{1-d_{t}}F_{\epsilon|\alpha}(\omega_{1}'+\gamma d_{t}+\alpha)\times\cdots\\
&  \times F_{\epsilon|\alpha}(w_{T}+\gamma d_{T-1}+\alpha)^{d_{T}%
}(1-F_{\epsilon|\alpha}(w_{T}+\gamma d_{T-1}+\alpha))^{1-d_{T}}dF_{\varpi
|(w_{s},w_{s+1},w_{t},w_{t+1})=(\omega_{0},\omega_{1},\omega_{0}^{\prime
},\omega_{1}^{\prime}),\alpha}.
\end{align*}
Given the exchangeability assumption, If $d_{s}=d_{t}$, we have
\[
P(C|(w_{s},w_{s+1},w_{t},w_{t+1})=(\omega_{0},\omega_{1},\omega_{0}^{\prime
},\omega_{1}^{\prime}),\alpha)=P(C|(w_{s},w_{s+1},w_{t},w_{t+1})=(\omega
_{0}^{\prime},\omega_{1}^{\prime},\omega_{0},\omega_{1}),\alpha),
\]
and if $d_{s}\neq d_{t}$, we have
\[
P(C|(w_{s},w_{s+1},w_{t},w_{t+1})=(\omega_{0},\omega_{1},\omega_{0}^{\prime
},\omega_{1}^{\prime}),\alpha)=P(\tilde{C}|(w_{s},w_{s+1},w_{t},w_{t+1}%
)=(\omega_{0}^{\prime},\omega_{1}^{\prime},\omega_{0},\omega_{1}),\alpha),
\]
where $\tilde{C}=\{y_{0}=d_{0},y_{1}=d_{1},...,y_{s-1}=1,y_{s}=d_{t}%
,y_{s+1}=1,...,y_{t-1}=1,y_{t}=d_{s},y_{t+1}=1,...,y_{T}=d_{T}\}$. Then,
adding up $P(C|(w_{s},w_{s+1},w_{t},w_{t+1})=(\omega_{0},\omega_{1},\omega
_{0}^{\prime},\omega_{1}^{\prime}),\alpha)$ across all possible events $C$ and
$\tilde{C}$ yields
\begin{align}
&  P(y_{s-1}=y_{t-1}=1,y_{s+1}=y_{t+1}=1|(w_{s},w_{s+1},w_{t},w_{t+1}%
)=(\omega_{0},\omega_{1},\omega_{0}^{\prime},\omega_{1}^{\prime}%
),\alpha)\nonumber\\
=  &  P(y_{s-1}=y_{t-1}=1,y_{s+1}=y_{t+1}=1|(w_{s},w_{s+1},w_{t}%
,w_{t+1})=(\omega_{0}^{\prime},\omega_{1}^{\prime},\omega_{0},\omega
_{1}),\alpha). \label{eq:A23}%
\end{align}

Invoke Bayes' theorem to deduce
\begin{align}
&  f_{w_{s},w_{s+1},w_{t},w_{t+1}|y_{s-1}=y_{t-1}=1,y_{s+1}=y_{t+1}=1,\alpha
}(\omega_{0},\omega_{1},\omega_{0}^{\prime},\omega_{1}^{\prime})\nonumber\\
=  &  \frac{P(y_{s-1}=y_{t-1}=1,y_{s+1}=y_{t+1}=1|(w_{s},w_{s+1},w_{t}%
,w_{t+1})=(\omega_{0},\omega_{1},\omega_{0}^{\prime},\omega_{1}^{\prime
}),\alpha)}{P(y_{s-1}=y_{t-1}=1,y_{s+1}=y_{t+1}=1|\alpha)}\nonumber\\
&  \times f_{w_{s},w_{s+1},w_{t},w_{t+1}|\alpha}(\omega_{0},\omega_{1}%
,\omega_{0}^{\prime},\omega_{1}^{\prime})\nonumber\\
=  &  \frac{P(y_{s-1}=y_{t-1}=1,y_{s+1}=y_{t+1}=1|(w_{s},w_{s+1},w_{t}%
,w_{t+1})=(\omega_{0}^{\prime},\omega_{1}^{\prime},\omega_{0},\omega
_{1}),\alpha)}{P(y_{s+1}=y_{t+1}=1|y_{s-1}=1,\alpha)}\nonumber\\
&  \times f_{w_{s},w_{s+1},w_{t},w_{t+1}|\alpha}(\omega_{0}^{\prime}%
,\omega_{1}^{\prime},\omega_{0},\omega_{1})\nonumber\\
=  &  f_{w_{s},w_{s+1},w_{t},w_{t+1}|y_{s-1}=y_{t-1}=1,y_{s+1}=y_{t+1}%
=1,\alpha}(\omega_{0}^{\prime},\omega_{1}^{\prime},\omega_{0},\omega_{1}),
\label{eq:A24}%
\end{align}
where the second equality follows from (\ref{eq:A23}) and the exchangeability assumption.

Applying similar arguments to obtain
\begin{equation}
f_{w_{s},w_{t}|y_{s-1}=y_{t-1}=1,y_{s+1}=y_{t+1}=1,\alpha}(\omega_{0}%
,\omega_{0}^{\prime})=f_{w_{s},w_{t}|y_{s-1}=y_{t-1}=1,y_{s+1}=y_{t+1}%
=1,\alpha}(\omega_{0}^{\prime},\omega_{0}). \label{eq:A25}%
\end{equation}
Combine (\ref{eq:A24}) and (\ref{eq:A25}) to deduce
\begin{align*}
&  f_{w_{s+1},w_{t+1}|(w_{s},w_{t})=(\omega_{0},\omega_{0}^{\prime}%
),y_{s-1}=y_{t-1}=1,y_{s+1}=y_{t+1}=1,\alpha}(\omega_{1},\omega_{1}^{\prime
})\\
=  &  f_{w_{s+1},w_{t+1}|(w_{s},w_{t})=(\omega_{0}^{\prime},\omega
_{0}),y_{s-1}=y_{t-1}=1,y_{s+1}=y_{t+1}=1,\alpha}(\omega_{1}^{\prime}%
,\omega_{1}).
\end{align*}
Then, the desired result follows from
\begin{align*}
&  f_{w_{s+1}|(w_{s},w_{t})=(\omega_{0},\omega_{0}^{\prime}),y_{s-1}%
=y_{t-1}=1,y_{s+1}=y_{t+1}=1,\alpha}(\omega_{1})\\
=  &  \int f_{w_{s+1},w_{t+1}|(w_{s},w_{t})=(\omega_{0},\omega_{0}^{\prime
}),y_{s-1}=y_{t-1}=1,y_{s+1}=y_{t+1}=1,\alpha}(\omega_{1},\omega_{1}^{\prime
})d\omega_{1}^{\prime}\\
=  &  \int f_{w_{s+1},w_{t+1}|(w_{s},w_{t})=(\omega_{0}^{\prime},\omega
_{0}),y_{s-1}=y_{t-1}=1,y_{s+1}=y_{t+1}=1,\alpha}(\omega_{1}^{\prime}%
,\omega_{1})d\omega_{1}^{\prime}\\
=  &  f_{w_{t+1}|(w_{s},w_{t})=(\omega_{0}^{\prime},\omega_{0}),y_{s-1}%
=y_{t-1}=1,y_{s+1}=y_{t+1}=1,\alpha}(\omega_{1}).
\end{align*}
\end{proof}

\section{Some Technical Details for Section \ref{SEC:inferences}}

\label{SEC:Tinfereneces}

\subsection{Numerical Bootstrap}\label{SEC:numericalB}

If $\varepsilon_{n}=n^{-1},\ $the numerical bootstrap is reduced to the
classic bootstrap. Numerical bootstrap excludes the case $\varepsilon
_{n}=n^{-1}$\ and requires $n\varepsilon_{n}\rightarrow\infty$. The idea of
numerical bootstrap is similar to the $m$-out-of-$n$ bootstrap; $\varepsilon
_{n}^{-1}$ plays a similar role as $m$. As was shown in \cite*{HongLi2020}, this
procedure is less general than the $m$-out-of-$n$ procedure. However, once it
works, it has better finite sample performance than the $m$-out-of-$n$
bootstrap. We refer to \cite*{HongLi2020} for the details.

Below is a heuristic illustration {of} why numerical bootstrap works for
$\hat{\beta}$. $\varepsilon_{n}^{-1/3}\left(  \hat{\beta}^{\ast}-\beta\right)
$ can be shown to be $O_{P}\left(  1\right)  $ similarly as in Section \ref{SEC:mOutOfn}. Note that
\begin{equation}
\varepsilon_{n}^{-1/3}\left(  \hat{\beta}^{\ast}-\hat{\beta}\right)
=\varepsilon_{n}^{-1/3}\left(  \hat{\beta}^{\ast}-\beta\right)  -\varepsilon
_{n}^{-1/3}\left(  \hat{\beta}-\beta\right)  =\varepsilon_{n}^{-1/3}\left(
\hat{\beta}^{\ast}-\beta\right)  +o_{P}\left(  1\right)  \label{EQ:epsilonn}%
\end{equation}
by $n\varepsilon_{n}\rightarrow\infty.$ Thus, the asymptotic distribution
$\varepsilon_{n}^{-1/3}\left(  \hat{\beta}^{\ast}-\hat{\beta}\right)  $ is the
same as that of $\varepsilon_{n}^{-1/3}\left(  \hat{\beta}^{\ast}%
-\beta\right)  $. Let%
\[
\mathcal{L}_{n,1}^{\ast}\left(  b\right)  \equiv n^{-1}\sum_{i=1}^{n}\xi
_{i}\left(  b\right)  +\left(  n\varepsilon_{n}\right)  ^{1/2}\cdot n^{-1}%
\sum_{j=1}^{n}\left(  \xi_{j}^{\ast}\left(  b\right)  -n^{-1}\sum_{i=1}^{n}%
\xi_{i}\left(  b\right)  \right)  .
\]
Then $\hat{\beta}^{\ast}=\arg\max_{b\in\mathcal{B}}\mathcal{L}_{n,1}\left(
b\right)  $. By equation (\ref{EQ:epsilonn}), the asymptotic distribution of
$\varepsilon_{n}^{-1/3}\left(  \hat{\beta}^{\ast}-\hat{\beta}\right)  $ can be
established if we can show the limiting distribution of $\varepsilon
_{n}^{-2/3}\mathcal{L}_{n,1}^{\ast}\left(  \beta+\boldsymbol{s}\varepsilon
_{n}^{1/3}\right)  $.

The previous results suggest that
\begin{align*}
&  \varepsilon_{n}^{-2/3}\cdot n^{-1}\sum_{i=1}^{n}\xi_{i}\left(
\beta+\boldsymbol{s}\varepsilon_{n}^{1/3}\right) \\
&  =\varepsilon_{n}^{-2/3}\mathbb{E}\left(  \xi_{i}\left(  \beta
+\boldsymbol{s}\varepsilon_{n}^{1/3}\right)  \right)  +\varepsilon_{n}%
^{-2/3}\cdot n^{-1}\sum_{i=1}^{n}\left[  \xi_{i}\left(  \beta+\boldsymbol{s}%
\varepsilon_{n}^{1/3}\right)  -\mathbb{E}\left(  \xi_{i}\left(  \beta
+\boldsymbol{s}\varepsilon_{n}^{1/3}\right)  \right)  \right] \\
&  =\varepsilon_{n}^{-2/3}\mathbb{E}\left(  \xi_{i}\left(  \beta
+\boldsymbol{s}\varepsilon_{n}^{1/3}\right)  \right)  +o_{P}\left(  1\right)
\\
&  \overset{P}{\rightarrow}\frac{1}{2}\boldsymbol{s}^{\prime}V_{1}%
\boldsymbol{s}%
\end{align*}
over a compact set of $\boldsymbol{s},$ where the second equality holds by
$n\varepsilon_{n}\rightarrow\infty.$ The following holds by the i.i.d.
sampling:%
\[
\varepsilon_{n}^{-2/3}\cdot\left(  n\varepsilon_{n}\right)  ^{1/2}\cdot
n^{-1}\sum_{j=1}^{n}\left(  \xi_{j}^{\ast}\left(  \beta+\boldsymbol{s}%
\varepsilon_{n}^{1/3}\right)  -n^{-1}\sum_{i=1}^{n}\xi_{i}\left(
\beta+\boldsymbol{s}\varepsilon_{n}^{1/3}\right)  \right)  \rightsquigarrow
W_{1}^{\ast}\left(  \boldsymbol{s}\right)  ,
\]
where $W_{1}^{\ast}\left(  \boldsymbol{s}\right)  $ is an independent copy of
$W_{1}\left(  \boldsymbol{s}\right)  .$ As a result,%
\[
\varepsilon_{n}^{-2/3}\mathcal{L}_{n,1}^{\ast}\left(  \beta+\boldsymbol{s}%
\varepsilon_{n}^{1/3}\right)  \rightsquigarrow\frac{1}{2}\boldsymbol{s}%
^{\prime}V_{1}\boldsymbol{s}+W_{1}^{\ast}\left(  \boldsymbol{s}\right)  ,
\]
as desired.

$\hat{\gamma}$ does not directly fit into the theoretical framework of
\cite*{HongLi2020}. More specifically, condition (vi) in Theorem 4.1 in
\cite*{HongLi2020} is not satisfied. The previous results suggest that
everything in \cite*{HongLi2020} can go through by modifying condition (vi) to
that%
\[
\Sigma\left(  s,t\right)  =\lim_{n\rightarrow\infty}\left(  nh_{n}\right)
^{1/3}\mathbb{E}\left(  h_{n}\varsigma_{ni}\left(  \gamma+s\left(
nh_{n}\right)  ^{-1/3},\beta\right)  \varsigma_{ni}\left(  \gamma+t\left(
nh_{n}\right)  ^{-1/3},\beta\right)  \right)
\]
exists for each $s,t$ in $\mathbb{R}$. This is true by Lemma \ref{LE:ita_ib}%
. In what follows, we illustrate why numerical bootstrap works for
$\hat{\gamma}.$

To concentrate on the key intuition, here we suppose that the effect of the first step
estimator $\hat{\beta}$ has been handled, and it does not affect the
asymptotics of $\hat{\gamma}^{\ast}$. Let
\[
\mathcal{L}_{n,2}^{\ast}\left(  r\right)  \equiv n^{-1}\sum_{i=1}^{n}%
\varsigma_{ni}\left(  r,\beta\right)  +\left(  n\varepsilon_{n}\right)
^{1/2}\cdot n^{-1}\sum_{j=1}^{n}\left(  \varsigma_{nj}^{\ast}\left(
r,\beta\right)  -n^{-1}\sum_{i=1}^{n}\varsigma_{ni}\left(  r,\beta\right)
\right)  ,
\]
where we use the same $h_{n}$ in $\varsigma_{ni}\left(  r,\beta\right)  $ and
$\varsigma_{nj}^{\ast}\left(  r,\beta\right)  $. The convergence rate of
$\hat{\gamma}_{n}^{\ast}$ to $\gamma$\ can be shown to be $\left(
\varepsilon_{n}^{-1}h_{n}\right)  ^{1/3}$. Thus, we only need to show the
limit of $\left(  \varepsilon_{n}^{-1}h_{n}\right)  ^{2/3}\mathcal{L}%
_{n,2}^{\ast}\left(  \gamma+s\left(  \varepsilon_{n}^{-1}h_{n}\right)
^{-1/3}\right)  $. Previous results suggest that
\begin{align*}
&  \left(  \varepsilon_{n}^{-1}h_{n}\right)  ^{2/3}\cdot n^{-1}\sum_{i=1}%
^{n}\varsigma_{ni}\left(  \gamma+s\left(  \varepsilon_{n}^{-1}h_{n}\right)
^{-1/3},\beta\right) \\
&  =\left(  \varepsilon_{n}^{-1}h_{n}\right)  ^{2/3}\mathbb{E}\left(
\varsigma_{ni}\left(  \gamma+s\left(  \varepsilon_{n}^{-1}h_{n}\right)
^{-1/3},\beta\right)  \right) \\
&  +\left(  \varepsilon_{n}^{-1}h_{n}\right)  ^{2/3}\cdot n^{-1}\sum_{i=1}%
^{n}\left(  \varsigma_{ni}\left(  \gamma+s\left(  \varepsilon_{n}^{-1}%
h_{n}\right)  ^{-1/3},\beta\right)  -\mathbb{E}\left(  \varsigma_{ni}\left(
\gamma+s\left(  \varepsilon_{n}^{-1}h_{n}\right)  ^{-1/3},\beta\right)
\right)  \right) \\
&  =\left(  \varepsilon_{n}^{-1}h_{n}\right)  ^{2/3}\mathbb{E}\left(
\varsigma_{ni}\left(  \gamma+s\left(  \varepsilon_{n}^{-1}h_{n}\right)
^{-1/3},\beta\right)  \right)  +o_{P}\left(  1\right) \\
&  \overset{P}{\rightarrow}\frac{1}{2}V_{2}s^{2},
\end{align*}
and%
\begin{align}
&  \left(  \varepsilon_{n}^{-1}h_{n}\right)  ^{2/3}\cdot\left(  n\varepsilon
_{n}\right)  ^{1/2}\cdot n^{-1}\sum_{j=1}^{n}\left(  \varsigma_{nj}^{\ast
}\left(  \gamma+s\left(  \varepsilon_{n}^{-1}h_{n}\right)  ^{-1/3}%
,\beta\right)  -n^{-1}\sum_{i=1}^{n}\varsigma_{ni}\left(  \gamma+s\left(
\varepsilon_{n}^{-1}h_{n}\right)  ^{-1/3},\beta\right)  \right) \nonumber\\
&  \rightsquigarrow W_{2}^{\ast}\left(  s\right)  \label{EQ:epsilonnW}%
\end{align}
by i.i.d. and the Central Limit Theorem, where $W_{2}^{\ast}\left(  s\right)
$ is an independent copy of $W_{2}\left(  s\right)  .$ To let equation
(\ref{EQ:epsilonnW}) hold, it additionally requires $\varepsilon_{n}^{-1}%
h_{n}\rightarrow\infty$ and $\varepsilon_{n}^{-1}h_{n}^{4}\rightarrow0,$
similar to the additional restriction on $m$.

\subsection{Classic Bootstrap\label{SEC:failBootstrap}}

The classic bootstrap estimators for $\hat{\beta}$ and $\hat{\gamma},$ denoted
as $\hat{\beta}^{\ast}$ and $\hat{\gamma}^{\ast},$ are constructed from%
\[
\hat{\beta}^{\ast}=\arg\max_{b\in\mathcal{B}}n^{-1}\sum_{j=1}^{n}\xi_{j}%
^{\ast}\left(  b\right)  ,\text{ and }\hat{\gamma}^{\ast}=\arg\max
_{r\in\mathcal{R}}n^{-1}\sum_{j=1}^{n}\varsigma_{nj}^{\ast}\left(
r,\hat{\beta}\right)  .
\]
Based on the proof in \cite*{AbrevayaHuang2005}, we have%
\[
n^{1/3}\left(  \hat{\beta}^{\ast}-\beta\right)  \overset{d}{\rightarrow}%
\arg\max_{\boldsymbol{s}\in\mathbb{R}^{K}}\left(  \frac{1}{2}\boldsymbol{s}%
^{\prime}V_{1}\boldsymbol{s}+W_{1}\left(  \boldsymbol{s}\right)  +W_{1}^{\ast
}\left(  \boldsymbol{s}\right)  \right)
\]
and%
\[
\left(  nh_{n}\right)  ^{1/3}\left(  \hat{\gamma}^{\ast}-\gamma\right)
\overset{d}{\rightarrow}\arg\max_{s\in\mathbb{R}}\left(  \frac{1}{2}V_{2}%
s^{2}+W_{2}\left(  s\right)  +W_{2}^{\ast}\left(  s\right)  \right)  ,
\]
where $W_{1}\left(  \boldsymbol{s}\right)  $ and $W_{1}^{\ast}\left(
\boldsymbol{s}\right)  $ are identical and independent Gaussian processes with
zero mean and covariance kernel $H_{1},$ and $W_{2}\left(  s\right)  $ and
$W_{2}^{\ast}\left(  s\right)  $ are identical and independent Gaussian
processes with zero mean and covariance kernel $H_{2}.$ $V_{1},V_{2},H_{1}$
and $H_{2}$ are the same as in Theorem \ref{TH:betahat}.

Therefore%
\begin{align*}
n^{1/3}\left(  \hat{\beta}^{\ast}-\hat{\beta}\right)   &  =n^{1/3}\left(
\hat{\beta}^{\ast}-\beta\right)  -n^{1/3}\left(  \hat{\beta}-\beta\right) \\
&  \overset{d}{\rightarrow}\arg\max_{\boldsymbol{s}\in\mathbb{R}^{K}}\left(
\frac{1}{2}\boldsymbol{s}^{\prime}V_{1}\boldsymbol{s}+W_{1}\left(
\boldsymbol{s}\right)  +W_{1}^{\ast}\left(  \boldsymbol{s}\right)  \right)
-\arg\max_{\boldsymbol{s}\in\mathbb{R}^{K}}\left(  \frac{1}{2}\boldsymbol{s}%
^{\prime}V_{1}\boldsymbol{s}+W_{1}\left(  \boldsymbol{s}\right)  \right)  ,
\end{align*}
and%

\begin{align*}
\left(  nh_{n}\right)  ^{1/3}\left(  \hat{\gamma}^{\ast}-\hat{\gamma}\right)
&  =\left(  nh_{n}\right)  ^{1/3}\left(  \hat{\gamma}^{\ast}-\gamma\right)
-\left(  nh_{n}\right)  ^{1/3}\left(  \hat{\gamma}-\gamma\right) \\
&  \overset{d}{\rightarrow}\arg\max_{s\in\mathbb{R}}\left(  \frac{1}{2}%
V_{2}s^{2}+W_{2}\left(  s\right)  +W_{2}^{\ast}\left(  s\right)  \right)
-\arg\max_{s\in\mathbb{R}}\left(  \frac{1}{2}V_{2}s^{2}+W_{2}\left(  s\right)
\right).
\end{align*}

Here, we provide a {sketch showing} the inconsistency of the classic bootstrap.

By similar arguments of Lemma 3 in \cite*{AbrevayaHuang2005}, the convergence
rate of $\hat{\beta}^{\ast}$ to $\beta$\ and $\hat{\gamma}^{\ast}$ to $\gamma$
can be shown be at $n^{-1/3}$ and $\left(  nh_{n}\right)  ^{-1/3}$ respectively.

Define
\[
Z_{n,1}^{\ast}\left(  \boldsymbol{s}\right)  \equiv n^{2/3}\cdot n^{-1}%
\sum_{j=1}^{n}\xi_{j}^{\ast}\left(  \beta+\boldsymbol{s}n^{-1/3}\right)  .
\]
Similar to Theorem 1 in \cite*{AbrevayaHuang2005}, one can show%
\begin{equation}
Z_{n,1}^{\ast}\left(  \boldsymbol{s}\right)  \rightsquigarrow\frac{1}%
{2}\boldsymbol{s}^{\prime}V_{1}\boldsymbol{s}+W_{1}\left(  \boldsymbol{s}%
\right)  +W_{1}^{\ast}\left(  \boldsymbol{s}\right)  , \label{EQ:Z_n1_limit}%
\end{equation}
where $W_{1}\left(  \boldsymbol{s}\right)  $ and $W_{1}^{\ast}\left(
\boldsymbol{s}\right)  $ are independent and identical Gaussian processes. The
intuition of this result can be seen from the following decomposition of
$Z_{n,1}^{\ast}\left(  \boldsymbol{s}\right)  :$
\begin{align*}
Z_{n,1}^{\ast}\left(  \boldsymbol{s}\right)   &  =n^{2/3}\cdot n^{-1}%
\sum_{i=1}^{n}\xi_{i}\left(  \beta+\boldsymbol{s}n^{-1/3}\right)
+n^{2/3}\cdot n^{-1}\sum_{j=1}^{n}\left(  \xi_{j}^{\ast}\left(  \beta
+\boldsymbol{s}n^{-1/3}\right)  -n^{-1}\sum_{i=1}^{n}\xi_{i}\left(
\beta+\boldsymbol{s}n^{-1/3}\right)  \right) \\
&  =Z_{n,1}\left(  \boldsymbol{s}\right)  +n^{2/3}\cdot n^{-1}\sum_{j=1}%
^{n}\left(  \xi_{j}^{\ast}\left(  \beta+\boldsymbol{s}n^{-1/3}\right)
-n^{-1}\sum_{i=1}^{n}\xi_{i}\left(  \beta+\boldsymbol{s}n^{-1/3}\right)
\right)  ,
\end{align*}
where the first term weakly converges to $\frac{1}{2}\boldsymbol{s}^{\prime
}V_{1}\boldsymbol{s}+W_{1}\left(  \boldsymbol{s}\right)  ,$ and the second
term weakly converges to $W_{1}^{\ast}\left(  \boldsymbol{s}\right)  $.

Since the convergence rate of $\hat{\beta}^{\ast}$ to $\beta$ is $n^{-1/3}$, (\ref{EQ:Z_n1_limit}) implies that
\[
n^{1/3}\left(  \hat{\beta}^{\ast}-\beta\right)  \overset{d}{\rightarrow}%
\arg\max_{\boldsymbol{s}\in\mathbb{R}^{K}}\left(  \frac{1}{2}\boldsymbol{s}%
^{\prime}V_{1}\boldsymbol{s}+W_{1}\left(  \boldsymbol{s}\right)  +W_{1}^{\ast
}\left(  \boldsymbol{s}\right)  \right),
\]
and%
\begin{align*}
n^{1/3}\left(  \hat{\beta}^{\ast}-\hat{\beta}\right)   &  =n^{-1/3}\left(
\hat{\beta}^{\ast}-\beta\right)  -n^{-1/3}\left(  \hat{\beta}-\beta\right) \\
&  \overset{d}{\rightarrow}\arg\max_{\boldsymbol{s}\in\mathbb{R}^{K}}\left(
\frac{1}{2}\boldsymbol{s}^{\prime}V_{1}\boldsymbol{s}+W_{1}\left(
\boldsymbol{s}\right)  +W_{1}^{\ast}\left(  \boldsymbol{s}\right)  \right)
-\arg\max_{\boldsymbol{s}\in\mathbb{R}^{K}}\left(  \frac{1}{2}\boldsymbol{s}%
^{\prime}V_{1}\boldsymbol{s}+W_{1}\left(  \boldsymbol{s}\right)  \right)  .
\end{align*}

For $\hat{\gamma}^{\ast},$ let
\begin{align*}
\hat{Z}_{n,2}^{\ast}\left(  s\right)   &  \equiv\left(  nh_{n}\right)
^{2/3}\cdot n^{-1}\sum_{j=1}^{n}\varsigma_{nj}^{\ast}\left(  \gamma+s\left(
nh_{n}\right)  ^{-1/3},\hat{\beta}\right)  ,\text{ and}\\
Z_{n,2}^{\ast}\left(  s\right)   &  \equiv\left(  nh_{n}\right)  ^{2/3}\cdot
n^{-1}\sum_{j=1}^{n}\varsigma_{nj}^{\ast}\left(  \gamma+s\left(
nh_{n}\right)  ^{-1/3},\beta\right)  .
\end{align*}
The equicontinuity of $\left(  nh_{n}\right)  ^{2/3}\cdot n^{-1}\sum_{j=1}%
^{n}\varsigma_{nj}^{\ast}\left(  r,b\right)  $ can be proved using similar
arguments as in Theorem 1 of \cite*{AbrevayaHuang2005}. By that,
\[
\hat{Z}_{n,2}^{\ast}\left(  s\right)  =Z_{n,2}^{\ast}\left(  s\right)
+o_{P}\left(  1\right)  ,
\]
holds uniformly over a compact set of $s.$ Thus we only need to establish the
asymptotics of $Z_{n,2}^{\ast}\left(  s\right)  .$ To that end, decompose
$Z_{n,2}^{\ast}\left(  s\right)  $ as%
\begin{align}
Z_{n,2}^{\ast}\left(  s\right)   &  =Z_{n,2}\left(  s\right)  +Z_{n,2}^{\ast
}\left(  s\right)  -Z_{n,2}\left(  s\right) \nonumber\\
&  =Z_{n,2}\left(  s\right)  +\left(  nh_{n}\right)  ^{2/3}\cdot n^{-1}%
\sum_{j=1}^{n}\left(  \varsigma_{nj}^{\ast}\left(  \gamma+s\left(
nh_{n}\right)  ^{-1/3},\beta\right)  -n^{-1}\sum_{i=1}^{n}\varsigma
_{ni}\left(  \gamma+s\left(  nh_{n}\right)  ^{-1/3},\beta\right)  \right)
\nonumber\\
&  =Z_{n,2}\left(  s\right)  +\left(  nh_{n}\right)  ^{2/3}\cdot n^{-1}%
\sum_{j=1}^{n}\left(  \varsigma_{nj}^{\ast}\left(  \gamma+s\left(
nh_{n}\right)  ^{-1/3},\beta\right)  -n^{-1}\sum_{i=1}^{n}\varsigma
_{ni}\left(  \gamma+s\left(  nh_{n}\right)  ^{-1/3},\beta\right)  \right)
.\nonumber
\end{align}

Using the facts that the re-sampling is i.i.d. and $n^{-1}\sum_{j=1}%
^{n}\varsigma_{nj}^{\ast}\left(  r,b\right)  $ is equicontinuous in $r,$ it
holds that%

\[
\left(  nh_{n}\right)  ^{2/3}\cdot n^{-1}\sum_{j=1}^{n}\left(  \varsigma
_{nj}^{\ast}\left(  \gamma+s\left(  nh_{n}\right)  ^{-1/3},\beta\right)
-n^{-1}\sum_{i=1}^{n}\varsigma_{ni}\left(  \gamma+s\left(  nh_{n}\right)
^{-1/3},\beta\right)  \right)  \rightsquigarrow W_{2}^{\ast}\left(  s\right)
,
\]
where $W_{2}^{\ast}\left(  s\right)  $\textbf{ }is identically distributed as
$W_{2}\left(  s\right)  .$

Lemmas \ref{LE:ita_ib} and \ref{LE:equi} imply that
\[
Z_{n,2}\left(  s\right)  \rightsquigarrow\frac{1}{2}V_{2}s^{2}+W_{2}\left(
s\right).
\]
The independence of $W_{2}\left(  s\right)  $ and $W_{2}^{\ast}\left(
s\right)  $ can be shown using the same arguments in the proof of Theorem 1 in
\cite*{AbrevayaHuang2005}.

Combing above results implies
\[
\hat{Z}_{n,2}^{\ast}\left(  s\right)  \rightsquigarrow\frac{1}{2}V_{2}%
s^{2}+W_{2}\left(  s\right)  +W_{2}^{\ast}\left(  s\right)  .
\]
Thus,
\[
\left(  nh_{n}\right)  ^{1/3}\left(  \hat{\gamma}^{\ast}-\gamma\right)
\overset{d}{\rightarrow}\arg\max_{s\in\mathbb{R}}\left(  \frac{1}{2}V_{2}%
s^{2}+W_{2}\left(  s\right)  +W_{2}^{\ast}\left(  s\right)  \right)  ,
\]
and
\begin{align*}
\left(  nh_{n}\right)  ^{1/3}\left(  \hat{\gamma}^{\ast}-\hat{\gamma}\right)
&  =\left(  nh_{n}\right)  ^{1/3}\left(  \hat{\gamma}^{\ast}-\gamma\right)
-\left(  nh_{n}\right)  ^{1/3}\left(  \hat{\gamma}-\gamma\right) \\
&  \overset{d}{\rightarrow}\arg\max_{s\in\mathbb{R}}\left(  \frac{1}{2}%
V_{2}s^{2}+W_{2}\left(  s\right)  +W_{2}^{\ast}\left(  s\right)  \right)
-\arg\max_{s\in\mathbb{R}}\left(  \frac{1}{2}V_{2}s^{2}+W_{2}\left(  s\right)
\right).
\end{align*}

\subsection{\texorpdfstring{$m$}{Lg}-out-of-\texorpdfstring{$n$}{Lg} Bootstrap\label{SEC:mOutOfn}}

Here $m\rightarrow\infty$ as $n\rightarrow\infty,$ but $m/n\rightarrow0$ as
$n\rightarrow\infty.$\ This procedure is as follows. Draw $(
y_{j}^{T\ast},x_{j}^{T\ast\prime})  ^{\prime},$ $j=1,...,m,$
independently from the collection of the sample values $\left(  y_{1}%
^{T},x_{1}^{T\prime}\right)  ^{\prime},$ $\left(  y_{2}^{T},x_{2}^{T\prime
}\right)  ^{\prime},$ $...,$ $\left(  y_{n}^{T},x_{n}^{T\prime}\right)
^{\prime}$ with replacement$.$ Let $\hat{\beta}^{\ast}$ and $\hat{\gamma
}^{\ast}$ be the estimator from the sampling observations, that is
\begin{equation}
\hat{\beta}^{\ast}=\arg\max_{b\in\mathcal{B}}m^{-1}\sum_{j=1}^{m}\xi_{j}%
^{\ast}\left(  b\right)  \text{ and }\hat{\gamma}^{\ast}=\arg\max
_{r\in\mathcal{R}}m^{-1}\sum_{j=1}^{m}\varsigma_{nj}^{\ast}\left(
r,\hat{\beta}\right)  , \label{EQ:betagammaMofN}%
\end{equation}
where the bandwidth used in $\varsigma_{nj}^{\ast}$ is $h_{n},$ for
simplicity. As the name suggests, this procedure only samples a small portion
($m$ observations) from the data ($n$ observations), with the hope of
\textquotedblleft correcting\textquotedblright\ the inconsistency of the
classic bootstrap.\ \cite*{LeePun2006} proved the consistency of $m$-out-of-$n$
bootstrap for non-standard M-estimators under mild conditions. After proving
the general result, they applied it to the maximum score estimator by verifying
the required technical conditions. We claim that these technical conditions
can be similarly verified for our estimator and
\[
m^{1/3}\left(  \hat{\beta}^{\ast}-\hat{\beta}\right)  \overset{d}{\rightarrow
}\arg\max_{\boldsymbol{s}\in\mathbb{R}^{K}}\left(  \frac{1}{2}\boldsymbol{s}%
^{\prime}V_{1}\boldsymbol{s}+W_{1}\left(  \boldsymbol{s}\right)  \right)
\]
and%
\begin{equation}
\left(  mh_{n}\right)  ^{1/3}\left(  \hat{\gamma}^{\ast}-\hat{\gamma}\right)
\overset{d}{\rightarrow}\arg\max_{s\in\mathbb{R}}\left(  \frac{1}{2}V_{2}%
s^{2}+W_{2}\left(  s\right)  \right)  . \label{EQ:mngamma}%
\end{equation}
To make equation (\ref{EQ:mngamma}) hold, we additionally require
$mh_{n}\rightarrow\infty,$\ $mh_{n}^{4}\rightarrow0,$ analogous to the
conditions in Assumption \ref{A:hn}. Because of the length limitations of the
paper, the details are not pursued here. Instead, we have provided a heuristic
illustration.

Note $\hat{\beta}^{\ast}$ and $\hat{\gamma}^{\ast}$
in this section are obtained from expression (\ref{EQ:betagammaMofN}).\ Let
\[
Z_{m,1}^{\ast}\left(  \boldsymbol{s}\right)  \equiv m^{2/3}\cdot m^{-1}%
\sum_{j=1}^{m}\xi_{j}^{\ast}\left(  \beta+\boldsymbol{s}m^{-1/3}\right)  .
\]
Rewrite $Z_{m,1}^{\ast}\left(  \boldsymbol{s}\right)  $ as
\begin{align*}
Z_{m,1}^{\ast}\left(  \boldsymbol{s}\right)   &  =m^{2/3}\cdot m^{-1}%
\sum_{j=1}^{m}\left(  \xi_{j}^{\ast}\left(  \beta+\boldsymbol{s}%
m^{-1/3}\right)  -n^{-1}\sum_{i=1}^{n}\xi_{i}\left(  \beta+\boldsymbol{s}%
m^{-1/3}\right)  \right)  +m^{2/3}\cdot n^{-1}\sum_{i=1}^{n}\xi_{i}\left(
\beta+\boldsymbol{s}m^{-1/3}\right) \\
&  =m^{2/3}\cdot m^{-1}\sum_{j=1}^{m}\left(  \xi_{j}^{\ast}\left(
\beta+\boldsymbol{s}m^{-1/3}\right)  -n^{-1}\sum_{i=1}^{n}\xi_{i}\left(
\beta+\boldsymbol{s}m^{-1/3}\right)  \right) \\
&  +m^{2/3}\mathbb{E}\left(  \xi_{i}\left(  \beta+\boldsymbol{s}%
m^{-1/3}\right)  \right)  +m^{2/3}\cdot n^{-1}\sum_{i=1}^{n}\left[  \xi
_{i}\left(  \beta+\boldsymbol{s}m^{-1/3}\right)  -\mathbb{E}\left(  \xi
_{i}\left(  \beta+\boldsymbol{s}m^{-1/3}\right)  \right)  \right]  .
\end{align*}
Intuitively, the first term in the above equation weakly converges to $W_{1}^{\ast
}\left(  \boldsymbol{s}\right)  $, the second term converges to $\frac{1}%
{2}\boldsymbol{s}^{\prime}V_{1}\boldsymbol{s},$ and the last term converges to
zero in probability. One can similarly show $\hat{\beta}^{\ast}-\beta
=O_{P}\left(  m^{-1/3}\right)  .$

Therefore,
\[
m^{1/3}\left(  \hat{\beta}^{\ast}-\beta\right)  \overset{d}{\rightarrow}%
\arg\max_{\boldsymbol{s}\in\mathbb{R}^{K}}\frac{1}{2}\boldsymbol{s}^{\prime
}V_{1}\boldsymbol{s}+W_{1}^{\ast}\left(  \boldsymbol{s}\right)  .
\]
Finally,%
\begin{align*}
m^{1/3}\left(  \hat{\beta}^{\ast}-\hat{\beta}\right)   &  =m^{1/3}\left(
\hat{\beta}^{\ast}-\beta\right)  -m^{1/3}\left(  \hat{\beta}-\beta\right) \\
&  =m^{1/3}\left(  \hat{\beta}^{\ast}-\beta\right)  +o_{P}\left(  1\right) \\
&  \overset{d}{\rightarrow}\arg\max_{\boldsymbol{s}\in\mathbb{R}^{K}}\frac
{1}{2}\boldsymbol{s}^{\prime}V_{1}\boldsymbol{s}+W_{1}^{\ast}\left(
\boldsymbol{s}\right)  .
\end{align*}
Note that the distribution of $W_{1}^{\ast}\left(  \boldsymbol{s}\right)  $ is
the same as that of $W_{1}\left(  \boldsymbol{s}\right)  ,$ and the claim is
proved for $\hat{\beta}^{\ast}.$  The asymptotic distribution of $\left(  mh_{n}\right)  ^{1/3}\left(
\hat{\gamma}^{\ast}-\hat{\gamma}\right)  $ can be similarly established.

\section{Additional Simulation Results}\label{app:moretables}
\subsection{Simulation Results of Designs 3--5}
\begin{center}
\begin{tabular}
[c]{rcccccc}%
\multicolumn{7}{c}{Table 3A: Design 3, Performance of $\hat{\beta}$ and
$\hat{\gamma}$}\\\hline\hline
& \multicolumn{3}{c}{$n=2500$} & \multicolumn{3}{c}{$n=5000$}\\\cline{2-7}
& $\hat{\beta}_{2}$ & $\hat{\beta}_{3}$ & $\hat{\gamma}$ & $\hat{\beta}_{2}$  & $\hat{\beta}_{3}$ & $\hat{\gamma}$ \\\hline
 OY BIAS & $0.7\%$ & $-0.1\%$ & $1.5\%$ & $0.5\%$ & $-0.5\%$ & $0.8\%$ \\
     STD& $15.0\%$ & $14.9\%$ & $26.5\%$ & $11.7\%$ & $11.2\%$ & $20.8\%$ \\
     MAD& $12.0\%$ & $11.9\%$ & $21.2\%$ & $9.3\%$ & $8.9\%$ & $16.7\%$ \\
    RMSE& $15.0\%$ & $14.9\%$ & $26.5\%$ & $11.7\%$ & $11.2\%$ & $20.8\%$ \\\hline

 HK1 BIAS & $-0.7\%$ & $-0.5\%$ & $5.1\%$ & $-0.3\%$ & $-0.1\%$ & $3.6\%$ \\
     STD& $10.2\%$ & $9.8\%$ & $21.4\%$ & $7.7\%$ & $7.9\%$ & $17.4\%$ \\
     MAD& $8.1\%$ & $7.8\%$ & $17.6\%$ & $6.3\%$ & $6.3\%$ & $14.4\%$ \\
    RMSE& $10.2\%$ & $9.8\%$ & $22.0\%$ & $7.7\%$ & $7.9\%$ & $17.8\%$ \\\hline

 HK2 BIAS & $-0.1\%$ & $-0.1\%$ & $6.3\%$ & $0.1\%$ & $0.7\%$ & $4.6\%$ \\
     STD& $17.2\%$ & $17.1\%$ & $37.3\%$ & $14.3\%$ & $14.9\%$ & $32.2\%$ \\
     MAD& $13.9\%$ & $13.7\%$ & $30.4\%$ & $11.6\%$ & $12.1\%$ & $26.1\%$ \\
    RMSE& $17.2\%$ & $17.1\%$ & $37.8\%$ & $14.3\%$ & $14.9\%$ & $32.5\%$ \\\hline

     & \multicolumn{3}{c}{$n=10000$} & \multicolumn{3}{c}{$n=20000$} \\\cline{2-7}
  & $\hat{\beta}_{2}$ & $\hat{\beta}_{3}$ & $\hat{\gamma}$ & $\hat{\beta}_{2}$  & $\hat{\beta}_{3}$ & $\hat{\gamma}$ \\\hline
OY BIAS & $0.5\%$ & $0.1\%$ & $1.3\%$ & $-0.1\%$ & $0.1\%$ & $1.5\%$ \\
     STD& $9.4\%$ & $9.6\%$ & $17.5\%$ & $7.3\%$ & $7.1\%$ & $14.0\%$ \\
     MAD& $7.6\%$ & $7.6\%$ & $14.1\%$ & $5.9\%$ & $5.6\%$ & $11.1\%$ \\
    RMSE& $9.4\%$ & $9.6\%$ & $17.5\%$ & $7.3\%$ & $7.1\%$ & $14.0\%$ \\\hline

 HK1 BIAS & $-0.4\%$ & $-0.1\%$ & $3.5\%$ & $-0.0\%$ & $-0.3\%$ & $3.6\%$ \\
     STD& $6.3\%$ & $6.3\%$ & $13.6\%$ & $5.0\%$ & $5.0\%$ & $11.2\%$ \\
     MAD& $5.1\%$ & $5.0\%$ & $11.3\%$ & $4.0\%$ & $4.0\%$ & $9.5\%$ \\
    RMSE& $6.3\%$ & $6.3\%$ & $14.0\%$ & $5.0\%$ & $5.0\%$ & $11.7\%$ \\\hline

 HK2 BIAS & $0.4\%$ & $-0.0\%$ & $3.8\%$ & $-0.0\%$ & $0.1\%$ & $2.5\%$ \\
     STD& $12.6\%$ & $12.2\%$ & $26.9\%$ & $11.1\%$ & $10.5\%$ & $23.5\%$ \\
     MAD& $10.2\%$ & $9.9\%$ & $21.8\%$ & $9.0\%$ & $8.4\%$ & $18.5\%$ \\
    RMSE& $12.6\%$ & $12.2\%$ & $27.2\%$ & $11.1\%$ & $10.5\%$ & $23.7\%$ \\
\hline\hline
\end{tabular}

\begin{tabular}
[c]{rcccccc}%
\multicolumn{7}{c}{Table 3B: Design 3, Numerical Bootstrap}\\\hline\hline
& \multicolumn{3}{c}{$n=2500$} & \multicolumn{3}{c}{$n=5000$}\\\cline{2-7}
& $\hat{\beta}_{2}$ & $\hat{\beta}_{3}$ & $\hat{\gamma}$ & $\hat{\beta}_{2}$  & $\hat{\beta}_{3}$ & $\hat{\gamma}$ \\\hline

 $c = 0.8$ COV & $89.7\%$ & $89.3\%$ & $86.8\%$ & $93.9\%$ & $94.5\%$ & $91.1\%$ \\
 LEN & $106.7\%$ & $106.7\%$ & $107.5\%$ & $90.2\%$ & $90.3\%$ & $92.1\%$ \\\hline
 $c = 0.9$ COV & $89.4\%$ & $88.9\%$ & $86.4\%$ & $93.7\%$ & $94.4\%$ & $91.1\%$ \\
 LEN & $104.0\%$ & $103.9\%$ & $104.5\%$ & $88.3\%$ & $88.1\%$ & $90.2\%$ \\\hline
 $c = 1.0$ COV & $88.9\%$ & $88.5\%$ & $85.9\%$ & $94.1\%$ & $93.9\%$ & $90.7\%$ \\
 LEN & $101.2\%$ & $101.5\%$ & $102.1\%$ & $86.6\%$ & $86.5\%$ & $88.6\%$ \\\hline
 $c = 1.1$ COV & $87.6\%$ & $88.0\%$ & $85.0\%$ & $93.9\%$ & $93.9\%$ & $90.1\%$ \\
 LEN & $99.3\%$ & $99.2\%$ & $99.8\%$ & $85.2\%$ & $85.1\%$ & $87.0\%$ \\\hline
 $c = 1.2$ COV & $87.6\%$ & $87.1\%$ & $84.6\%$ & $93.4\%$ & $93.7\%$ & $90.0\%$ \\
 LEN & $97.1\%$ & $97.1\%$ & $97.5\%$ & $83.7\%$ & $83.9\%$ & $85.5\%$ \\\hline

     & \multicolumn{3}{c}{$n=10000$} & \multicolumn{3}{c}{$n=20000$} \\\cline{2-7}
  & $\hat{\beta}_{2}$ & $\hat{\beta}_{3}$ & $\hat{\gamma}$ & $\hat{\beta}_{2}$  & $\hat{\beta}_{3}$ & $\hat{\gamma}$ \\\hline

$c = 0.8$ COV & $94.3\%$ & $92.8\%$ & $91.5\%$ & $94.1\%$ & $95.0\%$ & $90.5\%$ \\
 LEN & $75.8\%$ & $75.3\%$ & $77.6\%$ & $62.7\%$ & $62.5\%$ & $63.7\%$ \\\hline
 $c = 0.9$ COV & $94.2\%$ & $92.6\%$ & $90.6\%$ & $94.9\%$ & $95.0\%$ & $91.4\%$ \\
 LEN & $74.5\%$ & $74.0\%$ & $76.9\%$ & $61.7\%$ & $61.8\%$ & $63.7\%$ \\\hline
 $c = 1.0$ COV & $94.2\%$ & $92.9\%$ & $90.7\%$ & $94.6\%$ & $95.4\%$ & $91.3\%$ \\
 LEN & $73.3\%$ & $73.0\%$ & $75.9\%$ & $60.9\%$ & $60.9\%$ & $63.4\%$ \\\hline
 $c = 1.1$ COV & $94.0\%$ & $93.4\%$ & $91.0\%$ & $94.8\%$ & $96.1\%$ & $92.0\%$ \\
 LEN & $72.4\%$ & $72.0\%$ & $74.9\%$ & $60.0\%$ & $60.1\%$ & $63.1\%$ \\\hline
 $c = 1.2$ COV & $93.8\%$ & $93.3\%$ & $91.1\%$ & $95.0\%$ & $95.6\%$ & $91.5\%$ \\
 LEN & $71.5\%$ & $71.2\%$ & $74.1\%$ & $59.3\%$ & $59.5\%$ & $62.7\%$ \\

\hline\hline
\end{tabular}

\begin{tabular}
[c]{rcccccccc}%
\multicolumn{9}{c}{Table 4A: Design 4, Performance of $\hat{\beta}$ and
$\hat{\gamma}$}\\\hline\hline
& \multicolumn{4}{c}{$n=2500$} & \multicolumn{4}{c}{$n=5000$}\\\cline{2-9}
& $\hat{\beta}_{2}$ & $\hat{\beta}_{3}$ & $\hat{\beta}_{4}$ & $\hat{\gamma}$ & $\hat{\beta}_{2}$  & $\hat{\beta}_{3}$ & $\hat{\beta}_{4}$ & $\hat{\gamma}$ \\\hline

 OY BIAS & $0.3\%$ & $-0.2\%$ & $0.0\%$ & $1.7\%$ & $0.1\%$ & $-0.1\%$ & $-0.9\%$ & $0.4\%$ \\
     STD& $16.5\%$ & $17.1\%$ & $16.8\%$ & $29.3\%$ & $13.7\%$ & $13.6\%$ & $13.8\%$ & $24.3\%$ \\
     MAD& $13.2\%$ & $13.6\%$ & $13.4\%$ & $23.2\%$ & $10.9\%$ & $10.8\%$ & $11.1\%$ & $19.6\%$ \\
    RMSE& $16.5\%$ & $17.1\%$ & $16.7\%$ & $29.3\%$ & $13.7\%$ & $13.6\%$ & $13.8\%$ & $24.3\%$ \\\hline

 HK1 BIAS & $-1.0\%$ & $-0.9\%$ & $-1.5\%$ & $10.4\%$ & $-0.5\%$ & $-1.1\%$ & $-0.5\%$ & $8.7\%$ \\
     STD& $14.1\%$ & $14.1\%$ & $14.3\%$ & $29.5\%$ & $11.4\%$ & $11.4\%$ & $11.3\%$ & $23.5\%$ \\
     MAD& $11.1\%$ & $11.2\%$ & $11.4\%$ & $24.9\%$ & $9.1\%$ & $9.1\%$ & $8.9\%$ & $20.2\%$ \\
    RMSE& $14.1\%$ & $14.1\%$ & $14.3\%$ & $31.2\%$ & $11.4\%$ & $11.4\%$ & $11.3\%$ & $25.0\%$ \\\hline

 HK2 BIAS & $0.4\%$ & $-1.5\%$ & $-0.3\%$ & $11.2\%$ & $0.6\%$ & $-0.5\%$ & $0.1\%$ & $6.7\%$ \\
     STD& $20.5\%$ & $21.9\%$ & $21.8\%$ & $43.6\%$ & $18.9\%$ & $18.6\%$ & $18.6\%$ & $38.7\%$ \\
     MAD& $16.4\%$ & $17.7\%$ & $17.3\%$ & $35.9\%$ & $15.0\%$ & $14.9\%$ & $14.8\%$ & $31.2\%$ \\
    RMSE& $20.5\%$ & $21.9\%$ & $21.8\%$ & $45.0\%$ & $18.9\%$ & $18.6\%$ & $18.6\%$ & $39.2\%$ \\\hline

     & \multicolumn{4}{c}{$n=10000$} & \multicolumn{4}{c}{$n=20000$} \\\cline{2-9}
  & $\hat{\beta}_{2}$ & $\hat{\beta}_{3}$ & $\hat{\beta}_{4}$ & $\hat{\gamma}$ & $\hat{\beta}_{2}$  & $\hat{\beta}_{3}$ & $\hat{\beta}_{4}$ & $\hat{\gamma}$ \\\hline

 OY BIAS & $-0.7\%$ & $0.6\%$ & $0.2\%$ & $2.1\%$ & $-0.3\%$ & $-0.2\%$ & $0.2\%$ & $-0.2\%$ \\
     STD& $10.4\%$ & $10.1\%$ & $10.3\%$ & $19.8\%$ & $8.3\%$ & $8.0\%$ & $8.1\%$ & $15.5\%$ \\
     MAD& $8.3\%$ & $8.1\%$ & $8.3\%$ & $16.1\%$ & $6.8\%$ & $6.5\%$ & $6.4\%$ & $12.4\%$ \\
    RMSE& $10.4\%$ & $10.1\%$ & $10.3\%$ & $19.9\%$ & $8.3\%$ & $8.0\%$ & $8.1\%$ & $15.5\%$ \\\hline

 HK1 BIAS & $-0.3\%$ & $-0.7\%$ & $-0.1\%$ & $6.7\%$ & $0.1\%$ & $-0.8\%$ & $-0.4\%$ & $5.4\%$ \\
     STD& $8.9\%$ & $9.4\%$ & $9.4\%$ & $18.7\%$ & $7.7\%$ & $7.6\%$ & $7.4\%$ & $15.3\%$ \\
     MAD& $7.1\%$ & $7.6\%$ & $7.6\%$ & $15.9\%$ & $6.1\%$ & $6.1\%$ & $5.9\%$ & $12.8\%$ \\
    RMSE& $8.9\%$ & $9.5\%$ & $9.4\%$ & $19.9\%$ & $7.7\%$ & $7.6\%$ & $7.4\%$ & $16.3\%$ \\\hline

 HK2 BIAS & $0.6\%$ & $0.8\%$ & $0.2\%$ & $7.6\%$ & $0.6\%$ & $-0.2\%$ & $-0.0\%$ & $5.0\%$ \\
     STD& $16.1\%$ & $16.4\%$ & $16.5\%$ & $32.4\%$ & $14.2\%$ & $14.2\%$ & $14.1\%$ & $28.7\%$ \\
     MAD& $13.0\%$ & $13.2\%$ & $13.1\%$ & $26.4\%$ & $11.4\%$ & $11.4\%$ & $11.3\%$ & $23.1\%$ \\
    RMSE& $16.1\%$ & $16.4\%$ & $16.4\%$ & $33.2\%$ & $14.2\%$ & $14.2\%$ & $14.1\%$ & $29.1\%$ \\

\hline\hline
\end{tabular}

\begin{tabular}
[c]{rcccccccc}%
\multicolumn{9}{c}{Table 4B: Design 4, Numerical Bootstrap}\\\hline\hline
& \multicolumn{4}{c}{$n=2500$} & \multicolumn{4}{c}{$n=5000$}\\\cline{2-9}
& $\hat{\beta}_{2}$ & $\hat{\beta}_{3}$ & $\hat{\beta}_{4}$ & $\hat{\gamma}$ & $\hat{\beta}_{2}$  & $\hat{\beta}_{3}$ & $\hat{\beta}_{4}$ & $\hat{\gamma}$ \\\hline
 $c = 0.8$ COV & $89.1\%$ & $90.5\%$ & $89.3\%$ & $85.2\%$ & $93.2\%$ & $94.0\%$ & $92.1\%$ & $88.1\%$ \\
 LEN & $111.6\%$ & $111.6\%$ & $111.7\%$ & $111.2\%$ & $95.9\%$ & $95.9\%$ & $96.0\%$ & $97.0\%$ \\\hline
 $c = 0.9$ COV & $88.3\%$ & $88.7\%$ & $88.7\%$ & $84.1\%$ & $92.9\%$ & $93.1\%$ & $92.6\%$ & $86.0\%$ \\
 LEN & $108.1\%$ & $108.0\%$ & $108.0\%$ & $108.0\%$ & $93.7\%$ & $93.6\%$ & $93.6\%$ & $94.5\%$ \\\hline
 $c = 1.0$ COV & $86.7\%$ & $87.9\%$ & $87.9\%$ & $83.0\%$ & $92.0\%$ & $93.3\%$ & $91.5\%$ & $86.2\%$ \\
 LEN & $104.9\%$ & $104.9\%$ & $105.0\%$ & $104.9\%$ & $91.2\%$ & $91.4\%$ & $91.2\%$ & $92.3\%$ \\\hline
 $c = 1.1$ COV & $85.8\%$ & $87.6\%$ & $87.2\%$ & $81.9\%$ & $92.3\%$ & $93.0\%$ & $91.5\%$ & $85.0\%$ \\
 LEN & $102.1\%$ & $102.2\%$ & $102.3\%$ & $102.1\%$ & $89.3\%$ & $89.4\%$ & $89.3\%$ & $90.0\%$ \\\hline
 $c = 1.2$ COV & $86.0\%$ & $86.5\%$ & $86.4\%$ & $81.6\%$ & $91.4\%$ & $91.8\%$ & $91.0\%$ & $84.4\%$ \\
 LEN & $99.7\%$ & $99.6\%$ & $99.6\%$ & $99.6\%$ & $87.4\%$ & $87.3\%$ & $87.6\%$ & $88.2\%$ \\\hline

     & \multicolumn{4}{c}{$n=10000$} & \multicolumn{4}{c}{$n=20000$} \\\cline{2-9}
  & $\hat{\beta}_{2}$ & $\hat{\beta}_{3}$ & $\hat{\beta}_{4}$ & $\hat{\gamma}$ & $\hat{\beta}_{2}$  & $\hat{\beta}_{3}$ & $\hat{\beta}_{4}$ & $\hat{\gamma}$ \\\hline

 $c = 0.8$ COV & $95.7\%$ & $96.0\%$ & $96.0\%$ & $88.7\%$ & $94.6\%$ & $95.9\%$ & $96.6\%$ & $92.0\%$ \\
 LEN & $81.1\%$ & $80.9\%$ & $81.2\%$ & $83.4\%$ & $67.3\%$ & $67.4\%$ & $67.3\%$ & $70.7\%$ \\\hline
 $c = 0.9$ COV & $95.2\%$ & $95.8\%$ & $96.3\%$ & $88.1\%$ & $95.9\%$ & $95.7\%$ & $95.9\%$ & $92.1\%$ \\
 LEN & $79.5\%$ & $79.6\%$ & $79.8\%$ & $81.9\%$ & $66.3\%$ & $66.4\%$ & $66.5\%$ & $69.8\%$ \\\hline
 $c = 1.0$ COV & $95.2\%$ & $95.7\%$ & $95.1\%$ & $88.5\%$ & $96.1\%$ & $96.0\%$ & $96.0\%$ & $91.8\%$ \\
 LEN & $78.4\%$ & $78.3\%$ & $78.4\%$ & $80.5\%$ & $65.7\%$ & $65.7\%$ & $65.7\%$ & $68.9\%$ \\\hline
 $c = 1.1$ COV & $95.0\%$ & $95.6\%$ & $95.1\%$ & $87.9\%$ & $96.4\%$ & $96.5\%$ & $96.1\%$ & $91.5\%$ \\
 LEN & $76.9\%$ & $77.2\%$ & $77.0\%$ & $79.1\%$ & $64.6\%$ & $64.8\%$ & $64.7\%$ & $68.1\%$ \\\hline
 $c = 1.2$ COV & $95.0\%$ & $95.4\%$ & $94.7\%$ & $86.8\%$ & $95.8\%$ & $95.5\%$ & $95.8\%$ & $91.2\%$ \\
 LEN & $75.7\%$ & $75.9\%$ & $75.7\%$ & $77.4\%$ & $64.0\%$ & $64.1\%$ & $64.1\%$ & $66.9\%$ \\
\hline\hline
\end{tabular}

\begin{adjustbox}{width=\textwidth}
\begin{tabular}
[c]{rcccccccccc}%
\multicolumn{11}{c}{Table 5A: Design 5, Performance of $\hat{\beta}$ and
$\hat{\gamma}$}\\\hline\hline
& \multicolumn{5}{c}{$n=2500$} & \multicolumn{5}{c}{$n=5000$}\\\cline{2-11}
& $\hat{\beta}_{2}$ & $\hat{\beta}_{3}$ & $\hat{\beta}_{4}$  & $\hat{\beta}_{5}$ & $\hat{\gamma}$ & $\hat{\beta}_{2}$  & $\hat{\beta}_{3}$ & $\hat{\beta}_{4}$ & $\hat{\beta}_{5}$ & $\hat{\gamma}$ \\\hline
 OY BIAS & $-0.3\%$ & $0.4\%$ & $-0.5\%$ & $0.4\%$ & $1.9\%$ & $0.4\%$ & $-0.2\%$ & $0.3\%$ & $0.4\%$ & $0.3\%$ \\
     STD& $18.5\%$ & $18.5\%$ & $18.1\%$ & $18.9\%$ & $36.4\%$ & $14.5\%$ & $14.1\%$ & $14.7\%$ & $14.8\%$ & $28.6\%$ \\
     MAD& $14.8\%$ & $14.9\%$ & $14.6\%$ & $14.9\%$ & $28.9\%$ & $11.7\%$ & $11.2\%$ & $11.7\%$ & $11.7\%$ & $23.0\%$ \\
    RMSE& $18.5\%$ & $18.5\%$ & $18.1\%$ & $18.9\%$ & $36.5\%$ & $14.5\%$ & $14.1\%$ & $14.7\%$ & $14.8\%$ & $28.5\%$ \\\hline

 HK1 BIAS & $-1.3\%$ & $-1.7\%$ & $-2.2\%$ & $-1.3\%$ & $12.6\%$ & $-1.7\%$ & $-0.6\%$ & $-1.1\%$ & $-1.1\%$ & $12.1\%$ \\
     STD& $16.9\%$ & $17.5\%$ & $18.2\%$ & $18.0\%$ & $34.4\%$ & $14.4\%$ & $14.7\%$ & $14.6\%$ & $14.7\%$ & $29.0\%$ \\
     MAD& $13.5\%$ & $14.1\%$ & $14.4\%$ & $14.2\%$ & $29.2\%$ & $11.6\%$ & $11.8\%$ & $11.8\%$ & $11.8\%$ & $25.1\%$ \\
    RMSE& $17.0\%$ & $17.6\%$ & $18.4\%$ & $18.0\%$ & $36.6\%$ & $14.5\%$ & $14.7\%$ & $14.6\%$ & $14.8\%$ & $31.4\%$ \\\hline

 HK2 BIAS & $-0.5\%$ & $-0.8\%$ & $-1.5\%$ & $0.8\%$ & $14.2\%$ & $-0.7\%$ & $0.7\%$ & $-0.9\%$ & $0.3\%$ & $11.8\%$ \\
     STD& $23.3\%$ & $24.7\%$ & $24.7\%$ & $25.2\%$ & $47.7\%$ & $22.0\%$ & $22.2\%$ & $21.8\%$ & $21.4\%$ & $42.5\%$ \\
     MAD& $18.3\%$ & $19.7\%$ & $19.9\%$ & $20.1\%$ & $39.5\%$ & $17.8\%$ & $17.7\%$ & $17.4\%$ & $17.0\%$ & $35.7\%$ \\
    RMSE& $23.2\%$ & $24.7\%$ & $24.7\%$ & $25.2\%$ & $49.8\%$ & $22.0\%$ & $22.2\%$ & $21.8\%$ & $21.4\%$ & $44.1\%$ \\ \hline

     & \multicolumn{5}{c}{$n=10000$} & \multicolumn{5}{c}{$n=20000$} \\\cline{2-11}
  & $\hat{\beta}_{2}$ & $\hat{\beta}_{3}$ & $\hat{\beta}_{4}$  & $\hat{\beta}_{5}$ & $\hat{\gamma}$ & $\hat{\beta}_{2}$  & $\hat{\beta}_{3}$ & $\hat{\beta}_{4}$ & $\hat{\beta}_{5}$ & $\hat{\gamma}$ \\\hline
OY BIAS & $0.6\%$ & $0.1\%$ & $-0.0\%$ & $-0.0\%$ & $0.8\%$ & $-0.0\%$ & $-0.5\%$ & $0.5\%$ & $0.1\%$ & $0.5\%$ \\
     STD& $11.2\%$ & $11.2\%$ & $12.0\%$ & $11.5\%$ & $22.2\%$ & $9.5\%$ & $9.3\%$ & $9.3\%$ & $9.1\%$ & $18.1\%$ \\
     MAD& $9.0\%$ & $9.1\%$ & $9.7\%$ & $9.2\%$ & $17.9\%$ & $7.5\%$ & $7.4\%$ & $7.4\%$ & $7.4\%$ & $14.5\%$ \\
    RMSE& $11.2\%$ & $11.2\%$ & $12.0\%$ & $11.4\%$ & $22.3\%$ & $9.5\%$ & $9.3\%$ & $9.3\%$ & $9.1\%$ & $18.1\%$ \\\hline

 HK1 BIAS & $-0.4\%$ & $-0.7\%$ & $-0.6\%$ & $-0.9\%$ & $10.7\%$ & $-0.9\%$ & $-0.5\%$ & $-0.1\%$ & $-0.3\%$ & $9.2\%$ \\
     STD& $12.2\%$ & $12.3\%$ & $12.5\%$ & $12.2\%$ & $24.2\%$ & $9.9\%$ & $10.1\%$ & $9.9\%$ & $10.6\%$ & $20.4\%$ \\
     MAD& $9.8\%$ & $10.0\%$ & $9.9\%$ & $9.7\%$ & $21.2\%$ & $7.9\%$ & $8.0\%$ & $7.9\%$ & $8.4\%$ & $17.7\%$ \\
    RMSE& $12.2\%$ & $12.3\%$ & $12.5\%$ & $12.2\%$ & $26.4\%$ & $10.0\%$ & $10.1\%$ & $9.9\%$ & $10.6\%$ & $22.4\%$ \\\hline

 HK2 BIAS & $0.7\%$ & $-0.2\%$ & $0.1\%$ & $1.0\%$ & $8.3\%$ & $-0.6\%$ & $0.3\%$ & $0.8\%$ & $2.1\%$ & $9.3\%$ \\
     STD& $19.1\%$ & $19.3\%$ & $20.1\%$ & $19.4\%$ & $38.9\%$ & $17.1\%$ & $16.8\%$ & $16.6\%$ & $17.6\%$ & $34.6\%$ \\
     MAD& $15.4\%$ & $15.6\%$ & $16.0\%$ & $15.4\%$ & $32.0\%$ & $13.7\%$ & $13.4\%$ & $13.3\%$ & $14.0\%$ & $28.6\%$ \\
    RMSE& $19.1\%$ & $19.3\%$ & $20.1\%$ & $19.4\%$ & $39.8\%$ & $17.1\%$ & $16.8\%$ & $16.6\%$ & $17.7\%$ & $35.8\%$ \\

\hline\hline
\end{tabular}
\end{adjustbox}

\begin{adjustbox}{width=\textwidth}
\begin{tabular}
[c]{rcccccccccc}%
\multicolumn{11}{c}{Table 5B: Design 5, Numerical Bootstrap}\\\hline\hline
& \multicolumn{5}{c}{$n=2500$} & \multicolumn{5}{c}{$n=5000$}\\\cline{2-11}
& $\hat{\beta}_{2}$ & $\hat{\beta}_{3}$ & $\hat{\beta}_{4}$  & $\hat{\beta}_{5}$ & $\hat{\gamma}$ & $\hat{\beta}_{2}$  & $\hat{\beta}_{3}$ & $\hat{\beta}_{4}$ & $\hat{\beta}_{5}$ & $\hat{\gamma}$ \\\hline

 $c = 0.8$ COV & $86.0\%$ & $85.0\%$ & $86.3\%$ & $84.8\%$ & $77.0\%$ & $90.0\%$ & $90.2\%$ & $91.0\%$ & $89.2\%$ & $81.9\%$ \\
 LEN & $113.7\%$ & $113.9\%$ & $113.8\%$ & $113.7\%$ & $113.6\%$ & $99.9\%$ & $99.7\%$ & $99.8\%$ & $99.8\%$ & $100.4\%$ \\\hline
 $c = 0.9$ COV & $85.0\%$ & $84.2\%$ & $85.3\%$ & $83.9\%$ & $75.8\%$ & $89.5\%$ & $89.7\%$ & $90.0\%$ & $88.4\%$ & $81.5\%$ \\
 LEN & $109.7\%$ & $109.6\%$ & $109.7\%$ & $109.7\%$ & $109.8\%$ & $96.8\%$ & $96.5\%$ & $96.9\%$ & $96.8\%$ & $97.3\%$ \\\hline
 $c = 1.0$ COV & $83.6\%$ & $83.2\%$ & $85.0\%$ & $83.1\%$ & $74.8\%$ & $89.1\%$ & $89.2\%$ & $89.0\%$ & $87.7\%$ & $80.2\%$ \\
 LEN & $106.3\%$ & $106.2\%$ & $106.3\%$ & $106.2\%$ & $106.6\%$ & $93.9\%$ & $93.9\%$ & $93.9\%$ & $93.9\%$ & $94.6\%$ \\\hline
 $c = 1.1$ COV & $83.3\%$ & $83.1\%$ & $84.2\%$ & $82.5\%$ & $73.9\%$ & $87.4\%$ & $88.8\%$ & $88.4\%$ & $87.4\%$ & $79.3\%$ \\
 LEN & $103.2\%$ & $103.2\%$ & $103.3\%$ & $103.3\%$ & $103.4\%$ & $91.6\%$ & $91.6\%$ & $91.7\%$ & $91.6\%$ & $92.2\%$ \\\hline
 $c = 1.2$ COV & $82.2\%$ & $82.0\%$ & $83.9\%$ & $81.7\%$ & $73.1\%$ & $86.6\%$ & $87.4\%$ & $87.9\%$ & $86.6\%$ & $77.8\%$ \\
 LEN & $100.3\%$ & $100.3\%$ & $100.4\%$ & $100.4\%$ & $100.7\%$ & $89.2\%$ & $89.3\%$ & $89.1\%$ & $89.2\%$ & $90.0\%$ \\\hline

     & \multicolumn{5}{c}{$n=10000$} & \multicolumn{5}{c}{$n=20000$} \\\cline{2-11}
  & $\hat{\beta}_{2}$ & $\hat{\beta}_{3}$ & $\hat{\beta}_{4}$  & $\hat{\beta}_{5}$ & $\hat{\gamma}$ & $\hat{\beta}_{2}$  & $\hat{\beta}_{3}$ & $\hat{\beta}_{4}$ & $\hat{\beta}_{5}$ & $\hat{\gamma}$ \\\hline
$c = 0.8$ COV & $94.6\%$ & $95.2\%$ & $94.6\%$ & $94.9\%$ & $86.9\%$ & $96.0\%$ & $97.3\%$ & $97.0\%$ & $97.2\%$ & $90.0\%$ \\
 LEN & $85.7\%$ & $85.8\%$ & $85.8\%$ & $85.9\%$ & $88.0\%$ & $71.8\%$ & $71.9\%$ & $72.0\%$ & $71.9\%$ & $75.8\%$ \\\hline
 $c = 0.9$ COV & $94.3\%$ & $94.6\%$ & $94.2\%$ & $94.5\%$ & $86.7\%$ & $97.0\%$ & $97.1\%$ & $97.2\%$ & $96.4\%$ & $88.8\%$ \\
 LEN & $83.7\%$ & $83.7\%$ & $83.8\%$ & $83.6\%$ & $85.6\%$ & $70.7\%$ & $70.7\%$ & $70.9\%$ & $70.7\%$ & $74.4\%$ \\\hline
 $c = 1.0$ COV & $93.8\%$ & $93.5\%$ & $93.9\%$ & $94.4\%$ & $86.2\%$ & $96.3\%$ & $96.7\%$ & $97.2\%$ & $96.8\%$ & $89.3\%$ \\
 LEN & $81.9\%$ & $81.9\%$ & $81.9\%$ & $81.8\%$ & $83.6\%$ & $69.6\%$ & $69.5\%$ & $69.7\%$ & $69.6\%$ & $72.9\%$ \\\hline
 $c = 1.1$ COV & $93.2\%$ & $92.8\%$ & $93.6\%$ & $93.8\%$ & $86.1\%$ & $95.6\%$ & $96.8\%$ & $96.7\%$ & $96.5\%$ & $88.7\%$ \\
 LEN & $80.1\%$ & $80.1\%$ & $80.0\%$ & $80.0\%$ & $81.6\%$ & $68.6\%$ & $68.4\%$ & $68.6\%$ & $68.3\%$ & $71.3\%$ \\\hline
 $c = 1.2$ COV & $92.7\%$ & $92.4\%$ & $93.2\%$ & $93.1\%$ & $84.8\%$ & $95.6\%$ & $96.4\%$ & $96.1\%$ & $96.3\%$ & $88.8\%$ \\
 LEN & $78.4\%$ & $78.3\%$ & $78.4\%$ & $78.4\%$ & $79.8\%$ & $67.6\%$ & $67.5\%$ & $67.4\%$ & $67.4\%$ & $70.2\%$ \\

\hline\hline
\end{tabular}
\end{adjustbox}

\end{center}

\subsection{Additional Monte Carlo Experiments: Designs 6--8}\label{app:add_simulation}

To investigate the impact of serial correlations of $x_{it}$ on our estimator and the
proposed inference procedure, we conduct additional simulations for several designs.
These designs are the same as in Designs 3--5, except for that we
allow for $x$ to be auto-correlated, similar to Design 2, as opposed to
Design 1.

We consider Designs 6--8, which employ the same models as those in
Designs 3--5, respectively, but with serially dependent $x_{it}$. More
specifically:
\begin{eqnarray*}
x_{i0,j} &=&\frac{\sqrt{15}}{4}u_{it,j}+\frac{1}{4}u_{it,k+1},\text{ }%
j=1,2,...,k\text{ and} \\
x_{it,j} &=&0.25x_{it-1,j}+\sqrt{1-0.25^{2}}\left( \frac{\sqrt{15}}{4}%
u_{it,j}+\frac{1}{4}u_{it,k+1}\right) ,j=1,2,...,k\text{ for all }t\geq 1,
\end{eqnarray*}%
with $\left( u_{it,1},u_{it,2},...,u_{it,k+1}\right) $ distributed as $%
N\left( 0_{(k+1)\times 1},I_{(k+1)\times (k+1)}\right) $, and $\left(
u_{it,1},u_{it,2},...,u_{it,k+1}\right) $ being i.i.d. across $i$ and $t$.
The parameter $k$ is set to be 3, 4, or 5 in Designs 6 through 8, respectively.

To conserve space, we only report the inference results for $c=1.$ We
report the BIAS, STD, MAD, RMSE, COV, and LEN for our estimators.  All
results are collected into one table for each design. The tables are
numbered corresponding to the names of the designs. For instance, results
for Design 6 are presented in Table 6.

A brief summary of our findings is as follows: Our estimation results seem
to remain relatively unchanged with serially correlated $x.$ Importantly,
they are not significantly biased. The inference procedure performs well,
maintaining the same level of performance as designs with serially independent $x_{it}$.

\begin{center}
\begin{tabular}
[c]{rcccccc}%
\multicolumn{7}{c}{Table 6: Design 6}\\\hline\hline
& \multicolumn{3}{c}{$n=2500$} & \multicolumn{3}{c}{$n=5000$}\\\cline{2-7}
& $\hat{\beta}_{2}$ & $\hat{\beta}_{3}$ & $\hat{\gamma}$ & $\hat{\beta}_{2}$  & $\hat{\beta}_{3}$ & $\hat{\gamma}$ \\\hline
 OY BIAS & $-0.8\%$ & $0.5\%$ & $2.5\%$ & $0.8\%$ & $-0.4\%$ & $2.1\%$ \\
     STD& $15.4\%$ & $15.1\%$ & $25.2\%$ & $11.2\%$ & $11.6\%$ & $20.3\%$ \\
     MAD& $12.4\%$ & $12.2\%$ & $20.0\%$ & $9.0\%$ & $9.4\%$ & $16.3\%$ \\
    RMSE& $15.4\%$ & $15.1\%$ & $25.3\%$ & $11.2\%$ & $11.6\%$ & $20.4\%$ \\
COV & $90.7\%$ & $89.8\%$ & $87.4\%$ & $93.0\%$ & $92.5\%$ & $90.7\%$ \\
 LEN & $100.9\%$ & $101.3\%$ & $100.6\%$ & $86.5\%$ & $86.3\%$ & $87.1\%$ \\\hline

     & \multicolumn{3}{c}{$n=10000$} & \multicolumn{3}{c}{$n=20000$} \\\cline{2-7}
  & $\hat{\beta}_{2}$ & $\hat{\beta}_{3}$ & $\hat{\gamma}$ & $\hat{\beta}_{2}$  & $\hat{\beta}_{3}$ & $\hat{\gamma}$ \\\hline
 OY BIAS & $0.4\%$ & $-0.5\%$ & $2.8\%$ & $-0.2\%$ & $-0.1\%$ & $1.1\%$ \\
     STD& $9.4\%$ & $9.9\%$ & $16.5\%$ & $7.6\%$ & $7.6\%$ & $13.2\%$ \\
     MAD& $7.4\%$ & $8.0\%$ & $13.4\%$ & $6.1\%$ & $6.1\%$ & $10.5\%$ \\
    RMSE& $9.4\%$ & $9.9\%$ & $16.7\%$ & $7.6\%$ & $7.6\%$ & $13.2\%$ \\
COV & $95.7\%$ & $92.9\%$ & $92.7\%$ & $93.3\%$ & $93.6\%$ & $92.1\%$ \\
 LEN & $73.0\%$ & $72.7\%$ & $74.0\%$ & $60.6\%$ & $60.7\%$ & $61.2\%$ \\
\hline\hline
\end{tabular}

\begin{tabular}
[c]{rcccccccc}%
\multicolumn{9}{c}{Table 7: Design 7}\\\hline\hline
& \multicolumn{4}{c}{$n=2500$} & \multicolumn{4}{c}{$n=5000$}\\\cline{2-9}
& $\hat{\beta}_{2}$ & $\hat{\beta}_{3}$ & $\hat{\beta}_{4}$ & $\hat{\gamma}$ & $\hat{\beta}_{2}$  & $\hat{\beta}_{3}$ & $\hat{\beta}_{4}$ & $\hat{\gamma}$ \\\hline

 OY BIAS & $0.2\%$ & $-0.6\%$ & $0.6\%$ & $3.1\%$ & $-0.5\%$ & $-0.1\%$ & $0.3\%$ & $1.7\%$ \\
     STD& $16.1\%$ & $16.3\%$ & $16.9\%$ & $29.8\%$ & $13.4\%$ & $13.6\%$ & $13.5\%$ & $23.5\%$ \\
     MAD& $13.2\%$ & $13.1\%$ & $13.5\%$ & $23.9\%$ & $10.7\%$ & $10.9\%$ & $10.9\%$ & $19.1\%$ \\
    RMSE& $16.1\%$ & $16.3\%$ & $16.9\%$ & $29.9\%$ & $13.4\%$ & $13.6\%$ & $13.5\%$ & $23.6\%$ \\
COV & $87.0\%$ & $87.7\%$ & $86.7\%$ & $80.5\%$ & $90.8\%$ & $90.5\%$ & $90.9\%$ & $85.8\%$ \\
 LEN & $104.7\%$ & $104.8\%$ & $104.7\%$ & $104.2\%$ & $91.3\%$ & $91.3\%$ & $91.3\%$ & $91.4\%$ \\\hline

     & \multicolumn{4}{c}{$n=10000$} & \multicolumn{4}{c}{$n=20000$} \\\cline{2-9}
  & $\hat{\beta}_{2}$ & $\hat{\beta}_{3}$ & $\hat{\beta}_{4}$ & $\hat{\gamma}$ & $\hat{\beta}_{2}$  & $\hat{\beta}_{3}$ & $\hat{\beta}_{4}$ & $\hat{\gamma}$ \\\hline

 OY BIAS & $-0.1\%$ & $-0.1\%$ & $-0.1\%$ & $2.8\%$ & $-0.5\%$ & $0.5\%$ & $0.1\%$ & $2.1\%$ \\
     STD& $10.9\%$ & $10.7\%$ & $10.6\%$ & $19.2\%$ & $8.4\%$ & $8.2\%$ & $8.5\%$ & $15.0\%$ \\
     MAD& $8.8\%$ & $8.5\%$ & $8.4\%$ & $15.6\%$ & $6.8\%$ & $6.6\%$ & $6.9\%$ & $12.0\%$ \\
    RMSE& $10.9\%$ & $10.7\%$ & $10.6\%$ & $19.4\%$ & $8.5\%$ & $8.3\%$ & $8.5\%$ & $15.1\%$ \\
COV & $94.3\%$ & $93.5\%$ & $94.4\%$ & $89.8\%$ & $95.7\%$ & $95.9\%$ & $96.1\%$ & $91.9\%$ \\
 LEN & $78.5\%$ & $78.1\%$ & $78.2\%$ & $79.0\%$ & $65.4\%$ & $65.7\%$ & $65.6\%$ & $67.3\%$ \\

\hline\hline
\end{tabular}

\begin{adjustbox}{width=\textwidth}
\begin{tabular}
[c]{rcccccccccc}%
\multicolumn{11}{c}{Table 8: Design 8}\\\hline\hline
& \multicolumn{5}{c}{$n=2500$} & \multicolumn{5}{c}{$n=5000$}\\\cline{2-11}
& $\hat{\beta}_{2}$ & $\hat{\beta}_{3}$ & $\hat{\beta}_{4}$  & $\hat{\beta}_{5}$ & $\hat{\gamma}$ & $\hat{\beta}_{2}$  & $\hat{\beta}_{3}$ & $\hat{\beta}_{4}$ & $\hat{\beta}_{5}$ & $\hat{\gamma}$ \\\hline
 OY BIAS & $-1.7\%$ & $1.1\%$ & $0.1\%$ & $-0.1\%$ & $6.4\%$ & $0.1\%$ & $-0.0\%$ & $0.3\%$ & $-0.2\%$ & $2.2\%$ \\
     STD& $18.8\%$ & $18.6\%$ & $18.0\%$ & $18.1\%$ & $32.8\%$ & $14.7\%$ & $15.2\%$ & $14.6\%$ & $15.1\%$ & $27.3\%$ \\
     MAD& $15.0\%$ & $15.2\%$ & $14.2\%$ & $14.5\%$ & $26.4\%$ & $11.8\%$ & $12.1\%$ & $11.6\%$ & $12.2\%$ & $22.2\%$ \\
    RMSE& $18.8\%$ & $18.6\%$ & $18.0\%$ & $18.1\%$ & $33.4\%$ & $14.7\%$ & $15.2\%$ & $14.6\%$ & $15.1\%$ & $27.4\%$ \\
COV & $84.1\%$ & $82.5\%$ & $83.3\%$ & $83.8\%$ & $78.1\%$ & $88.5\%$ & $89.4\%$ & $89.6\%$ & $89.2\%$ & $83.0\%$ \\
 LEN & $106.2\%$ & $106.2\%$ & $106.0\%$ & $106.2\%$ & $106.0\%$ & $93.6\%$ & $93.7\%$ & $93.7\%$ & $93.7\%$ & $94.0\%$ \\ \hline

     & \multicolumn{5}{c}{$n=10000$} & \multicolumn{5}{c}{$n=20000$} \\\cline{2-11}
  & $\hat{\beta}_{2}$ & $\hat{\beta}_{3}$ & $\hat{\beta}_{4}$  & $\hat{\beta}_{5}$ & $\hat{\gamma}$ & $\hat{\beta}_{2}$  & $\hat{\beta}_{3}$ & $\hat{\beta}_{4}$ & $\hat{\beta}_{5}$ & $\hat{\gamma}$ \\\hline
OY BIAS & $0.3\%$ & $-0.2\%$ & $0.3\%$ & $-0.2\%$ & $2.9\%$ & $-0.2\%$ & $-0.1\%$ & $-0.2\%$ & $0.6\%$ & $1.3\%$ \\
     STD& $11.5\%$ & $11.7\%$ & $11.5\%$ & $11.8\%$ & $21.1\%$ & $9.4\%$ & $9.3\%$ & $8.9\%$ & $9.0\%$ & $16.2\%$ \\
     MAD& $9.2\%$ & $9.3\%$ & $9.1\%$ & $9.4\%$ & $16.9\%$ & $7.5\%$ & $7.4\%$ & $7.2\%$ & $7.2\%$ & $13.1\%$ \\
    RMSE& $11.5\%$ & $11.6\%$ & $11.5\%$ & $11.8\%$ & $21.3\%$ & $9.4\%$ & $9.3\%$ & $8.9\%$ & $9.0\%$ & $16.3\%$ \\
COV & $93.4\%$ & $94.6\%$ & $93.8\%$ & $92.3\%$ & $87.4\%$ & $95.7\%$ & $96.3\%$ & $96.8\%$ & $96.4\%$ & $92.2\%$ \\
 LEN & $81.8\%$ & $81.8\%$ & $81.6\%$ & $81.7\%$ & $82.5\%$ & $69.6\%$ & $69.5\%$ & $69.5\%$ & $69.5\%$ & $71.8\%$ \\
\hline
\hline
\end{tabular}
\end{adjustbox}
\end{center}

\bibliography{biblist}
\end{document}